\documentclass[oneside]{dissertation}

\pgfkeys{/dissertation/author={Louie Sinadjan}}
\pgfkeys{/dissertation/degree={BSc}}
\pgfkeys{/dissertation/unit={COMS30045}}
\pgfkeys{/dissertation/title={GPU Acceleration for Faster Evolutionary\\Spatial Cyclic Game Systems}}

\usepackage{tabularx} 
\usepackage{graphicx}
\usepackage{caption}
\usepackage{subcaption}
\usepackage{amsmath,amssymb}
\usepackage{tikz}
\usepackage{xcolor}
\usepackage{geometry}
\usepackage{float}
\usepackage{tcolorbox}
\tcbuselibrary{skins}
\usepackage{pdflscape}

\usepackage{listings}
\usepackage{xcolor}

\definecolor{metalgreen}{rgb}{0,0.5,0}
\definecolor{metalblue}{rgb}{0,0,0.7}
\definecolor{metalgrey}{rgb}{0.4,0.4,0.4}

\lstdefinestyle{metalstyle}{
    language=C++,
    basicstyle=\ttfamily\small,
    keywordstyle=\color{metalblue}\bfseries,
    commentstyle=\color{metalgreen}\itshape,
    stringstyle=\color{metalgrey},
    numbers=left,
    numberstyle=\tiny\color{gray},
    numbersep=5pt,
    showstringspaces=false,
    breaklines=true,
    frame=single,
    tabsize=4,
    morekeywords={device, thread, constant, kernel, uint, atomic_int}
}
\lstdefinestyle{cudastyle}{
    language=C++,
    basicstyle=\ttfamily\small,
    keywordstyle=\color{blue}\bfseries,
    commentstyle=\color{gray}\itshape,
    stringstyle=\color{orange},
    numberstyle=\tiny\color{gray},
    numbers=left,
    numbersep=5pt,
    tabsize=4,
    showspaces=false,
    showstringspaces=false,
    showtabs=false,
    frame=single,
    breaklines=true,
    breakatwhitespace=true,
    captionpos=b,
    morekeywords={__global__, __device__, __shared__, __host__, __syncthreads, curandState, curand_uniform, atomicAdd, atomicExch, cudaMemcpy, cudaMalloc, cudaFree, cudaStream_t, cudaMemcpyDeviceToHost, cudaMemcpyHostToDevice, cudaStreamCreate, cudaStreamSynchronize, cudaError_t}
}

\graphicspath{ {./figures/} }
\usepackage[width=0.8\textwidth, font=small, labelfont=bf]{caption}
\begin{document}





\maketitle

\vspace{2cm}
\begin{center}
    Copyright \textcopyright~2025 Louie Sinadjan
\end{center}


\frontmatter




\chapter*{Abstract}
This dissertation presents the design, implementation and evaluation of GPU-accelerated simulation frameworks for Evolutionary Spatial Cyclic Games (ESCGs), a class of agent-based models used to study ecological and evolutionary dynamics. Traditional single-threaded ESCG simulations are computationally expensive and scale poorly. To address this, high-performance implementations were developed using Apple's Metal and Nvidia's CUDA, with a validated single-threaded C++ version serving as a baseline comparison point. \\

Benchmarking results show that GPU acceleration delivers significant speedups, with the CUDA \texttt{maxStep} implementation achieving up to a $\approx28\times$ improvement. Larger system sizes, up to $3200\times3200$, became tractable, while Metal faced scalability limits. The GPU frameworks also enabled replication and critical extension of recent ESCG studies, revealing sensitivities to system size and runtime not fully explored in prior work. \\

Overall, this project provides a configurable ESCG simulation platform that advances the computational toolkit for this field of research. This dissertation forms the basis for two papers in preparation for submission to international conferences and journals.


\chapter*{Dedication and Acknowledgements}

I would like to thank my family and friends, for supporting me throughout my university journey. \\

\noindent
I would also like to thank my supervisor, Dave Cliff, for his guidance and inspiration. \\

\vspace{1cm} 

\noindent



\makedecl



\makeaidecl



\tableofcontents
\listoffigures
\listoftables


\chapter*{Ethics Statement}

    \begin{itemize}
        \item This project did not require ethical review, as determined by my supervisor
        \item This project fits within the scope of ethics application 0026, as reviewed by my supervisor
        \item An ethics application for this project was reviewed and approved by the faculty research ethics committee as application.
    \end{itemize}
    

\chapter*{Supporting Technologies}

\begin{quote}
\noindent
\begin{itemize}
    \item I used the \texttt{matplotlib-cpp} library to generate runtime graphs and visualisations directly from C++ code, particularly for plotting species densities and extinction probabilities.
    \item I used Nvidia's CUDA platform to implement GPU-accelerated versions of the ESCG simulation, utilising CUDA libraries such as \texttt{cuRAND} for pseudorandom number generation and \texttt{CUDA Streams} for asynchronous kernel execution and concurrency.
    \item I used the C++ Standard Library’s \texttt{<random>} module, specifically the \texttt{std::mt19937} Mersenne Twister engine, as the pseudorandom number generator in the single-threaded ESCG simulation, and to generate initial seeds for pseudorandom number generators in the GPU-accelerated Metal and CUDA implementations.
    \item I used the Metal Shading Language (MSL) and the Metal API to implement GPU-accelerated ESCG simulations on Apple Silicon devices, writing custom shaders for compute operations and pipelines. 
    \item I used an Apple MacBook Pro (M1 Pro, 2021) to develop and test Metal-based simulations, and an Nvidia RTX A2000 GPU for CUDA-based simulation development and benchmarking.
    \item For post-processing and advanced visualisation, I used Python’s \texttt{matplotlib} and \texttt{seaborn} library to create high-quality plots from simulation outputs exported in CSV format.
\end{itemize}
\end{quote}


\chapter*{Notation and Acronyms}

\begin{tabbing}
ESCG \hspace{1cm} \= : \hspace{0.5cm} \= Evolutionary Spatial Cyclic Game \\
RPS \> : \> Rock Paper Scissors \\
RPSLS \> : \> Rock Paper Scissors Lizard Spock \\

\\

MCS \> : \> Monte Carlo Step \\

\\

GPU \> : \> Graphics Processing Unit \\
CPU \> : \> Central Processing Unit \\
ALU \> : \> Arithmetic Logic Unit \\
SIMD \> : \> Single Instruction Multiple Data \\
SIMT \> : \> Single Instruction Multiple Threads \\
JIT \> : \> Just-in-time \\

\\

HPC \> : \> High-performance computing \\

\\

CUDA \> : \> Compute Unified Device Architecture \\
PXT \> : \> Parallel Thread Execution \\
CUBIN \> : \> CUDA Binary \\

\\

MSL \> : \> Metal Shading Language \\
.air \> : \> Apple Intermediate Representation Files \\
PSO \> : \> Pipeline State Object \\

\\

PRNG \> : \> Pseudorandom Number Generation \\

\end{tabbing}




%

\mainmatter

\chapter{Introduction}
\label{chap:context}
\section{Context}

Evolutionary Spatial Cyclic Games (ESCGs) are simulations designed to model interactions among competing species within a spatially structured environment. In these simulations, species engage in cyclic dominance relationships, similar to the familiar game of rock-paper-scissors, where each species can dominate at least one other species but is vulnerable to being dominated by at least one other species. Researchers have increasingly utilised ESCGs to investigate the underlying mechanisms responsible for ecosystem stability and biodiversity. Park et al. \cite{Park2023CompetitionPopulation} posed a fundamental question: 

\begin{center}
    \textit{ If a species is stronger and beats the other one then how can the "weaker" species survive?} 
\end{center}

Questions of similar nature have catalysed extensive research, resulting in an expanding body of literature dedicated to exploring spatial evolutionary dynamics, species interactions, and the conditions that promote or undermine ecological balance. Historically, this field was predominantly theoretical, relying on stochastic calculus, the Gause-Lotka-Volterra (GLV) equations, and Ginzburg-Landau equations to identify stable equilibrium points \cite{Reichenbach2007MobilityGames}. However, the emergence of spatially explicit models, first pioneered in 1975 by Laird and Schamp~\cite{Laird2006CompetitiveCoexistence}, has enabled a significant shift toward more experimental and computational approaches in ecological research. \textit{Spatially explicit models} describe systems in which individuals occupy discrete locations on a structured grid (or lattice), with each cell representing either an empty site or an individual agent. The species identity of each agent governs its behavioural dynamics and interactions with neighbouring agents, allowing complex spatial and ecological patterns to emerge from local rules. Building on these foundations, extended models have been developed to capture the complexity of real-world ecosystems. Notably, extended variations such as the Rock-Paper-Scissors-Lizard-Spock (RPSLS) model, popularised by the television series \textit{The Big Bang Theory}, and other formulations incorporating variable species-counts and modified dominance networks better represent the complexities of nature and have significantly enriched our understanding of competitive coexistence, ecological stability, and biodiversity. \newline

\begin{figure}[h]
    \centering
    \begin{subfigure}[]{0.32\textwidth}
        \includegraphics[width=\textwidth]{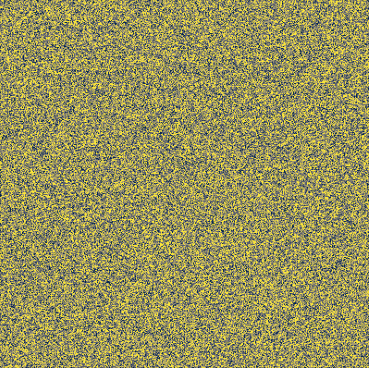}
        \caption{MCS = 0}
        \label{fig:spirals1}
    \end{subfigure}
    \hfill
    \begin{subfigure}[]{0.32\textwidth}
        \includegraphics[width=\textwidth]{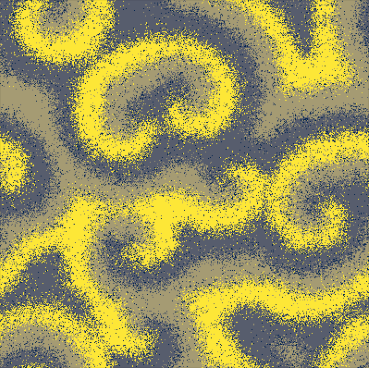}
        \caption{MCS = 20000}
        \label{fig:spirals2}
    \end{subfigure}
    \hfill
    \begin{subfigure}[]{0.32\textwidth}
        \includegraphics[width=\textwidth]{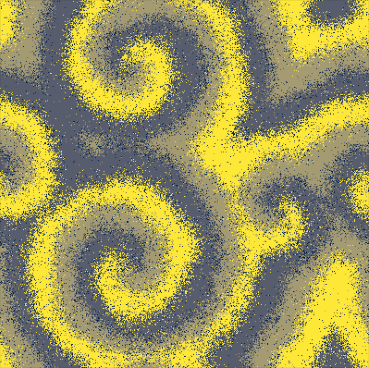}
        \caption{MCS = 95000}
         \end{subfigure}
    
    \caption{Snapshots from accelerated lattice simulation replicating the ESCG of Reichenbach, Mobilia, and Frey~\cite{Reichenbach2007MobilityGames}. 
    Shown at different Monte Carlo Steps (MCS). Parameters: lattice size = $400\times400$, initial empty cell probability = $0.1$, mobility = $3\times10^{-5}$, von-Neumann neighbourhood, circulant dominance of 3 species, periodic boundary conditions. 
    Code available at: \url{<https://github.com/louiesinadjan/escg>}.}
    \label{fig:spirals}
\end{figure}

For example, Reichenbach, Mobilia, and Frey \cite{Reichenbach2007MobilityGames} investigated an ESCG consisting of three cyclically dominant species interacting via the classic Rock-Paper-Scissors model. It is demonstrated through their lattice simulations, and later replicated and accelerated in this dissertation, that at sufficiently low levels of individual mobility, species coexistence emerges through distinct spiral-wave patterns (see \textbf{Figure \ref{fig:spirals}}). Such spatial arrangements closely mirror phenomena observed in biological systems, including myxobacterial aggregations \cite{Igoshin2004WavesAggregation} and spiral wave formations in \textit{Dictyostelium} mounds \cite{Siegert1995SpiralMounds}.

\section{Motivation}
\label{motivation}
    
ESCG researchers frequently use Monte Carlo simulations due to their effectiveness in capturing the inherent randomness of ecological processes. These stochastic simulations enable realistic modelling of species interactions, including reproduction, migration, and competition, and allow for statistical analysis of emergent spatial patterns and coexistence dynamics. Typically, a Monte Carlo step (MCS) consists of $N$ \textit{elementary steps}, where $N$ denotes the total number of cells in the system, and an elementary step involves selecting an individual agent and updating it according to predefined rules. To achieve statistically robust results, researchers commonly conduct multiple independent and identically distributed (IID) simulation trials. For instance, a modest lattice size of $100\times100$ involves $10,000$ elementary steps per MCS. Simulations frequently require durations of up to 100,000 MCS for stable dynamics to emerge. Conducting numerous IID trials to ensure statistical reliability further multiplies computational demands. Exploring different parameter configurations can increase the computational workload by an additional order of magnitude, potentially reaching as high as $10^{11}$ elementary steps or more for comprehensive experimentation. Such computational demands quickly become prohibitive even at relatively small lattice sizes, calling into question the reliability of conclusions drawn from limited-duration simulations. Zhong et al.\cite{Zhong2022SpeciesSpecies} performed ESCG simulations for up to $10^{5}$ MCS, concluding this duration was sufficient to reveal asymptotic behaviour. However, subsequent work by Cliff demonstrated these conclusions were premature \cite{Cliff2025OnNetworks, CliffGraphicalNetworks}, identifying purported steady-states as transient phenomena that decayed when simulations extended to $10^{7}$ MCS. Cliff’s findings illustrate critical shortcomings of contemporary ESCG experiments, specifically inadequate simulation durations and insufficient lattice sizes, highlighting the urgent need for improved computational approaches. \newline

This dissertation addresses the computational limitations inherent in traditional single-threaded ESCG simulations by leveraging GPU acceleration to significantly enhance performance. GPUs, being designed and optimised for massively parallel tasks, are particularly well-suited to the repetitive, lattice-based computations that define ESCGs. By exploiting this parallelism, the research facilitates extensive exploration of larger lattice sizes and longer simulation durations, overcoming previous computational constraints and enabling richer, more statistically significant analyses of ESCG phenomena. \newline

In pursuing this goal, careful consideration was given to the choice of GPU architectures. This dissertation explores two distinct platforms: Apple Silicon GPUs, programmed using the Metal Shading Language (MSL), and Nvidia GPUs, programmed using the CUDA framework. This multi-platform approach enables a broader evaluation of GPU-accelerated ESCG performance across different hardware.

\noindent
\textbf{The primary aims and objectives of this research are summarised as follows:}
\begin{enumerate}
    \item To implement and validate single-threaded (in C++), Metal, and CUDA-based ESCG simulation frameworks, ensuring accuracy and consistency across all platforms through replication of known ecological dynamics from existing literature.
    \item To deliver a comprehensive, flexible, and user-configurable simulation framework capable of supporting future research and experimentation in ESCGs, facilitating deeper insights into complex ecological and evolutionary dynamics.
    \item To evaluate and quantify the performance gains achieved through GPU acceleration compared to conventional single-threaded simulations, assessing scalability and throughput across varying lattice sizes and parameter configurations.
    \item To challenge existing literature by extending previously published ESCG research to significantly longer simulation durations and larger lattice sizes, enabling the examination of ecological dynamics beyond currently documented limits.
\end{enumerate}

As will be demonstrated in the remainder of this dissertation, these primary aims and objectives have been achieved. Details of the relevant technical and historical background are presented in \textbf{Chapter~\ref{chap:background}}, and \textbf{Chapter~\ref{chap:execution}} then explains the full details of my implementations of ESCGs on both Apple and Nvidia GPUs. In \textbf{Chapter~\ref{chap:evaluation}}, I present a critical evaluation of my GPU-accelerated ESCG implementations, including replication and extension of simulation results previously reported by other researchers and published in leading journals for nonlinear science such as \emph{Chaos, Solitons \& Fractals}. \\

The key findings of my comparative evaluations show that my GPU implementations are capable of achieving up to a $\sim28\times$ speedup compared to a directly comparable single-threaded baseline. These GPU systems represent a clear advance over the current state of the art in this field. A paper summarising their design, implementation, and evaluation is in preparation and will be submitted to the \emph{European Modelling and Simulation Symposium} (EMSS) to be held in Fez, Morocco, in September 2025 (submission deadline: 15\textsuperscript{th} May 2025). \\

Furthermore, my GPU-accelerated implementations have enabled the replication and critical extension of a recent ESCG study published by Park, Chen, and Szolnoki~\cite{Park2023CompetitionPopulation} in \emph{Chaos, Solitons \& Fractals}. In collaboration with my supervisor, Professor Dave Cliff, we have identified a major methodological vulnerability in Park et al.'s study. We are currently in the late stages of preparing a joint-authored paper that will be submitted to \emph{Chaos, Solitons \& Fractals} before the end of May 2025. \textbf{Appendices~\ref{emss-paper}} and \textbf{\ref{csf-paper}} present the abstracts of these two forthcoming publications. \\

Professor Dave Cliff has expressed full confidence that both papers will be accepted for publication in due course. Consequently, we expect that this dissertation will form the basis of two peer-reviewed international conference and journal papers:

\begin{itemize}
    \item L. Sinadjan \& D. Cliff (2025), ``GPU-Based Simulation of Evolutionary Spatial Cyclic Games: Nvidia vs Apple Silicon'', in \emph{Proceedings of the 2025 European Modelling \& Simulation Symposium}.
    \item D. Cliff \& L. Sinadjan (2025), ``Mobility Matters in a Cyclically Dominant Eight-Species Model of Competing Alliances'', in \emph{Chaos, Solitons \& Fractals}.
\end{itemize}


\chapter{Background}
\label{chap:background}
\section{Current Implementations of ESCGs}

The existing literature in this field offers limited insight into programmatic implementation, with most publications describing simulations through purely verbal descriptions of the relevant algorithms. This often introduces ambiguity in parameter settings, hindering reliable replication and accurate reproduction of published results. Cliff provided a detailed pseudocode algorithmic description of an ESCG \cite{Cliff2024NeverGames}, for which he also made available source code written in Python. While this represents a significant advancement in transparency and reproducibility within ESCG research, there is substantial qualitative difference of simulation descriptions  across publications, complicating direct comparisons between studies - which is exemplified by Cliff's paper proving previously published results to be wrong \cite{Cliff2025OnNetworks}. However, it is widely known that Python is fundamentally a slow executing programming language due to its interpreted nature and dynamic typing, which adds type checking at runtime. As previously discussed, a single ESCG experiment involving generation of reliable results can require $10^{11}$ elementary steps. Despite the low complexity of each individual step, the immense total of computation can require multiple of days of continuous runtime in a Python-implemented system. Other compiled languages such as C++ generally outperform Python as they are compiled directly into machine code, allowing for optimised execution without the overhead of interpretation. C++ achieves its speed through a combination of features, including manual memory management, which gives programmers precise control over resource allocation and deallocation. Moreover, C++ supports low-level programming close to the hardware while still offering high-level abstractions, enabling developers to write efficient and optimised code. Modern C++ compilers also apply aggressive optimisations, such as inlining, loop unrolling, and constant folding, which significantly reduce runtime overhead. \newline

C++ further distinguishes itself by offering extensive GPU programming capabilities through APIs such as CUDA \cite{Cook2012CUDAGPUs}, OpenCL \cite{Banger2013OpenCLExample}, and Metal \cite{appleMetalCPP}. CUDA, developed by Nvidia, allows developers to write highly parallelised code that executes directly on Nvidia GPUs. Metal serves a similar purpose within the Apple ecosystem, providing low-level access to GPU compute and graphics pipelines. By interfacing closely with these APIs, C++ enables precise control over memory management, thread scheduling, and data transfer between host and device. This fine-grained control is crucial for optimising latency and throughput, which is why C++ is widely adopted in domains such as high-frequency trading, scientific simulation, machine learning, and real-time rendering, where performance demands are exceptionally high. \newline

For this research, I chose C++ as the foundational programming language, leveraging Metal and CUDA for GPU acceleration on Apple and Nvidia hardware respectively. This choice aimed to achieve substantial performance improvements, enabling more extensive and efficient simulations than previously feasible.

\section{GPUs}

Historically, Graphics Processing Units (GPUs) mainly served the obvious purpose - graphics. Tasks in computer graphics such as rasterisation, 3D to 2D transformations, shading techniques such as Phong and Gouraud shading \cite{Watt1992AdvancedPractice}, and many others are inherently matrix-heavy, involving numerous linear algebra operations to manipulate geometry, lighting, and textures efficiently. Consequently, GPUs were engineered with highly parallel architectures to perform these computations swiftly and efficiently. Over time, however, the demand for GPU-driven parallel processing has significantly expanded beyond graphics applications. Today, the GPU market is among the fastest-growing sectors in technology, driven largely by their integral role in accelerating advancements in artificial intelligence, machine learning, and even cryptocurrency mining, as well as their broader adoption for general-purpose parallel computing tasks. \newline

To effectively leverage GPUs for ESCG acceleration, understanding some of their fundamental features is essential. The following subsections discuss key GPU concepts relevant to this dissertation:

\subsection{Parallel Processing}

A defining trait of GPUs is their capability to complete parallel tasks. While Central Processing Units (CPUs) typically feature a low number of powerful cores that are optimised for sequential tasks, a single GPU can consist of thousands of lightweight cores that operate simultaneously. This architecture is highly effective for workloads that can be decomposed into numerous smaller, independent tasks, a property known as data parallelism. \newline

In the context of GPU computing, the CPU is referred to as the \textit{host}, responsible for launching and managing GPU tasks, while the GPU is referred to as the \textit{device}, which performs the actual data-parallel computations. The term \textit{kernel} refers to the function that is executed on the GPU in parallel by many threads. Kernels are launched from the host code, which specifies how many threads should execute the kernel and how they should be grouped. This division of labour enables the host to handle control flow, data transfer, and task orchestration, while the device focusses on intensive parallel computation. \newline

At the heart of GPU parallelism are \textit{cores} and \textit{threads}. The word \textit{core} is not universally defined, and different companies advertise their GPU cores with different meaning. For example, the Nvidia Jetson Nano, which will be discussed in more detail later in this dissertation, is equipped with 128 CUDA cores, while the Nvidia RTX 5090 (the most recent flagship in Nvidia’s GPU lineup) boasts an immense 24,575 CUDA cores. By contrast, the Apple Silicon MacBook Pro (2021, M1 Pro) features 14 GPU cores. At first glance, these figures may suggest that the Jetson Nano would vastly outperform the M1 Pro in computational tasks. However, this assumption does not hold true. The discrepancy lies in the differing definitions and implementations of the term \textit{core}. In the context of Nvidia GPUs, a CUDA core is a lightweight execution unit that processes a single thread. This contributes to the GPU’s SIMD-style (Single Instruction, Multiple Data) parallel execution model, which Nvidia refers to as SIMT (Single Instruction, Multiple Threads). Conversely, Apple’s GPU cores, while fewer in number, are more akin to compute clusters, each comprising multiple arithmetic logic units (ALUs) or execution units. Apple does not publicly disclose the exact number of these lower-level components, making direct comparisons difficult. Thus, it becomes evident that core count alone, particularly when comparing across architectures from different manufacturers, is not a reliable indicator of overall performance. \newline

In GPU programming, a \textit{thread} is the smallest unit of execution, responsible for performing a specific set of instructions on a subset of data. Threads are designed to operate independently, each with its own local memory and execution context. By default, threads do not share memory with one another. However, threads can be organised into groups, within which they can share a limited amount of memory and synchronise with each other to coordinate their work. A common pattern in GPU programming involves launching a large number of threads that all execute the same instructions in parallel, but on different pieces of data. This approach allows for significant speedup when dealing with large-scale problems, as the combined results of many threads can be aggregated efficiently. Such a model is particularly well-suited for simulations like ESCGs, where operations can be independently applied across a grid.

\subsection{Atomics}

In parallel programming, a common issue known as a data race arises when multiple threads or tasks access a shared resource concurrently without proper synchronisation, leading to undefined or unpredictable behaviour. This problem is particularly relevant in the context of ESCGs, where the outcome of an elementary step depends not only on the selected cell but also on the state of one of its neighbours. When attempting to parallelise this process by executing multiple elementary steps simultaneously, it becomes likely that different threads will operate on neighbouring or overlapping cells. This can result in concurrent reads and writes to the same memory locations, causing race conditions that compromise the accuracy and determinism of the simulation. \newline

An atomic operation is a low-level synchronisation primitive that ensures that a specific memory operation, such as a read-modify-write, completes entirely without interference from other threads. This guarantees that when multiple threads attempt to update the same memory location concurrently, each operation is executed in a serialised manner, preserving correctness and preventing data races. Atomics are typically used for tasks such as incrementing counters, updating shared values, or performing reductions where threads must safely coordinate access to shared data. While locks and mutexes were considered as a potential solution, they are generally unsuitable for GPU architectures due to their high overhead and the increased risk of thread contention and deadlocks in massively parallel workloads.\footnote{Locks and mutexes are synchronisation mechanisms that enforce exclusive access to shared resources by allowing only one thread to access a resource at a time. In massively parallel environments like GPUs, using locks can cause \textit{thread contention}, where multiple threads attempt to acquire the same lock simultaneously, resulting in delays. This can escalate into \textit{deadlocks}, situations where two or more threads become permanently blocked, each waiting for a resource held by the other.} Consequently, such synchronisation mechanisms are typically unsupported by GPU programming frameworks such as CUDA and Metal. In contrast, atomic operations are lightweight, hardware-supported instructions that guarantee indivisible memory transactions, ensuring correctness when multiple threads concurrently access the same memory address. Specifically, when several threads simultaneously attempt to read from or write to a shared memory location, atomic operations enforce sequential execution at the hardware level, preventing race conditions and inconsistent states. Due to their low overhead and efficiency, atomics were leveraged extensively in my code to safely coordinate concurrent access to grid cells without compromising performance or correctness.

\section{Apple Silicon \& Metal}

In 2014, Apple emerged in the GPU computing space with the introduction of Metal, a low-overhead graphics and compute API. Years later, in the 2020 Apple WWDC (Worldwide Developers Conference), Apple announced that they would from that year onwards be moving from Intel chips to their own ARM-powered silicon chips \cite{Warren2020AppleYear}. With this, Metal evolved to support general-purpose GPU computing across Apple's ecosystem, including macOS and iPadOS, and became the primary and exclusive GPU programming interface for Apple devices. In the context of this dissertation, Metal is used as the parallel computing interface for experiments conducted on Apple Silicon hardware. \newline

As Metal was originally designed for graphics rendering, its compute components inherit terminology and structure from graphics pipelines. In Metal Shading Language (MSL), compute functions are referred to as \textit{shaders}, and are written in \texttt{.metal} files using a strongly typed syntax similar to C++14. These shader files are compiled into intermediate \texttt{.air} (Apple Intermediate Representation) files, which are then linked into final \texttt{.metallib} binary libraries. These libraries can be loaded by host code via the Metal API. Metal’s API is accessible from host-side C++ or Objective-C++ (\texttt{.mm}) code, allowing seamless integration with C++ simulation backends. A typical workflow for integrating and executing Metal shaders within host-side C++ or Objective-C++ code involves the following steps:

\begin{center}
\begin{minipage}{0.85\linewidth}
\begin{enumerate}
    \item Obtain a reference to the GPU hardware using a \texttt{MTLDevice} object.
    \item Load compiled shaders from a \texttt{.metallib} binary file into a \texttt{MTLLibrary} object.
    \item Retrieve specific compute shader functions from the \texttt{MTLLibrary} as \texttt{MTLFunction} objects.
    \item Create a precompiled, optimised pipeline state object (\texttt{MTLComputePipelineState}) from the shader function.
    \item Instantiate a \texttt{MTLCommandQueue} to schedule and manage execution commands.
    \item For each shader invocation:
    \begin{enumerate}
        \item Create a \texttt{MTLCommandBuffer} from the command queue.
        \item Encode the shader invocation into a \texttt{MTLComputeCommandEncoder}, specifying pipeline state and GPU resources (buffers, textures).
        \item Dispatch threads by specifying thread grid and threadgroup sizes.
        \item Finalise encoding and commit the command buffer for GPU execution.
    \end{enumerate}
\end{enumerate}
\end{minipage}
\end{center}

\section{Nvidia \& CUDA}

In 2006, Nvidia introduced CUDA (Compute Unified Device Architecture), a parallel computing platform and programming model that enabled developers to harness the general-purpose computational capabilities of Nvidia GPUs. Over time, CUDA has become the dominant platform for GPU computing, and Nvidia has established itself as the market leader in both consumer and data-centre-level parallel computing hardware. The company’s GPU product lines, notably the GeForce RTX 30, 40, and now 50 series, have continued to push the boundaries of throughput and efficiency in graphics and compute workloads. Additionally, Nvidia provides a range of developer-friendly devices, such as the Jetson Nano, Jetson Xavier, and Jetson Orin kits, which offer embedded access to CUDA-capable GPUs for edge computing, robotics, and AI applications. Thousands of applications developed with CUDA have been deployed to GPUs in embedded systems, workstations, data centres, and cloud environments, including high-profile industry software such as Adobe Premiere Pro and Autodesk Maya \cite{NvidiaCUDAZone}, which utilise CUDA to accelerate video rendering, simulation, and real-time effects. In the context of this dissertation, CUDA is employed as the primary GPU programming interface for experiments conducted on Nvidia hardware and an Nvidia RTX A2000 was used for most CUDA experimentation. \newline

CUDA was originally developed as a low-level parallel computing architecture tailored specifically for Nvidia GPUs. Its compute functions are referred to as \textit{kernels}, and are written in \texttt{.cu} files using an extended subset of C/C++ with Nvidia-specific qualifiers (e.g., \texttt{\_\_global\_\_}, \texttt{\_\_device\_\_}, \texttt{\_\_shared\_\_}). These source files are compiled by Nvidia’s \texttt{nvcc} compiler into either \textit{PTX} (Parallel Thread Execution) intermediate representation or directly into \textit{CUBIN} (CUDA binary) format. PTX can be linked into fat binaries or dynamically loaded by host code using the CUDA Runtime API or the CUDA Driver API. CUDA host code is typically written in C++ and manages GPU resources, memory transfers, and kernel launches. A standard CUDA execution flow involves:

\noindent
\begin{center}
\begin{minipage}{0.85\linewidth}
\begin{enumerate}
    \item Obtaining a reference to the desired GPU via cudaSetDevice().
    \item Allocating and transferring memory between the CPU and GPU using cudaMalloc() and cudaMemcpy().
    \item Compiling and loading kernel binaries.
    \item Launching the kernel with an explicit grid/block configuration using the \texttt{<<<...>>>} syntax.
\end{enumerate}
\end{minipage}
\end{center}

\section{Metal \& CUDA Differences}

While both Metal and CUDA enable high-performance parallel computing on GPUs, they are built for different ecosystems and follow distinct design philosophies. CUDA, developed by Nvidia, is a mature and widely adopted general-purpose GPU platform, while Metal, designed by Apple, began as a graphics API and later expanded to support compute tasks within Apple’s ecosystem. This section outlines some of the core differences between them.

\subsection{Memory Management}

In Metal, memory allocation is handled via the \texttt{newBufferWithLength()} method, which abstracts hardware-specific details and simplifies resource creation. On Apple Silicon, the unified memory architecture allows the CPU and GPU to share physical memory, reducing the need for explicit data transfers and enabling efficient access to shared data. However, resource behaviour depends on the selected storage mode. Metal provides three main storage modes: \texttt{MTLStorageModeShared}, \texttt{MTLStorageModePrivate}, and \texttt{MTLStorageModeMemoryless}. \texttt{Shared} mode allows both CPU and GPU access and is ideal for data updated by the CPU. \texttt{Private} mode restricts access to the GPU and is preferred for render targets or resources written exclusively by GPU workloads. \texttt{Memoryless} mode, used only for transient textures such as depth or stencil buffers, stores data in fast, low-power tile memory within the GPU. These modes offer flexibility in performance tuning and are made possible by Apple’s unified memory model. However, despite shared memory, developers must still initialise GPU-accessible buffer objects with appropriate sizes and explicitly populate them using functions like \texttt{std::memcpy(..., buffer->contents(), ...)} to ensure data is correctly prepared for GPU access \cite{Apple2023MetalStorage}. \newline

CUDA requires explicit and manual memory management. Developers must use \texttt{cudaMalloc()} to allocate device memory and \texttt{\-cudaMemcpy()} to transfer data between the host and GPU. This workflow is especially relevant on platforms such as the Jetson Nano and other Nvidia devices, which feature a dedicated GPU with its own memory, separate from the CPU. While this architecture enables high-bandwidth memory access for GPU computations, it introduces additional latency for host-device transfers and limits the size of available GPU memory, a notable drawback compared to unified memory. Consequently, although CUDA offers greater control and optimisation potential, it also increases development complexity and the risk of performance bottlenecks in memory-intensive applications. \newline

A key difference in the way memory is managed programmatically in Metal is its use of an \texttt{@autoreleasepool} to handle the lifecycle of short-lived GPU resources. Wrapping Metal command buffer creation and submission in an \texttt{@autoreleasepool} ensures timely deallocation of temporary objects like \texttt{MTLCommandBuffer}, which can otherwise accumulate and cause increased memory consumption over time. CUDA, by contrast, requires developers to manage memory explicitly using functions such as \texttt{cudaMalloc()} for allocation and \texttt{cudaFree()} for deallocation. Proper use of \texttt{cudaFree()} is essential to avoid memory leaks and resource exhaustion over time, especially in iterative or long-running GPU workloads. While this offers greater control, it also increases the complexity of memory management and the potential for programming errors. \newline

In summary, Metal’s memory management model prioritises developer simplicity and automatic handling of memory across compute units, whereas CUDA offers more explicit control and optimisation potential at the cost of increased complexity. Metal is well-suited to applications where ease of development and tight hardware integration are critical, while CUDA is preferred in scenarios demanding fine-grained memory control and maximum performance on heterogeneous GPU workloads.

\subsection{Threading \& Grid Striding}

In the CUDA thread hierarchy, the highest level unit of abstraction is the \textit{thread grid}, which are comprised of \textit{blocks}, where each thread block contains a 1D, 2D, or 3D arrangement of threads, identified by \texttt{threadIdx}, and blocks themselves are indexed by \texttt{blockIdx}. The total number of threads launched is defined by \texttt{gridDim} (grid dimension) and \texttt{blockDim} (block dimension). This hierarchical design enables fine-grained control over how computation and memory are structured. Shared memory is explicitly declared within a thread block using the \texttt{\_\_shared\_\_} keyword, allowing fast memory exchange between threads in the same block and global memory is accessible to all threads, but incurs higher latency. \newline

Metal adopts a simpler threading model. Threads are grouped into \textit{threadgroups}, which are similar to CUDA thread blocks. Each thread within a group has a unique \texttt{thread\_position\_in\_threadgroup}, and threadgroups themselves are indexed using \texttt{threadgroup\_position\_in\_grid}. MSL allows the use of the \texttt{threadgroup} address space to allocate memory that is shared within a threadgroup, offering functionality similar to CUDA’s shared memory. However, Metal abstracts more of the low-level configuration, which simplifies deployment at the cost of fine-grained control.\newline

A common technique used in both CUDA and Metal is \textit{grid striding}, which allows threads to process data sets larger than the number of threads dispatched. In CUDA, this is typically expressed as:

\begin{center}
\begin{verbatim}
for (int i = threadIdx.x + blockIdx.x * blockDim.x;
         i < N;
         i += blockDim.x * gridDim.x) {
    // process element i
}
\end{verbatim}
\end{center}

This pattern ensures that each thread starts at a unique index and steps forward by the total number of threads in the grid, enabling efficient coverage of large data sets. An equivalent can be achieved in Metal using \texttt{thread\_position\_in\_grid} and looping with a stride equal to the total thread count. \newline

It is also possible to synchronise threads within blocks (CUDA) or threadgroups (MSL). This is made possible using \texttt{\_\_syncthreads()} in CUDA and \texttt{threadgroup\_barrier()} in MSL, which are blocking functions that wait until all threads have reached that point in the code. These synchronisation points can be crucial when implementing algorithms that rely on shared memory or staged computation.

\subsection{Programming Model and Shader Integration}

In CUDA, both the host (CPU) and device (GPU) code can be written within the same \texttt{.cu} source file. This unified development model allows developers to write kernel functions marked with qualifiers such as \texttt{\_\_global\_\_} or \texttt{\_\_device\_\_}, and invoke them directly from the host code using the triple-angle-bracket syntax (e.g., \texttt{kernel<<<...>>>()}). This close coupling simplifies project structure and facilitates tighter coordination between CPU and GPU logic. CUDA extends standard C++ with additional features to support parallel execution, including vector types, memory space qualifiers, and built-in thread/block indexing variables.\newline

Metal adopts a more modular and separated approach. GPU code must be written in separate \texttt{.metal} files using the MSL, which is a strongly typed, C++14-like language tailored for GPU programming. These files must be compiled ahead of time into intermediate \texttt{.air} representations and then linked into \texttt{.metallib} binaries that the host application loads at runtime. This separation enforces a clearer distinction between CPU and GPU responsibilities but also requires additional build and resource management steps. MSL supports compute, vertex, and fragment functions, and uses attributes like \texttt{[[buffer(n)]]}, \texttt{[[thread\_position\_in\_grid]]}, and \texttt{[[threadgroup]]} to specify memory and execution behaviour. \newline

Additionally, CUDA allows for just-in-time (JIT) compilation of kernels via NVRTC or PTX for dynamic runtime control, whereas Metal is more restrictive, requiring shader code to be precompiled and bundled with the application. This can affect flexibility in applications where dynamic GPU code generation is required.\footnote{NVRTC (Nvidia Runtime Compilation) is a library that enables the compilation of CUDA C++ device code at runtime, allowing applications to generate and compile GPU kernels dynamically. PTX (Parallel Thread Execution) is an intermediate low-level virtual machine and assembly language for CUDA, which can also be JIT-compiled into executable GPU code at runtime.} Overall, CUDA provides a more integrated and flexible development model suited to scientific and high-performance computing (HPC) workloads, while Metal encourages structured separation of concerns.

\subsection{Concurrency and Execution Control}

Both CUDA and Metal support concurrent GPU execution and provide mechanisms for low-latency, high-throughput workloads through the use of command queues, synchronisation primitives, and atomic operations. \newline

Metal uses a different concurrency model based on command buffers and encoders. During execution, commands are encoded into a \texttt{MTLCommandBuffer} and submitted to a \texttt{MTLCommandQueue}, which is conceptually similar to a CUDA stream. Each command buffer encapsulates a sequence of GPU tasks, which are executed in the order submitted. Compute operations are defined using a \texttt{MTLComputeCommandEncoder}, and synchronisation is typically handled implicitly by the ordering of command buffers, though explicit synchronisation options exist. Pipeline State Objects (PSOs) are a core optimisation mechanism in Metal. A \texttt{MTLComputePipelineState} encapsulates the compiled configuration for a particular compute function, avoiding runtime compilation overhead. According to Apple's 2023 WWDC \cite{AppleInc.2023Target3}, PSOs are cached internally by Metal, allowing subsequent calls to reuse precompiled binaries and significantly reduce dispatch latency. Metal also supports atomic operations via the \texttt{atomic\_} types in the Metal Shading Language, enabling safe updates to shared memory locations across threads within a threadgroup or on global memory. \newline

CUDA offers fine-grained concurrency control via \textit{streams}, which serve as queues for asynchronous operations on the GPU. A stream allows kernels, memory transfers, and other commands to execute without blocking other streams. Developers can launch multiple kernels or memory operations in different streams to overlap computation and data movement. Synchronisation can be enforced using \texttt{cudaStreamSynchronize()} to wait for a specific stream to complete, or \texttt{cudaDeviceSynchronize()} to block until all GPU tasks finish. CUDA also supports atomic operations on shared or global memory, enabling coordination between threads during parallel execution. While CUDA does not employ pipeline state objects (PSOs) in the same way as Metal, it achieves similar performance optimisations through just-in-time compilation, kernel caching, and the reuse of persistent driver-level compilation artefacts across program executions. \newline

Overall, both frameworks provide robust support for concurrency, where CUDA offers more explicit control over kernel scheduling and synchronisation through streams, and Metal favours a command-buffer-based approach.

\subsection{Ease of Use}

The development experience differs notably between CUDA and Metal, particularly in terms of tooling support, debugging ease, and available documentation. CUDA benefits from mature tooling and broad adoption across academia and industry. Its integration with widely used integrated development environments (IDEs) and build systems, along with extensive documentation and community support, makes it relatively approachable for developers entering the space of general-purpose GPU computing. Conversely, Metal’s ecosystem is tightly integrated with Apple’s development tools, especially Apple's proprietary IDE - \emph{Xcode}. While this can offer a streamlined experience, such as automatic linking of frameworks and well-integrated project templates, it also leads to friction outside of the Apple ecosystem. Metal development is poorly supported in third-party IDEs, and its command-line tooling and error messaging can be difficult to interpret without deep familiarity with Apple’s development stack. \newline

A key usability limitation in Metal is the absence of built-in debugging features such as console output within shaders. Unlike CUDA, where \texttt{printf()} is available in device code (with some limitations), Metal shaders do not support direct logging. Developers must instead create buffers, assign variables to those buffers within \texttt{.metal} files, and read them back on the CPU side, a tedious process for debugging simple values. Furthermore, Metal’s documentation and online community support remains heavily oriented toward graphics and rendering pipelines, making it difficult to find relevant guidance or examples for general-purpose compute applications. \newline

Both frameworks share certain challenges inherent to low-level GPU programming. There is no automatic bounds check when accessing buffers or shared memory, and improper memory allocation, such as failing to allocate sufficient space or transferring excessive data, can silently lead to undefined behaviour or difficult-to-diagnose performance issues. In CUDA, while explicit memory management allows for fine control, it also increases the risk of bugs, memory leaks, and developer overhead. In Metal, memory handling is abstracted but can still cause issues if resource limits are misunderstood or under-provisioned. \newline

Overall, CUDA offers a more mature, flexible, and well-supported environment for general-purpose GPU programming, particularly for scientific or high-performance workloads. Metal provides a more constrained but tightly integrated experience, better suited to developers working exclusively within Apple’s ecosystem and willing to trade some flexibility for streamlined deployment.

\newpage
\section{ESCG Performance Bottlenecks}

\textbf{The Rules} \newline

A typical elementary step in an ESCG, outlined in studies such as \cite{Zhong2022SpeciesSpecies, Reichenbach2007MobilityGames}, involves selecting a random cell and one of its neighbours, followed by randomly choosing one of three actions: interaction (i.e., competition), reproduction, or migration. The elementary step is described in greater detail in \textbf{Algorithm~\ref{alg:es}} within \textbf{Chapter~\ref{chap:execution}}. These actions are selected based on predefined probabilities $\mu$, $\sigma$, and $\epsilon$, respectively, via a roulette wheel mechanism. While each individual step requires minimal computation, the sheer scale of simulations, often involving billions or even trillions of elementary steps, makes the overall process computationally expensive. \newline

\noindent\textbf{Random Number Generation} \newline

Among the primary bottlenecks in this workflow is the overhead introduced by random number generation, which is needed for each cell, neighbour, and action selection at every step. To support the stochastic nature of ESCGs, simulations rely heavily on pseudorandom number generators (PRNGs). These algorithms produce sequences of numbers that approximate the properties of true randomness, which are essential for probabilistic cell, neighbour, and action selection. Among various PRNGs, the Mersenne Twister algorithm has been celebrated as simple yet elegant, and is one of the most widely used due to its long period ($2^{19937} - 1$), high equidistribution, and efficient generation of high-quality random numbers \cite{matsumoto1998mersenne}. Its robustness and speed make it a common choice in scientific simulations, including ecological and evolutionary models like ESCGs. Another widely utilised PRNG is CURAND, a library provided as part of Nvidia’s CUDA toolkit. CURAND offers a suite of high-performance, GPU-accelerated random number generators that are optimised for parallel execution, making it particularly well-suited for large-scale simulations like ESCGs. It provides a simple and convenient API for generating uniform, normal, and other distributions directly on the GPU, reducing data transfer overhead and enabling efficient random number generation at scale. \newline

In current implementations, random numbers are generated on demand in a serial fashion. While this approach is straightforward, it becomes a significant performance bottleneck when simulations require billions of samples. A GPU-accelerated implementation can generate large batches of random numbers at runtime in parallel and store them in arrays. During the simulation, values can be retrieved from these arrays as needed, effectively reducing the computational overhead of generating each number on demand. This strategy leverages the massive parallelism of GPUs to produce random values quickly and shifts the cost of generation to a single efficient step. As a result, simulations such as ESCGs can benefit from reduced latency and significantly faster execution while maintaining the stochastic fidelity of their underlying models. \newline

\noindent\textbf{Take More Than One Step at a Time} \newline

As it stands, all papers exploring ESCGs describe one elementary step as potentially updating one cell and its neighbour. This naturally leads to all implementations processing one step at a time, one cell and its neighbour at a time. However, this approach incurs extensive runtimes for simulations of large lattices reaching high MCS. Due to the random nature of cell selection, multiple disjoint random cells can be selected and processed simultaneously, effectively executing multiple elementary steps in parallel while preserving the stochastic properties of the simulation. This approach, however, prompts an obvious concern: what if overlapping cells are selected to be processed together?

My initial attempt to address this problem involved a purely algorithmic solution without leveraging any GPU-specific programming constructs. As outlined in \textbf{Algorithm \ref{alg:independentcells}}, my approach involved iterating over a randomly generated sequence of cell indices and constructing a set of candidate cells deemed safe for concurrent processing. For each candidate cell, its neighbouring indices were computed and checked against the current set. If neither the cell nor its neighbours existed in the set, it was added; otherwise, the iteration was terminated. While conceptually sound, this method introduced significant computational overhead due to the neighbour checks and set operations, ultimately rendering its execution times slower than the baseline serial implementation.

\begin{center}
    \begin{algorithm}[H]
    \caption{Independent Cells Check}
    \label{alg:independentcells}
    \SetKwFunction{Independent}{independentCells}
    \SetKwProg{Fn}{Function}{:}{}
    \Fn{\Independent{}}{
        $row \gets nextCell \,/\,L$,\quad $col \gets nextCell \,\mod\, L$\;
        $conflictZone \gets \{\,nextCell\,\}$\tcp*[r]{Add current cell to set}
        $conflictZone \gets conflictZone \cup \{\,((row \pm 1) \,\mod\, L),\ ((col \pm 1) \,\mod\, L)\,\}$\tcp*[r]{Add VN neighbours}
        \ForEach{$n \in conflictZone$}{
            \If{$n \in neighbours$}{\Return \textbf{false}\;}
        }
        $neighbours \gets neighbours \cup conflictZone$\;
        \Return \textbf{true}\;
    }
    \end{algorithm}
\end{center}

Instead, atomics offer a practical and performant solution to handling overlapping updates. Atomics maintain the probabilistic rules governing the simulation without the need for complex synchronisation logic or conflict-resolution heuristics. Each thread independently attempts its update based on its own random draw of cell, neighbour, and action. The atomic instructions ensure that, even if multiple threads target the same region of the grid, the final state is still consistent with the stochastic model, as only one write will successfully complete for each contested memory address. Consequently, the simulation remains unbiased, and emergent spatial patterns evolve correctly over time. \newline

Moreover, this approach enables a substantial performance gain. By allowing thousands of GPU threads to simultaneously attempt updates, the simulation can scale to process tens or hundreds of thousands of elementary steps in a single kernel invocation. Although some threads may serialise due to atomic conflicts, the vast majority of non-overlapping updates proceed in parallel, yielding a net speed up over fully serial execution. Thus, atomic operations strike an effective balance between preserving model fidelity and leveraging the massive parallelism offered by GPU architectures.

 
\chapter{Project Execution}
\label{chap:execution}

\section{Single Threaded ESCG in C++}

To establish a reliable foundation for accelerating existing ESCG systems, a single-threaded ESCG simulation was first implemented in C++. This served both to reproduce results from current literature with accuracy and to provide a baseline structure and performance benchmark for subsequent GPU-accelerated versions.

\subsection{Implementation}

\begin{table}[]
\centering
\caption{Command-line Configurable Parameters and Default Values}
\label{tab:cli-vars}
\begin{tabular}{|l|l|l|}
\hline
\textbf{CLI Flag} & \textbf{Description} & \textbf{Default Value} \\
\hline
\texttt{--length}          & Length of the lattice                    & 200 \\
\texttt{--height}          & Height of the lattice                    & 200 \\
\texttt{--mcs}             & Monte Carlo Step Limit                   & 100000 \\
\texttt{--neighbourhood}   & Neighbourhood type (4 or 8-way)          & 4 \\
\texttt{--printFrequency}  & MCS interval to print density counts     & 200 \\
\texttt{--mobility}        & Mobility of an individual                & $3 \times 10^{-5}$ \\
\texttt{--species}         & Number of species in the simulation      & 3 \\
\texttt{--flux}            & Wrap boundary condition                  & true \\
\texttt{--empty}           & Initial empty cell probability           & 0.0 \\
\texttt{--save}            & Export snapshots to \texttt{.png}        & false \\
\texttt{--dominance}       & Import dominance matrix from \texttt{.csv} & false \\
\hline
\end{tabular}
\end{table}

The program architecture begins with a dedicated configuration header file, \texttt{config.hpp}, which defines a globally accessible \texttt{struct} containing all parameters configurable at runtime via the command line interface. These parameters, summarised in Table~\ref{tab:cli-vars}, allow users to control the simulation setup, including lattice dimensions (\texttt{--length}, \texttt{--height}), MCS limits (\texttt{--mcs}), neighbourhood type (\texttt{--neighbourhood}), and output frequency to the console (\texttt{--printFrequency}). Biological parameters such as mobility rate (\texttt{--mobility}), number of species (\texttt{--species}), and initial empty cell probability (\texttt{--empty}) are also customisable. Additional flags configure boundary conditions (\texttt{--flux}), enable periodic snapshot saving (\texttt{--save}), and specify whether a dominance matrix should be imported (\texttt{--dominance}). \\

Upon execution, these parameters are populated through command-line arguments using the \texttt{parseArgs()} function . This function utilises the \texttt{getopt\_long} interface to efficiently map each CLI flag to the corresponding field in the configuration structure. Default values are assigned where arguments are omitted, enabling both flexibility and reproducibility in experiments. The motivation behind designing a highly configurable program was to enable users to reproduce results from a wide range of published ESCG studies within a single system, while also supporting custom experimentation with minimal code modification. Once initialised, the parameters are passed into core simulation functions, influencing key behaviours such as lattice size, neighbourhood interaction type, mobility, species count, and MCS limits. \newline

Due to the highly configurable nature of the program, where lattice dimensions are set at runtime, the grid is stored as a 1D pointer (\texttt{int*}). Unlike some C compilers, C++ does not support variable-length arrays on the stack, necessitating the use of dynamic allocation and although the grid represents a 2D spatial structure, storing it as a 1D array provides significant practical advantages. First, it ensures a contiguous memory layout, improving cache performance and enabling more efficient memory operations. In contrast, 2D arrays, especially those implemented as pointer-to-pointer constructs (\texttt{int**}), often result in non-contiguous memory, introducing performance penalties and complicating memory transfers. Additionally, components of the system including visualisation utilities (and later -- MSL kernels), are designed to accept flat buffers. Despite being linear in memory, 2D indexing and neighbour referencing is still supported by mapping row-column coordinates to a 1D index using the formula:

\[
\begin{aligned}
    \texttt{index} = \texttt{row} \times L + \texttt{col}
\end{aligned}
\]

\noindent where \(L\) is the length of the lattice. Neighbouring cells can then be accessed by applying modular arithmetic to wrap around the edges, as shown below:

\[
\begin{aligned}
    \texttt{up} &= ((\texttt{row} - 1 + L) \mod L) \times L + \texttt{col} \\
    \texttt{down} &= ((\texttt{row} + 1) \mod L) \times L + \texttt{col} \\
    \texttt{left} &= \texttt{row} \times L + ((\texttt{col} - 1 + L) \mod L) \\
    \texttt{right} &= \texttt{row} \times L + ((\texttt{col} + 1) \mod L)
\end{aligned}
\]

\noindent This approach ensures correct neighbour indexing under periodic boundary conditions. \newline

At the start of the simulation, the grid is initialised with a uniform distribution of species, where each occupied cell is randomly assigned a species label from the range \([1, s]\), where \(s\) is the user-specified number of species. Additionally, the user may define the probability that a given cell is initialised as empty, in which case the cell is assigned the value $0$. Each species is represented as an integer \(i \in \{1, \dots, s\}\), while empty cells are denoted by $0$, allowing for straightforward encoding and efficient processing within the simulation logic. \newline

Like the grid, the dominance relationships between species are conceptually represented as a 2D adjacency matrix but are stored as a 1D pointer (\texttt{int*}) in memory. This design mirrors the grid’s storage approach and offers similar benefits. While one could argue that storing the dominance rules is unnecessary, since they can be hardcoded using conditional logic, this design choice enables greater flexibility. By parameterising the dominance matrix, the program allows users to define arbitrary interaction rules at runtime without modifying the source code. This supports reproducibility, facilitates comparison between studies, and promotes custom experimentation with varying ecological dynamics. \newline

Each entry in the flattened dominance matrix indicates whether one species dominates another. Since species values are 1-indexed in the simulation (i.e., the first row in the matrix is index 0 but refers to Species 1 as 0 refers to an empty site), the outcome of a species interaction is determined by \textbf{Algorithm \ref{alg:dominates}} (where \textit{species} denotes the species of the randomly selected cell, \textit{neighbour} denotes the species of the randomly selected neighbouring cell, \textit{speciesNum} specifies the total number of species in the simulation, and \textit{dominance} represents the flattened dominance matrix that defines the interaction outcomes between species)..\newline

\begin{algorithm}[H]
\caption{Dominance Check}
\label{alg:dominates}
\SetKwFunction{Dominates}{dominates}
\SetKwProg{Fn}{Function}{:}{}
\Fn{\Dominates{species, neighbour, speciesNum, dominance}}{
    \If{species = 0 \textbf{or} neighbour = 0}{
        \Return \textbf{false} \tcp*[r]{Empty cells cannot dominate}
    }
    \Return \texttt{dominance}$[((species - 1) \times speciesNum + (neighbour - 1))] == 1$\;
}
\end{algorithm}

\noindent
\newline
If the user opts to import dominance data, a matrix is loaded from a \texttt{dominance.csv} file and mapped onto the 1D dominance pointer described previously. If no file is provided, the program defaults to generating a circulant dominance matrix. A circulant graph is characterised by an adjacency matrix in which each row is a cyclic permutation of the one preceding it. In the context of ESCGs, this results in a structured and repeatable pattern of species interactions, where dominance relationships are defined by a specified set of positional offsets rather than a single fixed value. \newline

Formally, for a system with \( S \) species and a set of dominance offsets \( K = \{k_1, k_2, \dots, k_m\} \), the dominance matrix \( D \) is defined as:

\[
\begin{aligned}
D[i][j] =
\begin{cases}
1, & \text{if } (j - i + S) \bmod S \in K \\
0, & \text{otherwise}.
\end{cases}
\end{aligned}
\]

\noindent Here, \( D[i][j] = 1 \) indicates that species \( i \) dominates species \( j \), with each offset in the set \( K \) defining a forward cyclic dominance relation. The resulting dominance matrix \( D \) can be represented as a circulant graph \( C(S, K) \). \\

\begin{figure}[h]
\centering
\begin{tikzpicture}[
    node distance=3cm,
    every node/.style={circle, draw=black, fill=white, font=\small, minimum size=1.4cm},
    ->, >=stealth
]

\node (R) at (90:2cm) {Rock};
\node (S) at (162:2cm) {Scissors};
\node (L) at (234:2cm) {Lizard};
\node (Sp) at (306:2cm) {Spock};
\node (P) at (18:2cm) {Paper};

\draw (R) -- (S);    
\draw (R) -- (L);    

\draw (S) -- (P);    
\draw (S) -- (L);    

\draw (P) -- (R);    
\draw (P) -- (Sp);   

\draw (L) -- (Sp);   
\draw (L) -- (P);    

\draw (Sp) -- (S);   
\draw (Sp) -- (R);   

\end{tikzpicture}
\caption{Circulant dominance graph for Rock-Paper-Scissors-Lizard-Spock (RPSLS)}
\label{fig:rpsls-circulant}
\end{figure}
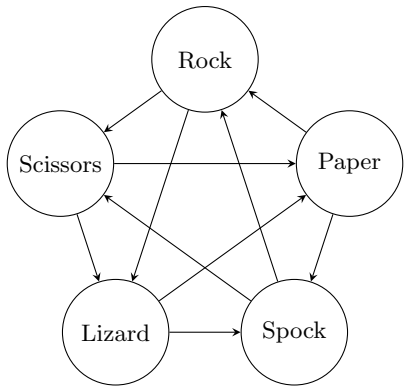

This generalised formulation allows each species to dominate multiple others, not necessarily in consecutive order, and accommodates both symmetric and asymmetric dominance structures. For instance, the classic Rock-Paper-Scissors interaction corresponds to the circulant graph \( C(3, \{1\}) \), where each species dominates exactly one other in a closed cycle. Similarly, the Rock-Paper-Scissors-Lizard-Spock interaction corresponds to \( C(5, \{1,2\}) \), where each species dominates two others (as illustrated in \textbf{Figure~\ref{fig:rpsls-circulant}}).

\noindent
Further pre-simulation initialisation includes setting the values of the action probabilities: $\mu$, $\sigma$, and $\epsilon$, which correspond to \textbf{interaction}, \textbf{reproduction}, and \textbf{migration}, respectively. While terminology and notation may vary slightly across the literature, the semantics of these game rules generally remain consistent. In this system, these actions are implemented as follows:

\begin{itemize}
    \item \textbf{Interaction ($\mu$)}: If both the selected cell and its neighbour are non-empty and belong to different species, they engage in a competitive interaction. The losing species' site becomes empty (represented by 0), based on the predefined dominance rules.
    
    \item \textbf{Reproduction ($\sigma$)}: If exactly one of the two sites (either the selected cell or its neighbour) is empty, the non-empty cell reproduces into the empty one.
    
    \item \textbf{Migration ($\epsilon$)}: The selected cell and its neighbour swap positions, regardless of species or emptiness.
\end{itemize}

\noindent
These rules can be expressed formally using the notation introduced by Reichenbach, Mobilia \& Frey.~\cite{Reichenbach2007MobilityGames}, adapted here for the well-known Rock-Paper-Scissors (RPS) model. Let $R$, $P$, and $S$ denote rock, paper, and scissors respectively, and let $\square$ denote an empty site. The following transitions illustrate a subset of the possible dynamics:

\[ 
\begin{aligned}
    RP &\xrightarrow{\mu} \square P &\quad \text{(Rock loses to Paper)} \\
    RR &\xrightarrow{\mu} RR &\quad \text{(No interaction between same species)} \\
    S\square &\xrightarrow{\sigma} SS &\quad \text{(Scissors reproduces)} \\
    PP &\xrightarrow{\sigma} PP &\quad \text{(No reproduction without empty site)} \\
    R\square &\xrightarrow{\epsilon} \square R &\quad \text{(Migration)} \\
    RP &\xrightarrow{\epsilon} PR &\quad \text{(Migration)}
\end{aligned}
\]

\noindent
The probability variables were initialised as follows, where $N$ is the total number of cells in the lattice (i.e., $N = L \times H$), and $M$ represents the mobility parameter, defined as the typical area explored by an individual per unit time. The default value for $M$ is set to $10^{-6}$, but can be configured at runtime:

\[
\begin{aligned}
    \mu &= 1 \\
    \sigma &= 1 \\
    \epsilon &= 2 M N
\end{aligned}
\]

\noindent
These probabilities are subsequently normalised to form a valid probability distribution, from which one action is randomly selected during each elementary step. \newline

To execute an elementary step, pseudo-random number generation is handled using the \texttt{<random>} header from the C++ Standard Library, which provides high-quality, flexible random engines and distributions. A Mersenne Twister engine (\texttt{std::mt19937}) is seeded once using \texttt{std::random\_device} and reused across steps for efficient number generation. Three distributions are instantiated: a uniform integer distribution for selecting a random grid cell, another for selecting a neighbouring direction (adapted based on 4- or 8-way neighbourhood configuration), and a uniform real distribution for probabilistically choosing between interaction, reproduction, or migration actions based on the normalised values of $\mu$, $\sigma$, and $\epsilon$. A general overview of the step logic is detailed in \textbf{Algorithm \ref{alg:es}}, where the argument $p$ represents the set of parameters configured by the user at runtime. \newline

To complete the simulation, the program performs $N$ elementary steps per MCS, where $N= L × H$ is the total number of cells in the lattice. This process is repeated for the number of MCS specified by the user (default: 100,000). At each MCS, density values are recorded for visualisation, and snapshots may be saved at designated intervals if enabled. The complete control flow is outlined in \textbf{Algorithm~\ref{alg:escg}}, where \textit{saveIntervals} refers to a set of MCS values, defined within the code, at which the simulation state is saved and exported to \texttt{.csv} files.

\begin{algorithm}[H]
\caption{Elementary Step}
\label{alg:es}
\SetKwFunction{Step}{step}
\SetKwProg{Fn}{Function}{:}{}
\Fn{\Step{$p$, grid, dominance, $\mu$, $\sigma$, $\epsilon$}}{
    $i \gets$ random cell index from $[0, L \times H)$\; \tcp*{Select a random cell}
    \texttt{species} $\gets$ \texttt{grid[$i$]}\
    
    $dir \gets$ random direction from  $[0, p.neighbourhood)$\; \tcp*{von-Neumann or Moore} 
    
    $(row, col) \gets (i \div L,\ i \bmod L)$\\
    $(n\_row,\ n\_col) \gets$ updated using $dir$ and boundary conditions\; \tcp*{Apply flux or reflect}
    $n_i \gets n\_row \times L + n\_col$\\
    \texttt{neighbour} $\gets$ \texttt{grid[$n_i$]}\
    
    $r \gets$ random float in $[0,\ \mu + \sigma + \epsilon]$\; \tcp*{Sample action}

    \uIf{\texttt{species == neighbour}}{
        \Return\; \tcp*{Skip same species}
    }

    \uIf{$r < \epsilon$}{
        Swap \texttt{grid[$i$]} and \texttt{grid[$n_i$]}\; \tcp*{Migration}
    }
    \uElseIf{$r < \epsilon + \mu$}{  \tcp*{Interaction}
        \uIf{\texttt{neighbour} $\neq 0$}{
            \uIf{\texttt{dominates(species, neighbour)}}{
                \texttt{grid[$n_i$]} $\gets 0$\; \tcp*{Neighbour dies}
            }
            \uElseIf{\texttt{dominates(neighbour, species)}}{
                \texttt{grid[$i$]} $\gets 0$\; \tcp*{Self dies}
            }
        }
    }
    \ElseIf{$r < \epsilon + \mu + \sigma$}{  \tcp*{Reproduction}
        \uIf{\texttt{neighbour} $== 0$}{
            \texttt{grid[$n_i$]} $\gets$ \texttt{species}\; \tcp*{Reproduce to neighbour}
        }
        \ElseIf{\texttt{species} $== 0$}{
            \texttt{grid[$i$]} $\gets$ \texttt{neighbour}\; \tcp*{Reproduce to self}
        }
    }
}
\end{algorithm}

\noindent\\

\begin{algorithm}[H]
\caption{Overall ESCG Loop}
\label{alg:escg}
\SetKwFunction{Simulate}{escg}
\SetKwProg{Fn}{Function}{:}{}
\Fn{\Simulate{$p$, grid, dominance, $\mu$, $\sigma$, $\epsilon$}}{
    $N \gets p.L \times p.H$\; \tcp*{Total number of cells}
    \For{$mcs \gets 0$ \KwTo $p.MCS$}{
        \texttt{densities()}\; \tcp*{Track population densities}

        \If{$p.save\ \textbf{and}\ mcs \in \texttt{saveIntervals}$}{
            \texttt{plot\_snapshot(grid, mcs, p)}\; \tcp*{Save snapshot}
        }

        \For{$n \gets 0$ \KwTo $N - 1$}{
            \texttt{step(p, grid, dominance, $\mu$, $\sigma$, $\epsilon$)}\; \tcp*{Perform elementary step}
        }
    }
}
\end{algorithm}

\noindent
\\
In addition to the core simulation logic, auxiliary functions were developed to support post-simulation analysis and result exploration. A dedicated \texttt{GridContext} struct, defined in \texttt{config.hpp}, was created to store all data necessary for visualisation. The \texttt{densities()} function, illustrated in \textbf{Algorithm \ref{alg:escg}}, computes the relative density of each species (including empty sites) at each MCS. These values are printed at regular intervals for real-time feedback and appended to the \texttt{GridContext}’s internal vectors (\texttt{steps} and \texttt{speciesDensities}) to enable subsequent visualisation of population dynamics over time. \newline

Simulation results were visualised using \texttt{matplotlib\-cpp}, a lightweight C++ wrapper around Python’s \texttt{matplotlib} library. The \texttt{plot\_densities()} function reads from the populated \texttt{GridContext} to generate semilogarithmic plots depicting species density evolution across the simulation. Additionally, the \texttt{plot\_snapshot()} function creates colour-mapped lattice images at specified MCS checkpoints. Both functions automatically include encoded simulation parameters in the plot titles and filenames, ensuring clarity and reproducibility of experimental results.

\subsection{Results}

To validate the correctness of this system, the work of Zhong et al.~\cite{Zhong2022SpeciesSpecies} was used as a reference benchmark. Their paper investigates the concept of ablated circulant graphs -- circulant dominance networks with selectively removed edges, and analyses how such modifications influence biodiversity and spatial dynamics. One of their key examples involves an altered RPSLS model, in which the edge representing Rock crushing Scissors is removed. This structural change was shown to significantly affect coexistence patterns and long-term species distributions. By importing a dominance matrix reflecting this ablation, simulations were run under equivalent conditions to reproduce these results and assess the fidelity of this implementation. \newline

\begin{figure}[h]
    \centering
    \begin{subfigure}[]{0.8\textwidth}
        \includegraphics[width=\textwidth]{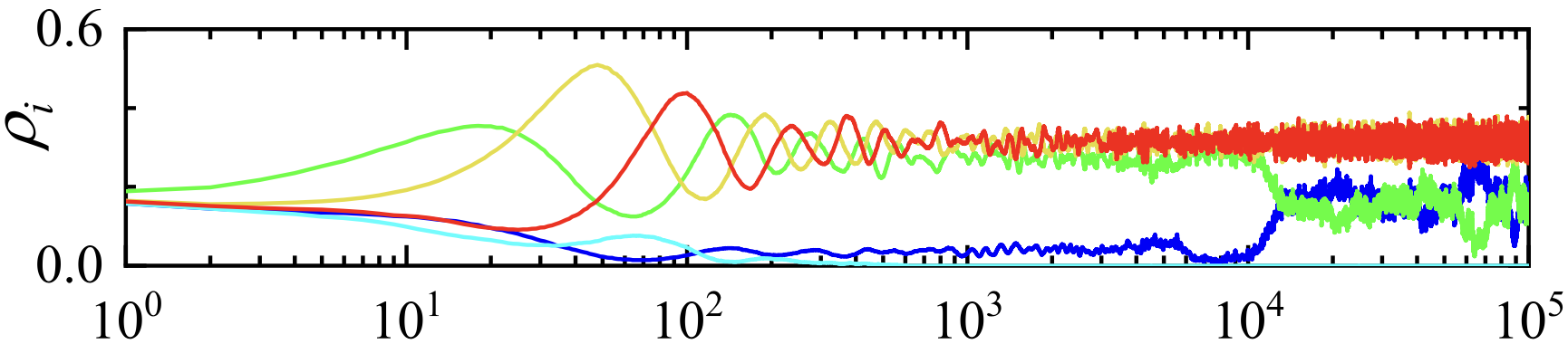}
        \caption{Zhong et al. Figure 2a}
        \label{fig:zhong2a}
    \end{subfigure}
    \hfill
    \begin{subfigure}[]{0.8\textwidth}
        \includegraphics[width=\textwidth]{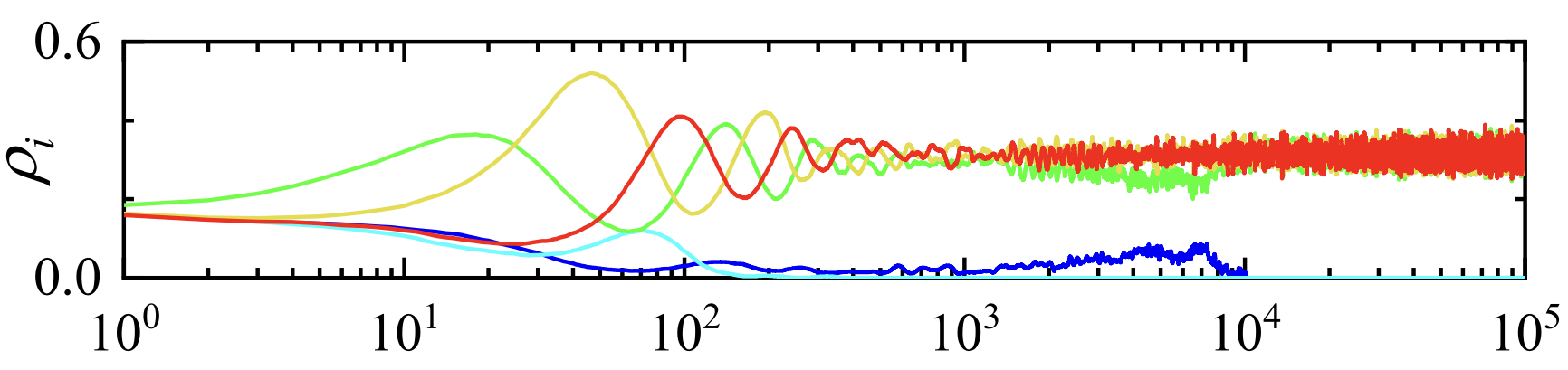}
        \caption{Zhong et al. Figure 2c}
        \label{fig:zhong2c}
    \end{subfigure}
    \hfill   
    \caption{Snapshots replicating \textbf{Figure 2} from Zhong et al.~\cite{Zhong2022SpeciesSpecies}, illustrating spatial population distributions of species over time.}
    \label{fig:zhong-fig2}
\end{figure}

Zhong et al. demonstrate that when the Rock–Scissors interaction is removed from the RPSLS model, the Paper species quickly goes extinct, typically between 200 and 600 MCS. As the simulation progresses, the absence of this interaction introduces a bifurcation in the system's long-term dynamics. By approximately 10,000 MCS, the simulation evolves into one of two distinct asymptotic states: in some cases, Rock thrives and spreads, while in others, it disappears entirely. This behaviour is visualised in \textbf{Figure \ref{fig:zhong-fig2}}, which presents the original results from Zhong et al., showing how species densities evolve under this modified interaction rule. \newline

The extinction of Paper leaves behind two overlapping sub-cycles: Rock–Lizard–Spock and Scissors–Lizard–Spock. Whether Rock survives depends on the local spatial configuration. Without the ability to suppress Scissors, Rock becomes vulnerable, especially since its only remaining prey, Lizard, is also contested by other species. If Scissors gains an early advantage, it often leads to Rock’s collapse. However, if Rock initially dominates local regions rich in Lizard and Spock, it can persist and even suppress Scissors through indirect competitive advantages. \newline

\begin{figure}[H]
    \centering
    \begin{subfigure}[]{0.75\textwidth}
        \includegraphics[width=\textwidth]{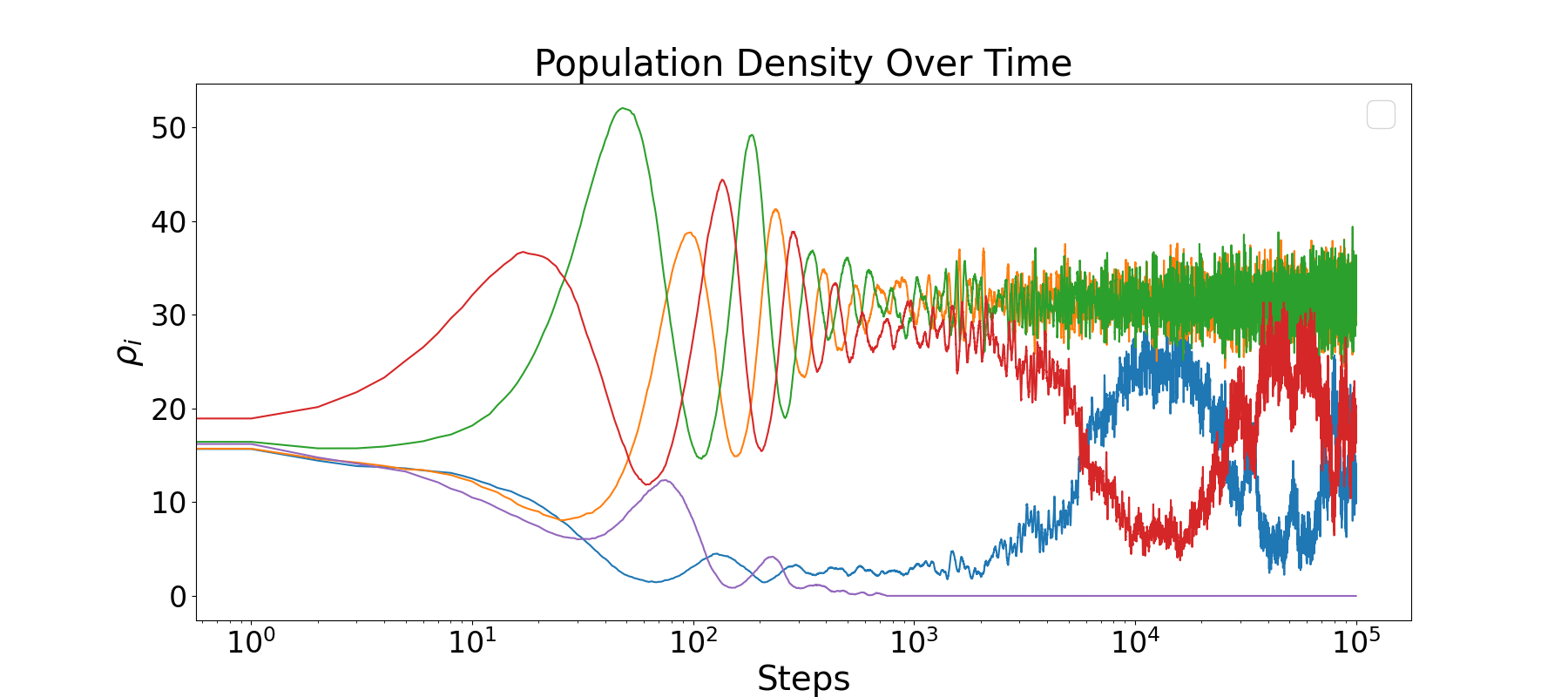}
        \caption{C++ single-threaded replication of Zhong et al. Figure 2a}
        \label{fig:replication-zhong2a}
    \end{subfigure}
    \hfill
    \begin{subfigure}[]{0.75\textwidth}
        \includegraphics[width=\textwidth]{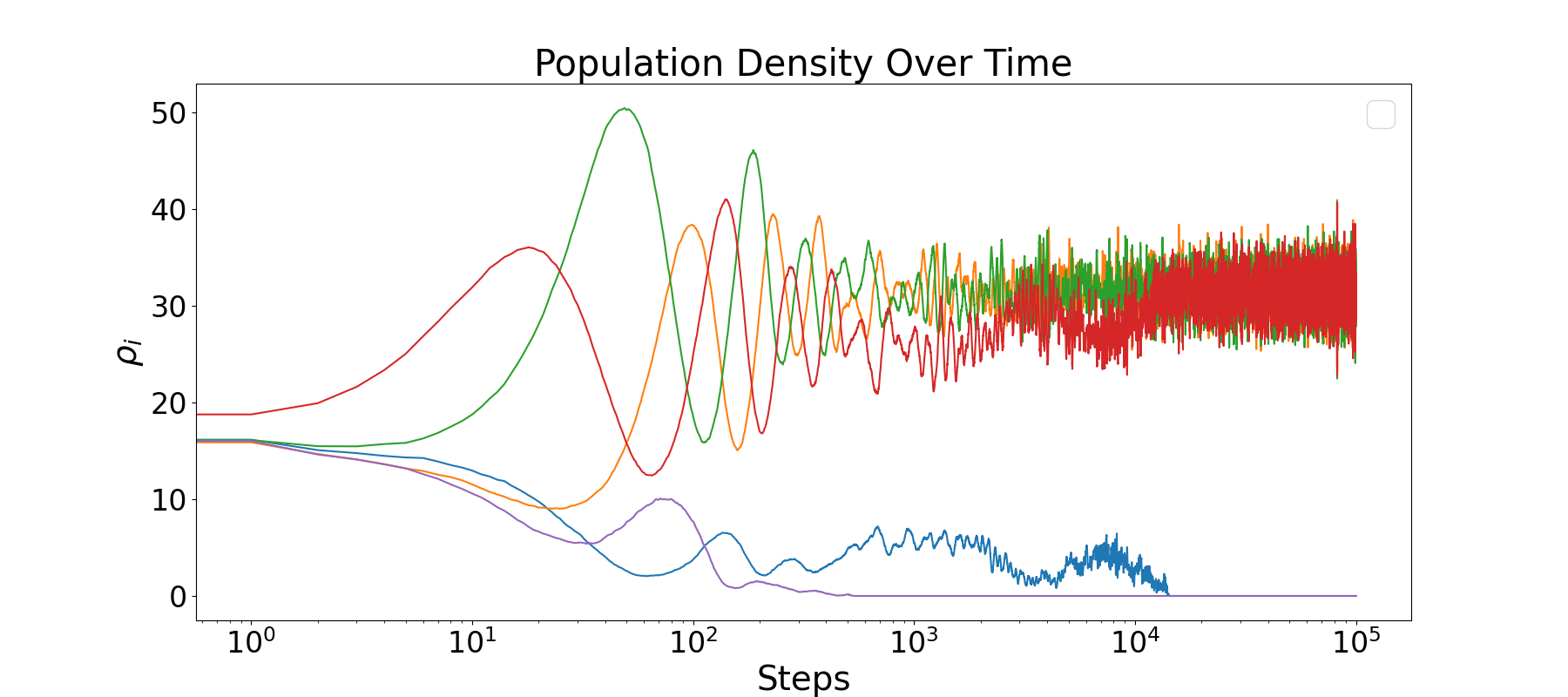}
        \caption{C++ single-threaded replication of Zhong et al. Figure 2c}
        \label{fig:replication-zhong2c}
    \end{subfigure}
    \hfill
    
    \caption{Density plots produced by the single-threaded C++ ESCG implementation, replicating \textbf{Figure 2} from Zhong et al. Ran with \texttt{--save true} and \texttt{--dominance true}.}
    \label{fig:zhong-replication}
\end{figure}

\textbf{Figure \ref{fig:zhong-replication}} presents the corresponding density evolution plots generated by this single-threaded C++ implementation. These results closely replicate the findings from \textbf{Figure \ref{fig:zhong-fig2}}, which reproduces the original population densities shown by Zhong et al. The similarity in the overall shape of the curves, key inflection points, and the timing of transitions (such as the early extinction of Paper and the bifurcating long-term behaviour of Rock around $10^4$ MCS) strongly validates the correctness of the implementation. Both simulations show the same characteristic oscillations during early-stage coexistence and converge toward similar long-term asymptotic dynamics, demonstrating the system's ability in capturing the stochastic, spatial, and cyclic behaviours fundamental to ESCGs.

\newpage
\section{Metal ESCGs}

Following the successful development and verification of the single-threaded C++ ESCG implementation, the next step was to extend the system using GPU acceleration. The first target platform for parallelisation was Apple Silicon, utilising Apple’s Metal API. As discussed earlier, the primary motivation for leveraging GPU hardware was twofold: to parallelise the computationally serial step logic, and to improve performance by precomputing large buffers of pseudo-random numbers. These precomputed values are then accessed during simulation steps via lookup operations, significantly reducing the overhead associated with serially generating random numbers on demand. However, challenges emerged early in the development of the ESCG implementation using Metal, most notably, the absence of a built-in PRNG shader within the MSL standard library. To address this limitation, a custom implementation of the Mersenne Twister algorithm was integrated directly into the Metal compute pipeline, enabling efficient, high-quality random number generation within GPU-executed kernels.

\subsection{Metal-Mersenne Twister}

The Mersenne Twister algorithm, developed in 1997 by Makoto Matsumoto and Takuji Nishimura, derives its name from its exceptionally long period based on a Mersenne prime. Since its introduction, various versions and optimisations have been developed \cite{matsumoto1998mersenne}. At its core, the algorithm operates as follows:

\begin{center}
\begin{enumerate}
    \item \textbf{State Initialisation}
    \begin{itemize}
        \item Initialise a 624-element array of 32-bit integers, referred to as the \textit{state array}.
    \end{itemize}

    \item \textbf{Twisting (Core Update Step)}
    \begin{itemize}
        \item This is performed once every 624 outputs to refresh the state.
        \item For each $i \in [0, 623]$:
        \begin{itemize}
            \item Combine the upper bits of \texttt{state}[i] with the lower bits of \texttt{state}[(i+1) mod 624].
            \item Right shift the result by 1.
            \item If the least significant bit is 1, XOR the result with a constant $A = \texttt{0x9908B0DF}$.
            \item XOR the final result with \texttt{state}[(i + 397) mod 624].
        \end{itemize}
        \item This operation is referred to as \textit{twisting} due to its reshuffling and scrambling of the internal state.
    \end{itemize}

    \item \textbf{Tempering (Output Transformation)}
    \begin{itemize}
        \item When generating a random number, the next element in the state array undergoes a series of transformations:
        \[
        \begin{aligned}
        y &= y \oplus (y \gg u) \\
        y &= y \oplus ((y \ll s) \& b) \\
        y &= y \oplus ((y \ll t) \& c) \\
        y &= y \oplus (y \gg l)
        \end{aligned}
        \]
        \item The constants $u$, $s$, $t$, $l$, $b$, and $c$ are empirically selected to improve statistical properties and eliminate correlations.
    \end{itemize}
\end{enumerate}
\end{center}

\noindent
The MSL implementation exploits GPU parallelism by assigning each thread its own instance of a local PRNG state. A custom struct named \texttt{MT19937} is instantiated per thread, encapsulating both the internal 624-element state array and the current index. Each thread initialises its \texttt{MT19937} instance independently using a unique seed from a device buffer, which is first mixed with the thread ID using a 32-bit finaliser inspired by the widely used MurmurHash3 algorithm, shown in \textbf{Algorithm \ref{alg:hash}}, to ensure greater randomness and decorrelation between threads. The threads then generate a fixed number of pseudorandom numbers in parallel, which are written to a global buffer for use during the ESCG simulation.

\begin{algorithm}[H]
\caption{32-bit Finaliser Inspired by MurmurHash3}
\label{alg:hash}
\SetKwFunction{Finalise}{hash}
\SetKwProg{Fn}{Function}{:}{}
\Fn{\Finalise{$x$}}{
    $x \gets x \oplus (x \gg 16)$\;
    $x \gets x \times \texttt{MURMUR\_CONST1}$\;
    $x \gets x \oplus (x \gg 13)$\;
    $x \gets x \times \texttt{MURMUR\_CONST2}$\;
    $x \gets x \oplus (x \gg 16)$\;
    \Return $x$\;
}
\end{algorithm} 

\noindent
\newline
Overall, the following constant definitions were used in the Metal-Mersenne Twister implementation:

\[
\begin{aligned}
\texttt{UPPER\_MASK}      &= 0x80000000  && \text{// Selects the most significant bit (MSB) of a 32-bit word} \\
\texttt{LOWER\_MASK}      &= 0x7FFFFFFF  && \text{// Selects the least significant 31 bits of a 32-bit word} \\
\texttt{TEMPERING\_MASK\_B} &= 0x9d2c5680 && \text{// Used in the tempering step to improve bit distribution} \\
\texttt{TEMPERING\_MASK\_C} &= 0xefc60000 && \text{// Also part of the tempering step, fine-tunes output randomness} \\
\texttt{MATRIX\_A}        &= 0x9908b0df  && \text{// Constant used during the twist transformation} \\
\texttt{STATE\_VECTOR\_LENGTH} &= 624     && \text{// Length of the MT19937 state array} \\
\texttt{STATE\_VECTOR\_M} &= 397         && \text{// Offset used in the twist step (good recurrence properties)} \\
\texttt{MURMUR\_CONST1}   &= 0x85ebca6b  && \text{// Mixing constant used in the MurmurHash3-inspired finaliser} \\
\texttt{MURMUR\_CONST2}   &= 0xc2b2ae35  && \text{// Second MurmurHash3 mixing constant for strong avalanche effect} \\
\texttt{MT\_INIT\_MULTIPLIER} &= 1812433253 && \text{// Used to initialise the state array (from MT19937 spec)} \\
\end{aligned}
\] 
\newline

To support the stochastic selection required by ESCGs, random numbers are extracted from each thread-local generator using simple modulus or scaling techniques. For example, a random integer within a lattice of size $N$ is produced by \texttt{extract(mt) \% N}, while a uniform random float in $[0, 1]$ is obtained via \texttt{extract(mt) / 4294967295.0f}. Specialised helper functions were defined to generate random cell indices, neighbour directions (in 4- or 8-way neighbourhoods), and continuous probabilities, enabling efficient sampling for cell selection, interaction types, and movement outcomes. \newline 

This parallelisation of PRNG allows random values to be efficiently precomputed and retrieved via indexed lookups, significantly reducing runtime overhead while ensuring statistical independence across threads. \newline

\noindent
The process of invoking this MSL kernel from the host is as follows:
\begin{itemize}
    \item Allocate an output array on the host to store the generated random numbers.
    \item Initialise a seed array, one per thread, with unique seeds (e.g., sequential integers).
    \item Transfer the seed array to a Metal \texttt{MTL::Buffer} using \texttt{ResourceStorageModeShared}.
    \item Allocate a second \texttt{MTL::Buffer} of sufficient size to hold all generated random numbers.
    \item Create a \texttt{MTL::CommandBuffer} and a corresponding \texttt{MTL::ComputeCommandEncoder}.
    \item Set the Metal pipeline state and bind both the seed and result buffers as kernel arguments using \texttt{setBuffer(...)}.
    \item Dispatch the threads using \texttt{dispatchThreads(...)}, where each thread generates a chunk of random numbers in parallel.
    \item End encoding and commit the command buffer.
    \item Wait for GPU execution to complete with \texttt{waitUntilCompleted()}.
    \item Map the result buffer back to host memory using \texttt{std::memcpy} with \texttt{resultBuffer->contents()} to access the generated random numbers.
\end{itemize}

\noindent
Unfortunately, even with additional hashing and per-thread state initialisation, it became apparent that the randomness of the PRNG was not sufficient for the simulation to evolve as expected. When the generated random numbers were applied to the ESCG, unusual artefacts emerged, most notably, vertical striping patterns on the lattice in place of the expected organic spatial domains. This suggested a lack of sufficient entropy or uniformity in the early outputs of the PRNG, a common issue known as poor initial dispersion. \newline

\begin{figure}[h]
    \centering
    \begin{subfigure}{0.32\textwidth}
        \includegraphics[width=\textwidth]{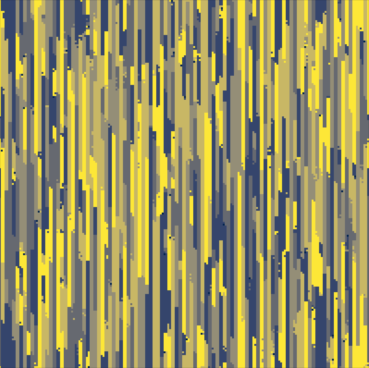}
        \caption{No burn-in applied}
        \label{fig:noburn}
    \end{subfigure}
    \hspace{0.02\textwidth}
    \begin{subfigure}{0.32\textwidth}
        \includegraphics[width=\textwidth]{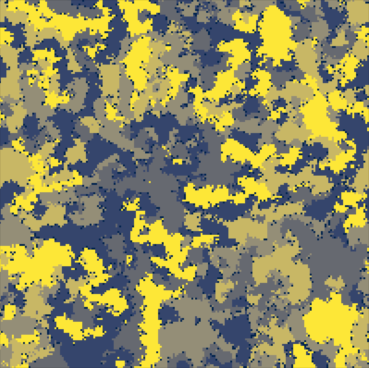}
        \caption{Burn-in of 50,000 applied}
        \label{fig:burn}
    \end{subfigure}
    
    \caption{Comparison of species distribution in the ESCG when using the Metal-Mersenne Twister PRNG. Without burn-in, vertical stripes emerge due to insufficient randomness. Applying a burn-in phase restores expected spatial domains by improving the statistical quality of generated numbers. (ESCG params: all default)}
    \label{fig:burns}
\end{figure}

To mitigate this, a \textit{burn-in} phase was introduced, where a predefined number of random numbers were generated and discarded before using the generator's output in the actual simulation. For operations such as random cell selection and continuous float generation, a burn-in of 50,000 was empirically found to be sufficient. However, neighbour direction selection, particularly when constrained to small discrete ranges such as \([0, 3]\) or \([0, 7]\) was more susceptible to bias. This sensitivity arises from the limited number of possible outcomes, which increases the impact of any non-uniformity in the underlying random number distribution due to modulus operations. \textbf{Figure \ref{fig:burns}} illustrates this difference, showing a comparison between a simulation run without burn-in and one with the appropriate burn-in phases applied. The latter exhibits the expected patch dynamics of a circulant dominance matrix with five species, validating the improved statistical quality of the generator's output post burn-in.

\subsection{Metal ESCG Implementation}

For GPU-accelerated simulations, several additional command-line flags were introduced to provide users with greater flexibility in configuring runtime behaviour, shown in \textbf{Table \ref{tab:gpu-vars}}. The \textbf{\texttt{--resume}} flag enables the simulation to resume from a previously saved grid state and parameter set at a specific MCS. To support this functionality, the \textbf{\texttt{--save}} flag was extended to export both the grid and simulation parameters as separate \texttt{.csv} files after execution (done in \texttt{io.cpp}). The \textbf{\texttt{--numRandoms}} flag enables users to specify the number of random numbers generated and stored per shader invocation—an important configuration due to the differing memory capacities across GPU devices. Internally, this value is immediately adjusted to ensure it is a multiple of the lattice size $N$, using the following expression:

\[
\texttt{numRandoms} = (\texttt{numRandoms} ~/~ N) \times N
\]

\noindent
This operation ensures alignment with MCS by leveraging integer division, which naturally truncates any remainder. As a result, the total number of random values becomes evenly divisible by $N$, allowing for consistent batching of complete simulation steps. This also ensures that no GPU threads generate excess random numbers or attempt to process additional cells beyond those required, which helps preserve memory efficiency and avoid redundant computation. Lastly, the \textbf{\texttt{--maxStep}} flag determines the granularity of kernel execution. When disabled, the system sends exactly $N$ random numbers to the step kernel, resulting in one full MCS per invocation. When enabled, all generated random numbers (as specified by \texttt{--numRandoms}) are consumed in a single call, performing multiple updates in one pass---effectively simulating $\texttt{numRandoms} \div N$ MCS. This mode enables faster progression through time steps, though at the cost of losing per-MCS observability in between.\newline

\begin{table}[]
\centering
\caption{Additional Command-line Configurable Parameters and Default Values}
\label{tab:gpu-vars}
\begin{tabular}{|l|l|l|l|}
\hline
\textbf{CLI Flag} & \textbf{Description} & \textbf{Default Value} \\
\hline
\texttt{--resume} & Resume simulation from a given state & false \\
\texttt{--numRandoms} & No. of random numbers to store and generate & 100,000,000 \\
\texttt{--maxStep} & Execute multiple MCS per kernel invocation & false \\
\hline
\end{tabular}
\end{table}

The Metal-extended implementation extends three additional structs in \texttt{config.hpp}: \texttt{MetalContext}, \texttt{StepContext}, and \texttt{RandomCommandBuffers}. These structures encapsulate the state and resources required for GPU-based compute workflows, streamlining the CPU-GPU communication and synchronisation. \newline

The \texttt{MetalContext} struct acts as the central state manager for all Metal-related operations. It contains references to key Metal components such as the \texttt{MTLDevice}, \texttt{MTLCommandQueue}, and various \texttt{MTLComputePipelineState} objects, which represent compiled shader functions. It also maintains GPU buffers for random numbers, the simulation grid, action selection, and dominance matrix. Each buffer corresponds to a different stage in the GPU pipeline, such as cell selection, neighbour direction computation, or the core step execution, allowing the host to configure and control GPU workflows with precision. Although it contains many fields, the primary purpose of \texttt{MetalContext} is to manage and persist Metal resources across kernel invocations while minimising overhead. \newline

In contrast, the \texttt{StepContext} struct serves a more focused purpose. It holds three pointers: \texttt{cells}, \texttt{neighbour\_dirs}, and \texttt{action\_probabilities}. These pointers are populated on the host using the pre-generated random numbers and are passed to the step kernel to dictate the precise updates to be performed. This design decouples the stochastic sampling phase from the simulation logic, leading to improved performance. If the \texttt{--maxStep} flag was enabled at runtime, this struct would never be instantiated as the buffers holding all of the random numbers would be sent to the step shader instead. \newline

Finally, the \texttt{RandomCommandBuffers} struct facilitates parallelism by enabling simultaneous random number generation and simulation execution. It maintains global access to the command buffers responsible for these tasks, allowing synchronisation or deferred execution when needed. \newline 

Following the parsing of command-line arguments, the first check determines whether the simulation is resuming from a previously saved state using the \texttt{--resume} flag. This flag enables the restoration of all relevant simulation data and parameters, allowing continuation from a specific MCS. \newline 

If the resume flag is set to true, the simulation attempts to reload its configuration and state from CSV files. Specifically, \texttt{importCSVToParams} populates the \texttt{Params} struct from disk, and \texttt{importCSVToGrid} reconstructs the lattice configuration into a 1D flattened grid. The grid is read from \texttt{grid.csv}, structured such that each line represents a row of the 2D lattice, and the final line contains a single integer indicating the last saved MCS (so that the program knows the MCS from which it is resuming). This approach ensures compatibility with arbitrary grid sizes and allows users to visualise or edit the saved state externally if required. \newline

Additionally, the output directory name is constructed using key simulation parameters—such as lattice dimensions, neighbourhood type, mobility , boundary condition (flux), and species count—to facilitate organised storage and comparison of results across different parameter configurations. If the simulation is not being resumed, it instead uses parameters directly from the CLI input (or default). A new grid is allocated and initialised from scratch, and the output directory is generated in the same manner as in resume mode. \newline

Whether resuming or starting fresh, the simulation next checks for the \texttt{--dominance} flag. If enabled, the dominance matrix is loaded from a CSV file using \texttt{importCSVToDominance}, which also updates the species count. If not provided, a default cyclic dominance matrix is generated programmatically using \texttt{generateCirculantAdjacencyMatrix}. If the \texttt{--save} flag is enabled, the system attempts to create the output directory (if it does not already exist). It then exports all relevant runtime data: simulation parameters (\texttt{params.csv}), current grid state (\texttt{grid.csv}), and the dominance matrix (\texttt{dominance.csv}). These files serve both as documentation for reproducibility and as input for future resumption. \newline

Alongside the initialisation of other system parameters, the \texttt{initMetalContext} function sets up the GPU environment by preparing all necessary Metal resources for simulation. It calculates the number of threads based on the total number of random values required, allocates an autorelease pool for memory management, and creates a reference to the Metal device and command queue. The precompiled shader library (\texttt{escg.metallib}) is then loaded, and its compute functions are compiled into executable pipelines. \newline

At this stage, the \texttt{GridContext} is instantiated, and (if the \textbf{\texttt{--maxStep}} flag is enabled) the associated \texttt{StepContext} buffers are also allocated. The compute buffers required by the \texttt{step} kernel are initialised independently from the rest of the \texttt{MetalContext} to account for the runtime setting of \texttt{--maxStep}. Depending on its value, the buffer sizes are initialised as either $N$ (one MCS per invocation) or \texttt{numRandoms} (multiple MCS per kernel call). \newline

\noindent
\textbf{Simulation} \newline

Subsequently, the program generally follows the structure of the single-threaded implementation until the main ESCG loop, where the execution logic diverges depending on the value of the \texttt{--maxStep} flag. \newline 

If \texttt{--maxStep} is set to \texttt{false}, the simulation processes exactly one MCS per kernel invocation, as shown by \textbf{Algorithm \ref{alg:metal-step}}. For each MCS, $N$ random numbers are consumed to drive $N$ elementary steps. These values are passed to the \texttt{step} kernel through a \texttt{StepContext} struct, which is populated on the host side prior to dispatch. An integer variable \texttt{index} is used to track the current position in random number buffers generated on the GPU. \newline

When \texttt{index} is zero, indicating that all previously consumed values have been used, the host invokes the \texttt{refreshRandomNumbers} function. This function triggers the random number generation kernels and returns a \texttt{RandomCommandBuffers} struct containing the associated command buffers for action probabilities, cell selections, and neighbour directions. The simulation continues to consume from the current random number batch until \texttt{index} reaches \texttt{numRandoms}, at which point the updated GPU buffers are copied to host memory. To ensure correctness, the program explicitly waits for the GPU to finish generating the new random values using \texttt{waitUntilCompleted()} on each of the relevant command buffers before copying. This staged, asynchronous approach enables concurrent random number generation and simulation execution while maintaining synchronisation where necessary.
                                                                                                  
\begin{algorithm}[H]
\caption{Simulation Loop (\texttt{--maxStep} disabled) in MSL}
\label{alg:metal-step}
\SetKwFor{For}{for}{do}{}
\SetKwIF{If}{ElseIf}{Else}{if}{then}{else if}{else}{end}
\For{$mcs \gets \texttt{currentMCS}$ \KwTo \texttt{MCS}}{
    \texttt{densities(\dots)}\;
    \;
    
    \If{$mcs \in$ saveInterval \textbf{and} \texttt{params.save}}{   
        \tcp*[l]{save snapshots and export \texttt{.csv} files}
    }
    \If{$mcs = \texttt{MCS}$}{
        \textbf{break}\;
    }
    \;

    \tcp*[l]{Fill StepContext pointers}
    \For{$i \gets 0$ \KwTo $N - 1$}{
        \If{\texttt{index} == 0}{
            \texttt{cmdBuffers} $\gets$ \texttt{refreshRandomNumbers(...)}\;
        }

        \If{\texttt{index} $\geq$ \texttt{params.numRandoms}}{
            \texttt{cmdBuffers.actionCommandBuffer->waitUntilCompleted()}\;
            \texttt{cmdBuffers.cellsCommandBuffer->waitUntilCompleted()}\;
            \texttt{cmdBuffers.neighboursCommandBuffer->waitUntilCompleted()}\;
            \texttt{memcpy(\dots)} \tcp*[l]{Copy all random buffers}
            \;
            \texttt{index} $\gets 0$\;
        }
        \;

        \texttt{stepCtx.cells[i]} $\gets$ \texttt{cells[index]}\;
        \texttt{stepCtx.neighbour\_dirs[i]} $\gets$ \texttt{neighbours[index]}\;
        \texttt{stepCtx.action\_probabilities[i]} $\gets$ \texttt{action\_probabilities[index]}\;
        \texttt{index++}\;
           
    }

    \;
    \texttt{memcpy(metalCtx.stepGridBuffer, grid)}\;
    \texttt{metalStep(...)}\;
    \texttt{memcpy(grid, metalCtx.stepGridBuffer)}\; 
    \;

    \If{\texttt{stasis()}}{
        \textbf{break}\;
    }
}
\end{algorithm} 

\noindent
\newline
Conversely, if \texttt{--maxStep} is set to \texttt{true}, the simulation takes a more aggressive approach to acceleration by processing multiple MCS per kernel invocation, as described in \textbf{Algorithm \ref{alg:metal-maxstep}}. In this mode, all \texttt{numRandoms} values are generated in a single batch and consumed in one dispatch of the \texttt{maxMetalStep} kernel. This means that $numRandoms / N$ full Monte Carlo Steps are executed within one GPU call. Unlike the standard mode, the \texttt{StepContext} struct is bypassed entirely; instead, the GPU directly accesses the global buffers holding all precomputed action probabilities, cell indices, and neighbour directions. \newline

Once the random number kernels are launched via \texttt{refreshRandomNumbers()}, the host copies the current state of the grid to the GPU buffer. The \texttt{metalStep} kernel is then invoked to process the entire batch of $numRandoms$ values. Since this constitutes multiple MCS steps, the main simulation loop increments \texttt{mcs} accordingly. After the kernel finishes execution, the host waits on the relevant command buffers to ensure all random number computations have completed. The updated grid is then copied back from the GPU to host memory. This bulk-processing approach eliminates repeated data transfers and kernel invocations, enabling faster simulation advancement, especially useful for exploring long-term behaviour. However, it comes at the cost of intermediate granularity, as simulation data is only collected every \texttt{step} MCS rather than every individual MCS. \newline                                                                                                         
\begin{algorithm}[H]
\caption{Simulation Loop (\texttt{--maxStep} enabled) in MSL}
\label{alg:metal-maxstep}
\SetKwFor{For}{for}{do}{}
\SetKwIF{If}{ElseIf}{Else}{if}{then}{else if}{else}{end}

$step \leftarrow \texttt{params.numRandoms} ~/ ~N$\;
\For{$mcs \leftarrow \texttt{currentMCS}$ to \texttt{MCS} in increments of $step$}{
    \texttt{densities(\dots)}\;
    \;
    
    \If{$mcs \in$ saveInterval \textbf{and} \texttt{params.save}}{   
        \tcp*[l]{save snapshots and export \texttt{.csv} files}
    }
    \If{$mcs = \texttt{MCS}$}{
        \textbf{break}\;
    }
    \;
    
    \texttt{cmdBuffers} $\leftarrow$ \texttt{refreshRandomNumbers(...)}\;
    \;
    
    \texttt{memcpy(metalCtx.stepGridBuffer, grid)}\;
    \texttt{maxMetalStep(\dots)}\;
    \texttt{memcpy(grid, metalCtx.stepGridBuffer)}\;
    \;
    
    \texttt{cmdBuffers.actionCommandBuffer->waitUntilCompleted()}\;
    \texttt{cmdBuffers.cellsCommandBuffer->waitUntilCompleted()}\;    
    \texttt{cmdBuffers.neighboursCommandBuffer->waitUntilCompleted()}\;
    \;
    
    \texttt{memcpy(\dots)} \tcp*[l]{Copy all random buffers}
    \;
    
    \If{\texttt{stasis()}}{
        \textbf{break}\;
    }
}
\end{algorithm}

\noindent
\newline
Both Metal-based simulation modes, whether \texttt{--maxStep} is enabled or not, invoke the same compute kernel defined in \texttt{step.metal}. This shader is designed to support both single and multi-MCS execution by dynamically adjusting the number of cells processed per thread at runtime. The distinction is handled using the \texttt{maxStep} boolean flag passed as a kernel argument. If \texttt{maxStep} is \texttt{true}, each thread processes a chunk of the full \texttt{numRandoms} buffer. If \texttt{false}, the thread handles a fraction of the grid proportional to $N$, the total number of cells. \newline

The logic of the step function remains consistent with the single-threaded implementation, but since all threads operate on a shared simulation grid, thread safety and correctness are maintained through the use of atomic operations on memory. These are crucial for ensuring consistent reads and writes to the grid when multiple threads may access overlapping regions. For example, the following operation:

\begin{lstlisting}[language=C++, style=metalstyle]
int species = atomic_load_explicit(&grid[cell_index], memory_order_relaxed);
\end{lstlisting}

\noindent
safely reads the current species value at a specific grid location, while:

\begin{lstlisting}[language=C++, style=metalstyle]
atomic_store_explicit(&grid[neighbour_index], 0, memory_order_relaxed);
\end{lstlisting}

\noindent
empties a site during an interaction event. For migration, a two-way swap between two cells is implemented via:

\begin{lstlisting}[language=C++, style=metalstyle]
atomic_exchange_explicit(&grid[cell_index], neighbour_specie, memory_order_relaxed);
atomic_exchange_explicit(&grid[neighbour_index], species, memory_order_relaxed);
\end{lstlisting}

\noindent
ensuring each exchange is conducted without race conditions. The use of \texttt{memory\_order\_relaxed} is particularly well-suited for the stochastic nature of ESCGs. While it avoids the performance cost of synchronisation barriers, it also enables the memory operations to complete in a non-deterministic order. This characteristic does not compromise correctness, since each atomic operation still guarantees data integrity, but it does allow for natural randomisation in how updates are interleaved across threads. Rather than being a limitation, this behaviour complements the randomness inherent in the simulation, helping to preserve the stochastic and decentralised dynamics of cell interactions. In this sense, \texttt{memory\_order\_relaxed} is not just an optimisation for data-parallel hardware, it is a conceptual match for the unpredictability and local chaos expected in spatially structured evolutionary games. \newline

In both GPU-accelerated simulation loops, each iteration concludes with a stasis check using the \texttt{stasis()} function to determine whether the system has reached a stable state.  A simulation is considered stable when only a single species remains active on the lattice. Even if multiple non-competing species existed on the grid, migration could still occur so the grid can not be considered stable. \newline

To support this, a \texttt{std::set} is initialised before the simulation begins, containing all unique species currently present on the grid. The \texttt{densities()} function is modified to accept a pointer to this set and, during each call, iterates through the population counts. If the count for any species reaches zero, that species is removed from the set. Consequently, the \texttt{stasis()} function simply checks whether the set’s size has been reduced to one, indicating that the ecosystem has collapsed into a monoculture. This lightweight check ensures minimal overhead while enabling early termination of the simulation in scenarios where competitive dynamics have effectively concluded. \newline

Additionally, density count computation has been offloaded to the GPU using the \texttt{densities.metal} kernel. This kernel is launched with $N$ threads, one for each cell in the grid. Each thread reads the species value at its assigned index and increments the count for that species in a shared atomic array. This array tracks the population of each species (including empty sites). The increment is performed atomically to prevent race conditions when multiple threads update the same species count. Though marginal for smaller lattice sizes, speedup still exists. The core operation looks like:

\begin{lstlisting}[language=C++, style=metalstyle]
int species = grid[id];
atomic_fetch_add_explicit(&result[species], 1, memory_order_relaxed);
\end{lstlisting}

\subsection{Results}
\begin{figure}[h]
    \centering
    \includegraphics[width=0.95\linewidth]{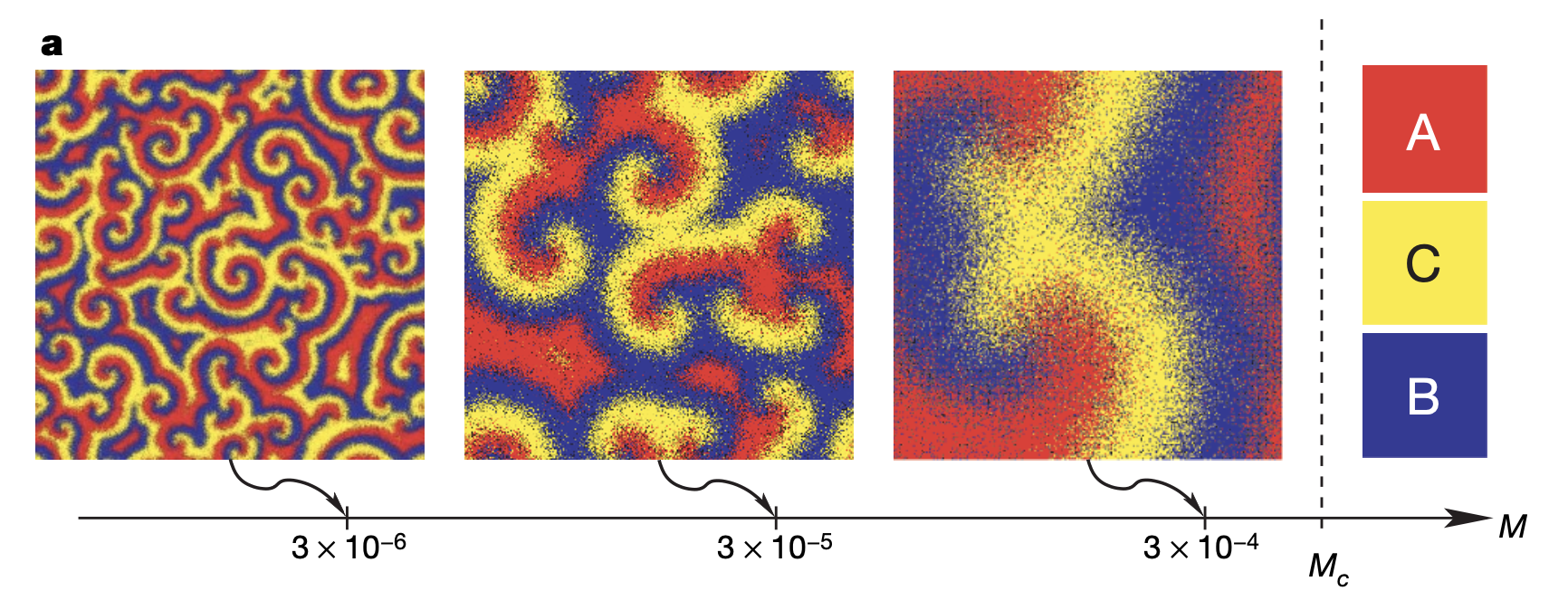}
    \caption{\textbf{Reichenbach, Mobilia and Frey Figure 2a}, showing threshold mobility values inducing spiral patterns on the grid \cite{Reichenbach2007MobilityGames}.}
    \label{fig:ReiMobFrey2a}
\end{figure}

\begin{figure}[h]
    \centering
    \begin{subfigure}[]{0.32\textwidth}
        \includegraphics[width=\textwidth]{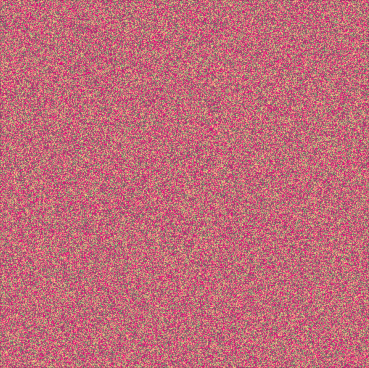}
        \caption{MCS = 0}
        \label{fig:spirals1}
    \end{subfigure}
    \hfill
    \begin{subfigure}[]{0.32\textwidth}
        \includegraphics[width=\textwidth]{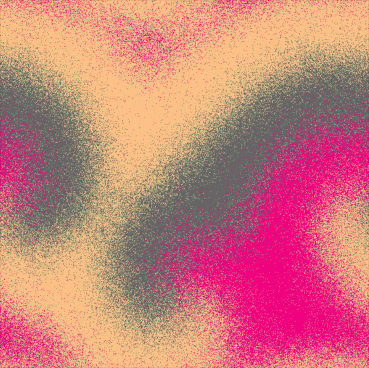}
        \caption{MCS = 50000}
        \label{fig:spirals2}
    \end{subfigure}
    \hfill
    \begin{subfigure}[]{0.32\textwidth}
        \includegraphics[width=\textwidth]{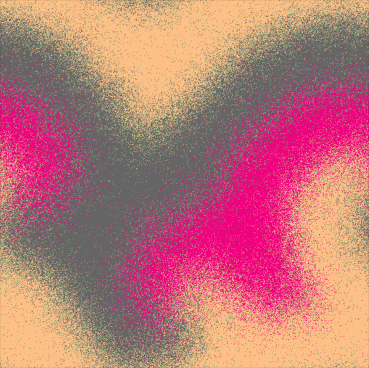}
        \caption{MCS = 100000}
         \end{subfigure}
    
    \caption{Snapshots from the Metal-ESCG implementation replicating Reichenbach, Mobilia, and Frey~\cite{Reichenbach2007MobilityGames} \textbf{Figure 2a}. 
    Shown at different Monte Carlo Steps (MCS). Parameters: lattice size = $600\times 600$, initial empty cell probability = $0.1$, mobility = $3\times10^{-4}$, von-Neumann neighbourhood, circulant dominance of 3 species, periodic boundary conditions. 
    Code available at: \url{<https://github.com/louiesinadjan/escg>}.}
    \label{fig:hazyspirals}
\end{figure}

\noindent
To validate the accuracy of the Metal-accelerated ESCG implementation, the system was benchmarked against results from existing literature. Reichenbach, Mobilia, and Frey presented a classic three-species ESCG governed by a circulant dominance matrix, demonstrating the emergence of spatial spiral patterns under varying mobility levels. As shown in \textbf{Figure \ref{fig:spirals}} (from \textbf{Section \ref{chap:context}}) and \textbf{Figure~\ref{fig:ReiMobFrey2a}}, this behaviour was successfully reproduced using the current system. In Reichenbach et al.'s original work, the grid with a mobility of $3 \times 10^{-5}$ displays sharp, well-defined spirals—replicated accurately in \textbf{Figure \ref{fig:spirals}}. At a higher mobility of $3 \times 10^{-4}$, the spirals become more diffuse and less pronounced, a phenomenon mirrored in this system’s output shown in \textbf{Figure \ref{fig:hazyspirals}}, which uses a larger lattice size to accommodate longer wavelength dynamics. \newline

These successful reproductions across different mobility regimes and lattice scales not only validate the correctness of the implementation but also demonstrate its robustness and scalability for high-resolution experimentation. 

\begin{figure}[h]
    \centering
    \includegraphics[width=0.95\linewidth]{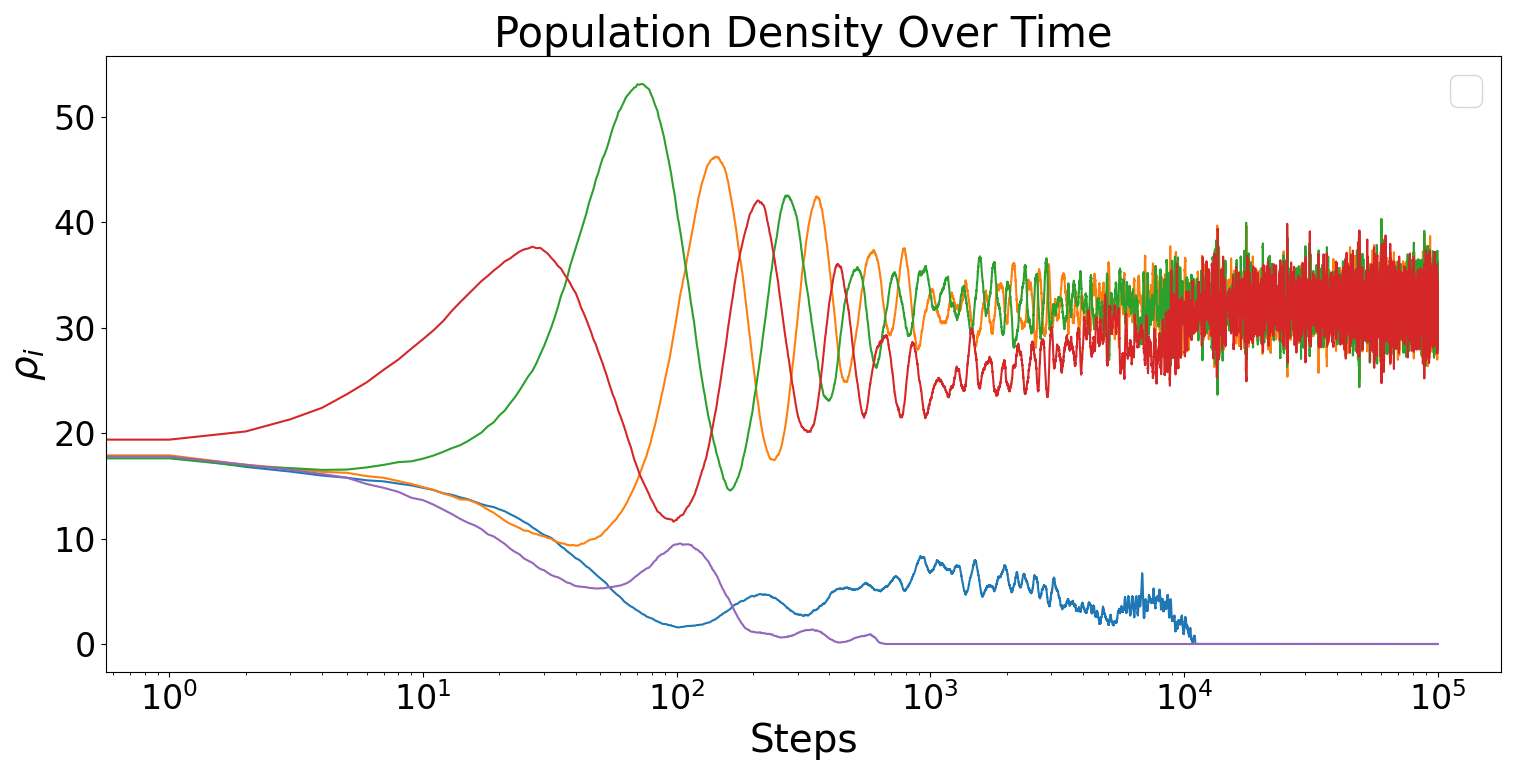}
    \caption{Metal-ESCG replication of Zhong et al. \textbf{Figure 2a}}
    \label{fig:metal-zhong}
\end{figure}

\noindent
To further reinforce consistency, the system was also used to replicate the experiment shown in Zhong et al.’s \textbf{Figure~2a}. As in the single-threaded version, the resulting density plots closely follow Zhong's original trends, showing matching inflection points in the same critical MCS time steps, shown in \textbf{Figure \ref{fig:metal-zhong}}. This confirms the simulation's fidelity to published behaviour across different configurations and implementations.

\newpage

\section{CUDA ESCG Implementation}

After completing the Metal ESCG implementation and proving its validity, attention turned toward its performance characteristics. Given that Metal is primarily designed for graphics rendering rather than general-purpose computation, questions emerged around whether greater performance gains could be achieved using a more compute-oriented platform. Motivated by this hypothesis, a Jetson Nano, an embedded system featuring an Nvidia GPU with 128 CUDA cores optimised for parallel processing, was acquired. This enabled the development of a CUDA-based ESCG implementation, tailored for high-throughput, general-purpose simulation tasks. \newline

Like the Metal-based version, the CUDA version uses a configuration header file (\texttt{config.cuh}) to store simulation parameters. However, the need for a dedicated \texttt{CUDAContext} struct is eliminated, since CUDA's \texttt{nvcc} compiler allows for host and device code to be compiled together. This enables GPU kernels to be directly invoked from the host without managing precompiled shader libraries, pipeline states, or Metal-specific dispatch semantics. \newline

This architectural transition introduced minimal semantic differences between the Metal and CUDA implementations, as most changes were syntactic. Random numbers for cell indices, neighbour directions, and action probabilities were precomputed in batches by invoking the GPU kernel \texttt{refreshRandomNumbers} specifically on the \texttt{stream\_numbers} CUDA stream. Each thread within this kernel independently generated multiple random values, directly writing these into global device memory buffers. These buffers were subsequently consumed by simulation kernels in both single-MCS mode (\texttt{cuda\_step}) and multi-MCS batch mode (\texttt{max\_cuda\_step}), dispatched separately on the \texttt{stream\_steps} CUDA stream. This clear separation of streams ensured asynchronous execution and efficient pipeline synchronisation between random number generation and simulation execution phases. CUDA utilises \texttt{cudaStream\_t} objects to manage asynchronous kernel execution and data transfers between host and device memory. Two streams were created to separately handle random number generation and simulation steps, allowing efficient parallel scheduling and synchronisation similar to Metal's command buffers:

\begin{lstlisting}[language=C++, style=cudastyle]
cudaStream_t stream_numbers, stream_steps;
cudaStreamCreate(&stream_numbers);
cudaStreamCreate(&stream_steps);
\end{lstlisting}

Explicit memory management is necessary in CUDA, requiring manual allocation of GPU buffers via \texttt{cudaMalloc()} for storing the simulation grid, dominance matrix, random number arrays, and density results:

\begin{lstlisting}[language=C++, style=cudastyle]
int* d_grid;
cudaMalloc(&d_grid, N * sizeof(int));
\end{lstlisting}

\noindent
Structurally, the CUDA implementation closely mirrors the Metal-based ESCG system. One difference, however, arises in the random number generation. CUDA benefits from the built-in \texttt{CURAND} library, offering well-tested, statistically robust pseudo-random number generation. Consequently, the CUDA version did not require the implementation of a custom Mersenne Twister algorithm, nor did it necessitate additional techniques such as seed hashing or burn-in periods, which had previously been essential in the Metal implementation. \newline

To ensure thread-safe operations on the shared lattice grid, CUDA atomic functions were used. Atomic operations permitted safe concurrent reading and writing of grid cells across thousands of parallel threads. In CUDA, a safe read was performed by atomically adding zero and assigning the result to a variable,  like so:

\begin{lstlisting}[language=C++, style=cudastyle]
int species = atomicAdd(&grid[cell_index], 0);
\end{lstlisting}

\noindent
Interaction and migration operations were executed using:

\begin{lstlisting}[language=C++, style=cudastyle]
atomicExch(&grid[neighbour_index], 0);                // Remove neighbour during interaction
atomicExch(&grid[cell_index], neighbour_specie);      // Migration or reproduction step
\end{lstlisting}

\noindent
Population densities were calculated on the GPU through the \texttt{compute\_densities} kernel, where each thread incremented an atomic species-count buffer to tally cell populations:

\begin{lstlisting}[language=C++, style=cudastyle]
atomicAdd(&result[species], 1);
\end{lstlisting}

\noindent
\newline
For the \texttt{max\_cuda\_step} mode, performance was further optimised by directly reusing memory-resident random number buffers. Since both the step logic and random number generation kernels operate on the same preallocated device memory, there is no need for intermediate \texttt{cudaMemcpy} operations. Instead, the random number pointers are passed directly to the \texttt{maxStep()} function, eliminating the overhead of copying potentially hundreds of millions of values between host and device. This results in more efficient memory usage and faster execution. \newline

The overall simulation loop for this mode is shown in \textbf{Algorithm~\ref{alg:cuda-maxstep}}. At each iteration, the number of Monte Carlo Steps (MCS) to simulate is calculated from the total number of random numbers generated. Random numbers are generated asynchronously on a separate CUDA stream (\texttt{stream\_numbers}), while the simulation step executes on \texttt{stream\_steps}, leveraging CUDA’s stream-based parallelism. The loop continues until either the specified MCS limit is reached or a stable state is detected via the \texttt{stasis()} function. \newline

\begin{algorithm}[H]
\caption{Simulation Loop (\texttt{--maxStep} enabled) in CUDA}
\label{alg:cuda-maxstep}
\SetKwFor{For}{for}{do}{}
\SetKwIF{If}{ElseIf}{Else}{if}{then}{else if}{else}{end}

$step \leftarrow \texttt{params.numRandoms} ~/ ~N$\;
\For{$mcs \leftarrow \texttt{currentMCS}$ to \texttt{MCS} in increments of $step$}{
    \texttt{densities(\dots)}\;
    \If{$mcs \in$ saveInterval \textbf{and} \texttt{params.save}}{   
        \tcp*[l]{save snapshots and export \texttt{.csv} files}
    }
    \If{$mcs = \texttt{MCS}$}{
        \textbf{break}\;
    }
    \texttt{generateRandomNumbers(d\_action\_probabilities, d\_cells, d\_neighbours, ...)}\;
    \texttt{maxStep(h\_grid, d\_grid, d\_dominance, d\_action\_probabilities, d\_cells, d\_neighbours, ...)} \tcp*[l]{Use the same RNG pointers}
    \If{\texttt{stasis()}}{
        \textbf{break}\;
    }
    \texttt{cudaStreamSynchronize(stream\_numbers)} \tcp*[l]{ Wait for RNG}\;
}
\end{algorithm}

\noindent
\newline
Meanwhile, the single-step simulation logic in CUDA retains a structure that closely mirrors the Metal implementation. However, where Metal relies on \texttt{commandBuffer} objects to manage kernel execution and synchronisation, CUDA substitutes these with explicit \texttt{cudaStream} synchronisation commands. This distinction is particularly evident in \textbf{Algorithm \ref{alg:cuda-step}}, where \texttt{cudaStreamSynchronize} is used to ensure that random number generation has completed before the results are copied and consumed by the simulation step. Despite the difference in API design, both approaches serve the same purpose: orchestrating asynchronous computation and data transfer between stages in a parallel processing pipeline. 
\begin{algorithm}[H]
\caption{Simulation Loop (\texttt{--maxStep} disabled) in CUDA}
\label{alg:cuda-step}
\SetKwFor{For}{for}{do}{}
\SetKwIF{If}{ElseIf}{Else}{if}{then}{else if}{else}{end}

\For{$mcs \gets \texttt{currentMCS}$ \KwTo \texttt{MCS}}{
    \texttt{densities(...)}\;
    
    \If{$mcs \in$ saveInterval \textbf{and} \texttt{params.save}}{
        \texttt{exportGridToCSV(...)}\;
    }

    \If{$mcs == \texttt{MCS}$}{
        \textbf{break}\;
    }

    \tcp*[l]{Populate StepContext with random numbers}
    \For{$i \gets 0$ \KwTo $N - 1$}{
        \If{\texttt{index} == 0}{
            \texttt{generateRandomNumbers(d\_action\_probabilities, d\_cells, d\_neighbours, ...)}\;
        }

        \If{\texttt{index} $\geq$ \texttt{params.numRandoms}}{
            \texttt{cudaStreamSynchronize(stream\_numbers)}\;
            \texttt{cudaMemcpy(...)} \tcp*[l]{Copy random buffers to host}
            \texttt{index} $\gets 0$\;
        }

        \texttt{stepCtx.cells[i]} $\gets$ \texttt{cells[index]}\;
        \texttt{stepCtx.neighbour\_dirs[i]} $\gets$ \texttt{neighbours[index]}\;
        \texttt{stepCtx.action\_probabilities[i]} $\gets$ \texttt{action\_probabilities[index]}\;
        \texttt{index++}\;
    }

    \texttt{cudaMemcpy(...)} \tcp*[l]{Transfer StepContext to device memory}
    \texttt{step(...)} \tcp*[l]{Invoke simulation step kernel}

    \If{\texttt{stasis()}}{
        \textbf{break}\;
    }
}
\end{algorithm}

\noindent
\newline
Due to compatibility issues linking \texttt{matplotlib-cpp} within a CUDA-based project, visualisations were not generated directly from the \textbf{io.cu} file. Instead, lattice snapshots and population densities were exported as CSV files, which were later visualised using a standalone Python script (\texttt{visualise.py}) that leveraged Python's \texttt{matplotlib} library. Visualisations are generated by providing the output folder name as a command-line argument to the script, which then automatically processes and plots all relevant \texttt{.csv} files within that directory.

\subsection{Results}

\begin{figure}[h]
    \centering
    \includegraphics[width=0.75\linewidth]{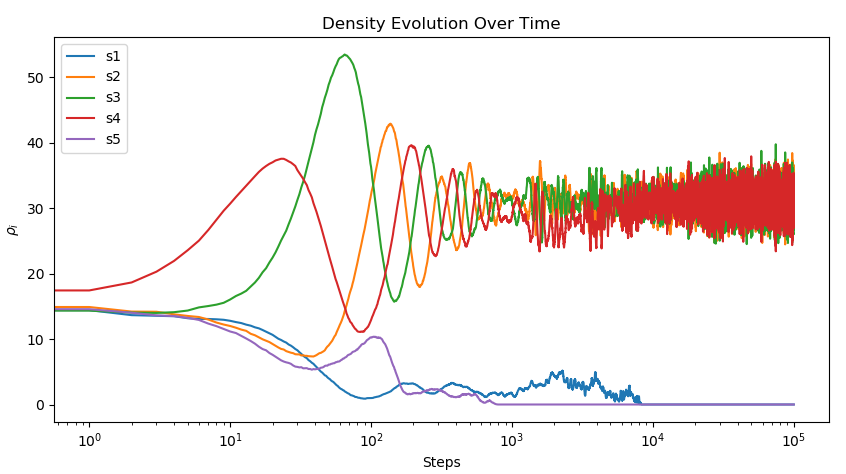}
    \caption{CUDA-based replication of Zhong et al.'s \textbf{Figure 2a}, illustrating the species density dynamics over time in an ablated RPSLS ESCG model.}
    \label{fig:cuda-zhong-densities}
\end{figure}

\begin{figure}[h]
    \centering
    \begin{subfigure}[]{0.32\textwidth}
        \includegraphics[width=\textwidth]{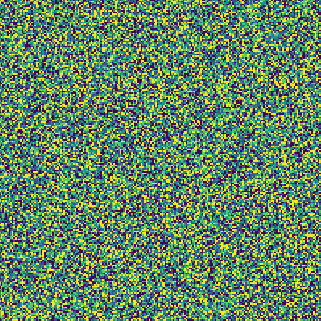}
        \caption{MCS = 0}
        \label{fig:cuda-zhong-0}
    \end{subfigure}
    \hfill
    \begin{subfigure}[]{0.32\textwidth}
        \includegraphics[width=\textwidth]{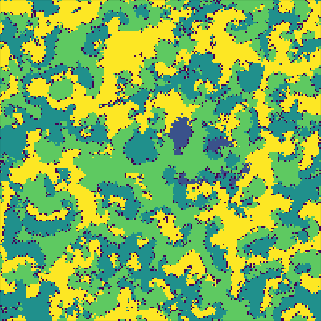}
        \caption{MCS = 6000}
        \label{fig:cuda-zhong-6000}
    \end{subfigure}
    \hfill
    \begin{subfigure}[]{0.32\textwidth}
        \includegraphics[width=\textwidth]{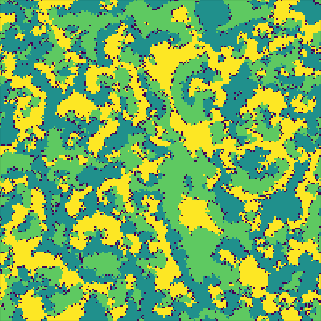}
        \caption{MCS = 50000}
        \end{subfigure}
    
    \caption{Snapshots from the CUDA-accelerated ESCG simulation of the ablated RPSLS system. At 6000 MCS, Rock species are still present, predominantly concentrated near the centre of the lattice. By 50000 MCS, however, the Rock species has gone extinct, highlighting a key phase in the long-term species dynamics.}
    \label{fig:cuda-zhong-ss}
\end{figure}

Once again, Zhong's \textbf{Figure 2 (\ref{fig:zhong2a})} serves as a benchmark for validating the correctness of this CUDA-based implementation. As with the previous systems, the resulting density plots produced by the CUDA simulation in \textbf{Figure \ref{fig:cuda-zhong-densities}} closely resemble those published by Zhong et al., reaffirming the system’s ability to capture the expected stochastic dynamics of ESCG models. In addition, spatial snapshots captured at corresponding MCS intervals, shown in \textbf{Figure \ref{fig:cuda-zhong-ss}}, visually support the trends observed in the density plots, further validating the accuracy and reliability of the simulation under CUDA. \newline

\begin{figure}[h]
    \centering
    \begin{subfigure}[]{0.32\textwidth}
        \includegraphics[width=\textwidth]{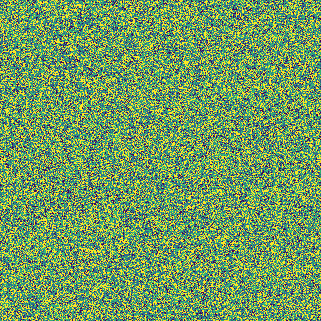}
        \caption{MCS = 0}
        \label{fig:cuda-spirals1}
    \end{subfigure}
    \hfill
    \begin{subfigure}[]{0.32\textwidth}
        \includegraphics[width=\textwidth]{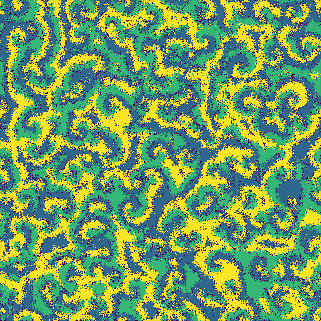}
        \caption{MCS = 50000}
        \label{fig:cuda-spirals2}
    \end{subfigure}
    \hfill
    \begin{subfigure}[]{0.32\textwidth}
        \includegraphics[width=\textwidth]{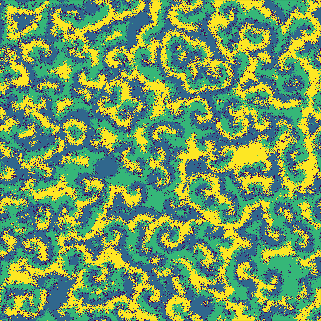}
        \caption{MCS = 100000}
         \end{subfigure}
    
    \caption{Snapshots from the CUDA-accelerated ESCG simulation replicating the spiral domain formations reported by Reichenbach, Mobilia, and Frey~\cite{Reichenbach2007MobilityGames} in their \textbf{Figure 2a}. Each frame corresponds to a different MCS, illustrating the emergence and evolution of spatial structures under low mobility conditions. Parameters: lattice size = $800 \times 800$, initial empty cell probability = $0.1$, mobility = $3 \times 10^{-6}$, von Neumann neighbourhood, circulant dominance among 3 species, and periodic boundary conditions. Source code available at: \url{<https://github.com/louiesinadjan/escg>}.}
    \label{fig:solid-spirals}
\end{figure}

The accuracy and reliability of the CUDA-based system are reinforced through successful replications of key results from Reichenbach, Mobilia, and Frey. As shown in \textbf{Figure \ref{fig:solid-spirals}}, the system captures the formation of solid spiral domains characteristic of low-mobility regimes in spatially structured ESCGs. These patterns, which emerge when mobility is set to $M = 3 \times 10^{-6}$, mirror those presented in \textbf{Figure 2a of Reichenbach et al.} (\textbf{Figure \ref{fig:ReiMobFrey2a}}), thereby validating the simulation's correctness. Furthermore, the use of a larger lattice size ($800 \times 800$) highlights the scalability of the GPU-accelerated approach and its potential to extend the scope of analysis beyond previously published work.

\chapter{Critical Evaluation}
\label{chap:evaluation}

\section{Stochastic Validity}

The first objective of this project was to implement and validate single-threaded (in C++), Metal, and CUDA-based ESCG simulation frameworks, ensuring accuracy and consistency across all platforms through replication of known ecological dynamics from existing literature. This objective has been successfully achieved. All three implementations were developed independently, yet consistently reproduced key spatial and stochastic patterns reported in prior studies, such as the emergence of spiral domains and population oscillations described by Reichenbach et al. \cite{Reichenbach2007MobilityGames} \textbf{(\ref{fig:spirals}, \ref{fig:hazyspirals}, \ref{fig:solid-spirals}}) and Zhong et al. \cite{Zhong2022SpeciesSpecies} (\textbf{\ref{fig:zhong-replication}, \ref{fig:metal-zhong}, \ref{fig:cuda-zhong-densities})}. Comparative visual analysis of lattice snapshots, as well as quantitative agreement in density evolution plots, demonstrated the functional equivalence of each system. These replications validate the correctness of each implementation and confirm that the underlying simulation logic was faithfully preserved across CPU and GPU architectures. As a result, it can be declared that each framework provides a sound and reliable foundation for subsequent benchmarking, scalability testing, and extended experimentation.

\section{Execution Efficiency}

This chapter presents an evaluation of the execution efficiency of the GPU-accelerated ESCG systems developed throughout this project. Beyond correctness and functionality, a key objective of this dissertation was to achieve significant speedups over the single-threaded baseline. To that end, a number of carefully considered compilation strategies were adopted across the different implementations. These decisions were made with the explicit aim of maximising runtime performance by fully leveraging the capabilities of the underlying hardware, and ensuring fair and reproducible comparisons between CPU, Metal, and CUDA backends.

\subsection{Compilation}

The single-threaded implementation was compiled using \texttt{clang++} with the \texttt{-std=c++17} and \texttt{-O3} flags. While the former enables modern language features that support clearer and potentially more efficient code structures, the latter plays a more critical role in performance. The \texttt{-O3} optimisation level activates advanced compiler transformations such as aggressive function inlining, loop unrolling, and vectorisation. These optimisations are particularly beneficial for the tight, iterative computations and memory-intensive patterns characteristic of ESCG simulations.\\ 

The Metal-accelerated implementation was also compiled with \texttt{clang++}, utilising the same \texttt{-O3} optimisation level along with the \texttt{-march=native} flag. The inclusion of \texttt{-march=native} enables the generation of instructions optimised for the specific host CPU architecture, further improving performance through better utilisation of vector units and cache hierarchies. Metal shader files were compiled using Apple’s \texttt{metal} compiler with \texttt{-O3}, producing intermediate \texttt{.air} files which were subsequently linked into a single \texttt{.metallib} binary. These steps ensured that GPU-side execution was optimised both at the compilation and linking stages, minimising overhead during runtime. \\ 

The CUDA implementation followed a similar optimisation strategy. Compilation was performed using \texttt{nvcc}, with \texttt{-O3} applied to both host and device code. This allowed the CUDA compiler to apply performance-critical transformations such as loop unrolling, constant folding, and fast math approximations. CUDA source files were compiled into a single monolithic binary, reducing dynamic linking overhead and facilitating faster kernel launch times. \\

Collectively, these compilation choices ensured that all implementations (CPU, Metal, and CUDA) were executed under conditions optimised for maximum throughput and minimal latency, enabling a fair and rigorous performance comparison.

\subsection{Random Number Generation}

To contextualise performance improvements, the single-threaded C++ implementations serve as a baseline benchmark against which the GPU-accelerated systems are compared. The Mersenne Twister pseudorandom number generation algorithm is first benchmarked independently across platforms. Identical implementations of the algorithm were deployed on each system: C++, Metal, and CUDA, to assess raw performance speedup outside the context of full ESCG execution. This provides insight into the computational gains achievable on more general-purpose tasks and reveals how individual components of the ESCG pipeline, such as stochastic sampling, can benefit from parallel acceleration. \\

\begin{figure}[h]
    \centering
    \includegraphics[width=1\linewidth]{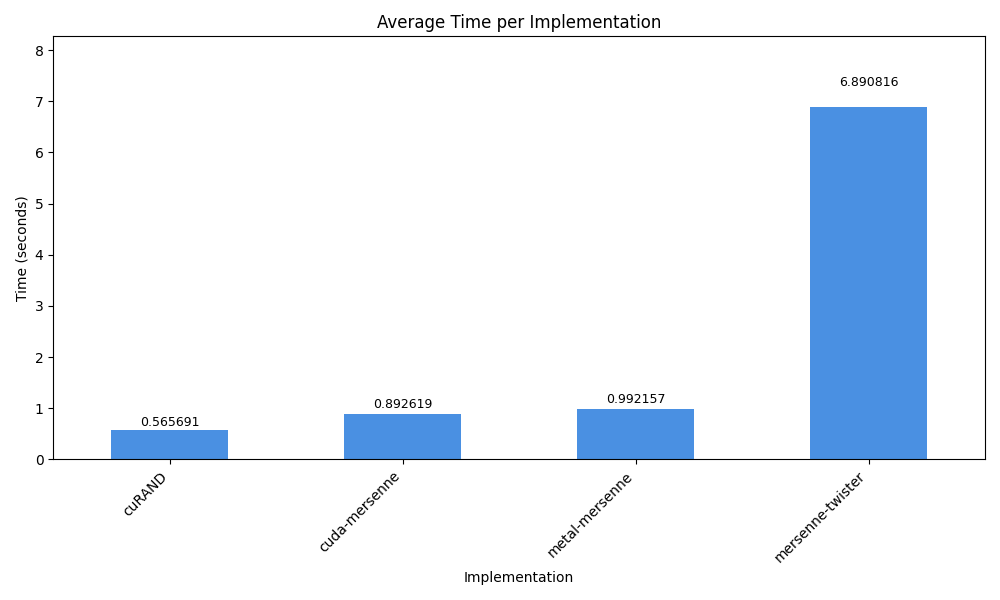}
    \caption{Average execution time (in seconds) for generating $10^{9}$ pseudorandom numbers using four different implementations: single-threaded C++ Mersenne Twister, Metal-based Mersenne Twister, CUDA-based Mersenne Twister, and CUDA’s native \texttt{cuRAND} library. 100 trials were run to compute these averages.}
    \label{fig:mersenne-comparison}
\end{figure}

\textbf{Figure \ref{fig:mersenne-comparison}} presents the average execution time for each pseudorandom number generator implementation when tasked with generating a total of one billion random numbers. Due to the memory constraints associated with this quantity (approximately 4GB of data) each benchmark trial was structured to generate 100 million random numbers per iteration, with the process repeated ten times. After each generation, the numbers were copied into a designated target array on the host device. Including memory transfers in the benchmark was a deliberate design choice, intended to more accurately reflect the operational context within ESCG simulations. Since these simulations frequently require copying large volumes of precomputed random numbers between memory buffers, this benchmark provides a realistic measure of end-to-end throughput and highlights potential performance bottlenecks beyond raw generation speed alone. \\

The benchmark results reveal substantial performance improvements achieved through GPU acceleration. As shown in \textbf{Figure~\ref{fig:mersenne-comparison}}, the baseline single-threaded C++ implementation of the Mersenne Twister algorithm required approximately 6.89 seconds to generate and store one billion random numbers, segmented into ten batches of 100 million values each. By comparison, the CUDA-based implementation of the same algorithm completed the task in just 0.89~seconds, yielding a speedup of approximately $6.89 \div 0.89 \approx 7.7\times$. Similarly, the original custom Metal-based Mersenne Twister variant achieved a runtime of 0.99~seconds, corresponding to a $6.89 \div 0.99 \approx 6.96\times$ improvement. \\

Among all tested implementations, Nvidia’s \texttt{cuRAND} library demonstrated the highest performance, completing the task in 0.57~seconds—resulting in a speedup of $6.89 \div 0.57 \approx 12.1\times$ relative to the original C++ version. Not only does \texttt{cuRAND} offer superior throughput and lower latency, but it also provides high-quality, immediate pseudorandomness. Unlike the custom Mersenne Twister generators used in the Metal and CUDA implementations, \texttt{cuRAND} does not require additional enhancements such as seed hashing or burn-in phases to improve stochastic fidelity. These characteristics ultimately motivated its selection as the default random number generation method in the final CUDA-based ESCG simulation framework. \\

It is important to note, however, that the speedups observed in pseudorandom number generation (PRNG) do not linearly translate to equivalent gains in overall ESCG simulation performance. The simplified relation:
\[
\texttt{(PRNG Speedup)} \times \texttt{(Step Logic Speedup)} \approx \texttt{Overall ESCG Speedup} \quad \leftarrow \text{Not necessarily valid}
\]
\noindent
does not generally hold, as it overlooks resource contention and shared computational constraints inherent to GPU execution. In the standalone PRNG benchmarks, all available GPU cores are fully dedicated to generating random numbers. In contrast, a full ESCG simulation must concurrently allocate those same computational resources to both stochastic number generation and parallel step logic processing. As a result, peak performance measured in isolation may not be achievable when both components are active within the same kernel execution timeline, limiting the practical realisation of idealised compound speedups.

\subsection{Optimal Max Step}

Following the evaluation of speedup achieved through GPU-accelerated pseudorandom number generation, the next benchmark investigates the performance impact of varying the number of random numbers generated per kernel invocation. This parameter is particularly critical in \texttt{--maxStep} simulation mode, where a larger batch of random numbers directly translates to more MCS being processed per invocation of the \texttt{step} kernel. However, generating excessively large batches may introduce significant delays before elementary updates are applied to the lattice, creating a trade-off between throughput and responsiveness. \\

To explore this balance, benchmark trials were conducted using ten CUDA-based ESCG simulations at three different system sizes: $L = 100$, $200$, and $400$, with \texttt{--maxStep} enabled and \texttt{--numRandoms} varied across configurations to simulate up to 100{,}000 MCS. The aim was to assess how the size of the random number batch influences runtime performance across increasing lattice scales. All other command-line parameters were considered irrelevant for this analysis, as they do not introduce measurable computational overhead or significantly affect runtime behaviour (other than the \texttt{--save} flag which was disabled due to the exporting computation). \\ 

It is important to note that in the default (non-maxStep) simulation mode, the value of \texttt{numRandoms} has little impact on overall runtime. This is because each kernel invocation processes exactly one MCS, and the system rarely experiences delays due to random number regeneration. By contrast, in \texttt{--maxStep} mode, performance becomes tightly coupled to the batching strategy. This experiment, therefore, isolates and evaluates the trade-offs associated with batch size selection under GPU-accelerated multi-MCS execution, especially as system complexity increases with lattice size. \\ 

\begin{figure}[h]
    \centering
    \includegraphics[width=1\linewidth]{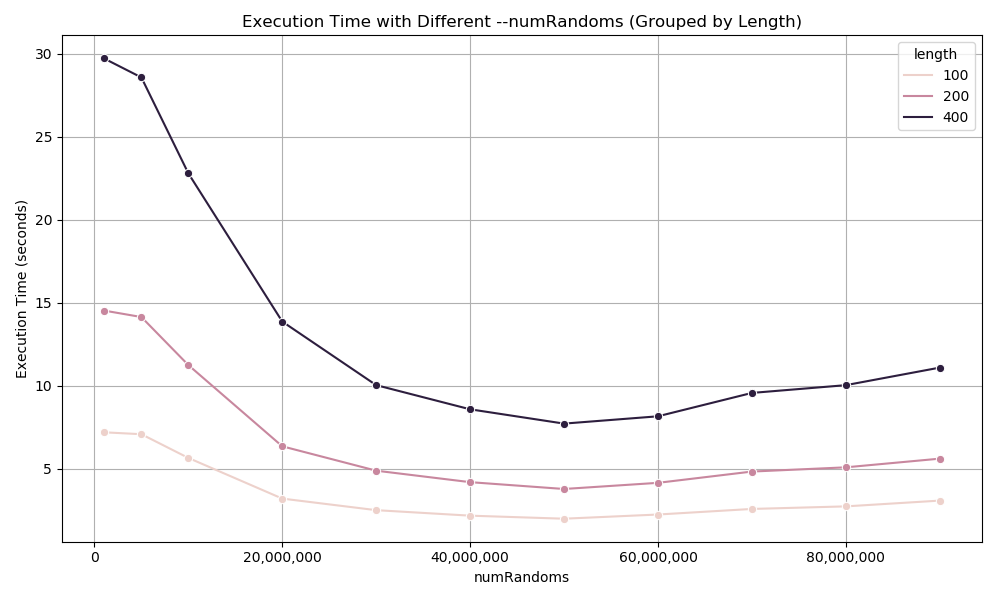}
    \caption{Execution time of the CUDA-based ESCG implementation as a function of the \texttt{--numRandoms} parameter, benchmarked across three lattice sizes: $100 \times 100$, $200 \times 200$, and $400 \times 400$. Each line represents the total time taken to complete a $100{,}000$ MCS simulation with \texttt{--maxStep} enabled.}
    \label{fig:numRandoms}
\end{figure}

As shown in \textbf{Figure \ref{fig:numRandoms}}, the runtime curves follow a broadly quadratic trend, with the minimum runtime observed around a \texttt{numRandoms} value of $50{,}000{,}000$ across all tested lattice sizes. This shape aligns with intuitive expectations: generating too many random numbers per kernel call increases latency due to prolonged pre-processing before simulation steps are executed, while generating too few leads to more frequent kernel launches and associated overheads—both of which degrade performance. \\

However, it may initially seem counterintuitive that each curve, regardless of system size, reaches its minimum at approximately the same \texttt{numRandoms} value. Since smaller lattices incur significantly lower per-MCS computational costs (scaling quadratically with lattice dimension) it would be reasonable to expect that they could accommodate larger batches of random numbers without encountering performance degradation. Nevertheless, the observed uniformity across system sizes is likely due to architectural factors: random number generation and grid update kernels are dispatched on separate CUDA streams, and the overall performance bottleneck arises not solely from simulation step execution but from a complex interplay between memory bandwidth saturation, kernel launch overhead, and inter-stream synchronisation latency. These overheads scale sublinearly—or in some cases, independently—of lattice size, leading to a hardware-constrained rather than simulation-size-constrained performance floor. It is also worth noting that these benchmarks were conducted on an Nvidia RTX A2000 GPU; more powerful devices featuring higher memory bandwidths and greater numbers of CUDA cores may be able to support larger optimal \texttt{numRandoms} values, thereby shifting the minima rightward in future experiments and conversely for less powerful devices.  Hence, it is recommended that a similar profiling trial be conducted prior to experimentation or results generation in order to identify the optimal batching threshold and maximise simulation throughput.

\newpage
\subsection{ESCG Simulations}

\begin{figure}[h]
    \centering
    \includegraphics[width=0.9\linewidth]{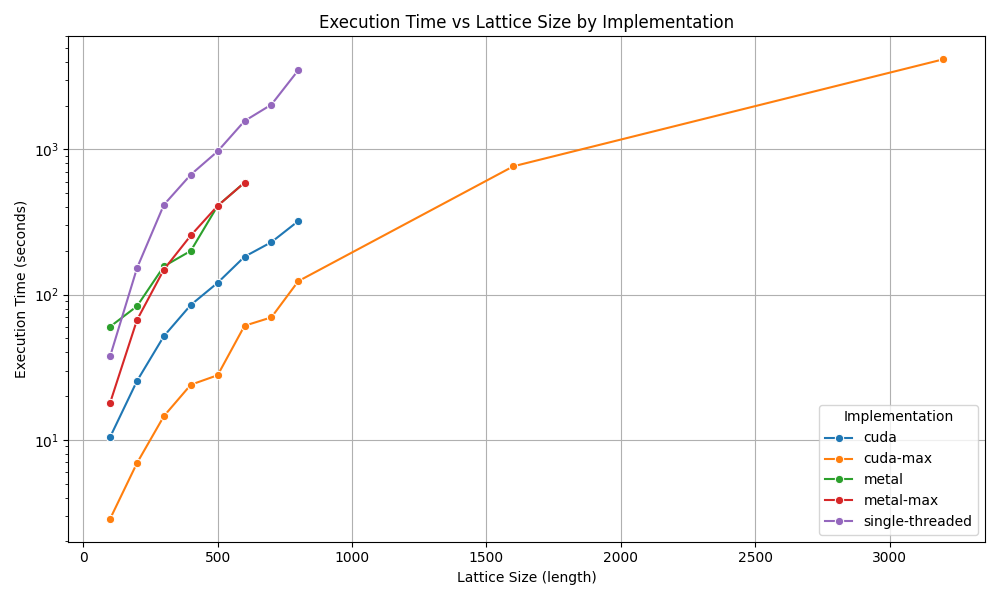}
    \caption{Comparison of execution times of different ESCG implementations across increasing lattice sizes to 100{,}000 MCS.}
    \label{fig:comparisons}
\end{figure}

\begin{figure}[h]
    \centering
    \includegraphics[width=1\linewidth]{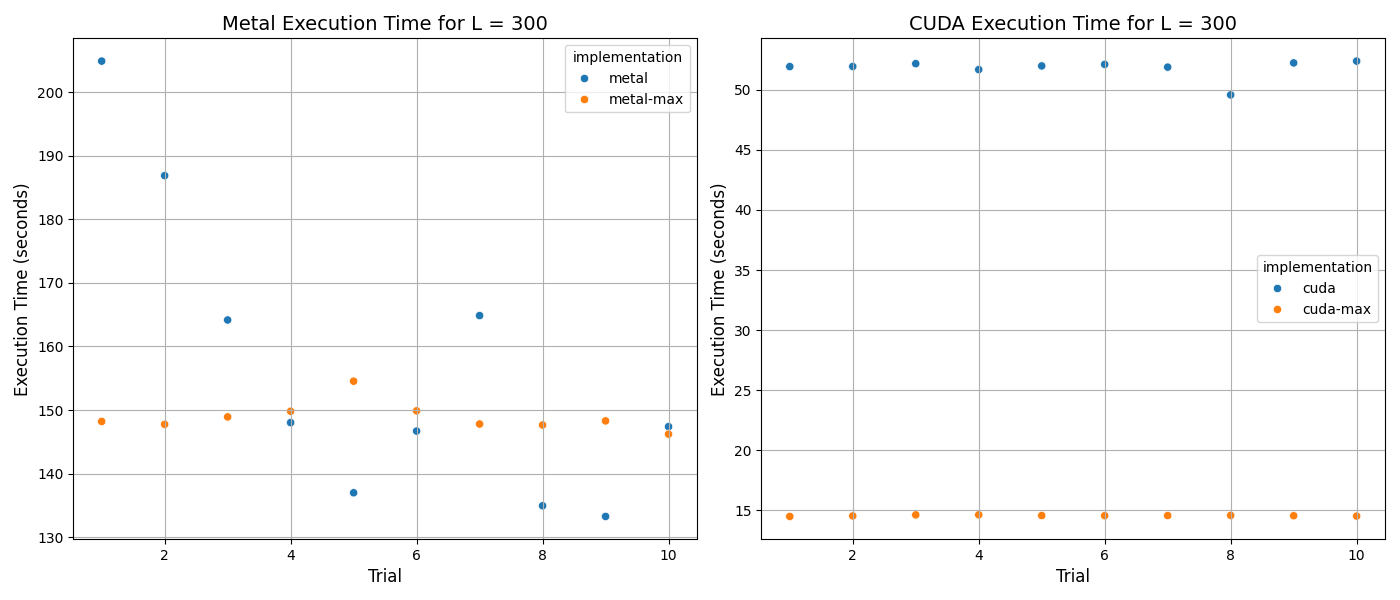}
    \caption{Execution time over 10 trials for both Metal and CUDA ESCG implementations at $L = 300$, comparing single-MCS and batched multi-MCS (\texttt{maxStep}) modes for both implementations.}
    \label{fig:warmups}
\end{figure}

\begin{landscape}
\begin{table}[h]
    \centering
    \caption{\normalsize Mean execution times (in seconds) with standard deviations for each implementation across increasing lattice sizes (\(L \times L\)). Results are averaged over multiple independent simulation runs. The fastest implementation for each lattice size is highlighted in bold. \textit{MS} refers to multi-MCS \textit{max step} implementations.}
    \label{tab:averages}
    \renewcommand{\arraystretch}{1.8} 
    \resizebox{\linewidth}{!}{%
    \begin{tabular}{|c|c|c|c|c|c|}
        \hline
        \textbf{Lattice Size} & \textbf{CUDA} & \textbf{CUDA-MS} & \textbf{Metal} & \textbf{Metal-MS} & \textbf{Single-threaded} \\
        \hline
        100   & 10.46 $\pm$ 0.19 & \textbf{2.86} $\pm$ 0.02 & 60.18 $\pm$ 1.35 & 18.02 $\pm$ 0.18 & 37.66 $\pm$ 0.68 \\
        200   & 25.46 $\pm$ 0.78 & \textbf{6.93} $\pm$ 0.04 & 83.26 $\pm$ 3.69 & 66.96 $\pm$ 0.49 & 152.48 $\pm$ 1.97 \\
        300   & 51.79 $\pm$ 0.80 & \textbf{14.57} $\pm$ 0.04 & 156.81 $\pm$ 23.62 & 148.91 $\pm$ 2.25 & 415.76 $\pm$ 0.92 \\
        400   & 84.88 $\pm$ 2.50 & \textbf{23.95} $\pm$ 0.08 & 200.13 $\pm$ 0.29 & 255.38 $\pm$ 31.96 & 670.20 $\pm$ 80.27 \\
        500   & 120.67 $\pm$ 1.00 & \textbf{27.86} $\pm$ 0.08 & 409.04 $\pm$ 2.33 & 408.87 $\pm$ 4.72 & 970.44 $\pm$ 1.91 \\
        600   & 182.44 $\pm$ 4.86 & \textbf{61.04} $\pm$ 0.11 & 591.12 $\pm$ 1.43 & 590.85 $\pm$ 1.63 & 1568.87 $\pm$ 1.02 \\
        700   & 229.78 $\pm$ 0.58 & \textbf{69.74} $\pm$ 0.16 & -- & -- & 2034.92 $\pm$ 0.94 \\
        800   & 321.12 $\pm$ 0.93 & \textbf{123.80} $\pm$ 0.50 & -- & -- & 3511.17 \\
        1600  & -- & \textbf{764.78} $\pm$ 3.62 & -- & -- & -- \\
        3200  & -- & \textbf{4173.50} & -- & -- & -- \\
        \hline
    \end{tabular}
    }
    \renewcommand{\arraystretch}{1.0}
\end{table}
\end{landscape}

Despite similar speedups observed during pseudorandom number generation, the CUDA implementation (executed on an Nvidia RTX A2000) significantly outperforms both the Metal implementation and the single-threaded baseline, as illustrated in \textbf{Figure \ref{fig:comparisons}}. At a lattice size of $L=100$, the standard CUDA implementation achieves an average execution time of 10.46 seconds, yielding a $\approx3.6\times$ speedup over the single-threaded version (37.66 seconds). The \texttt{--maxStep} variant improves this further, completing in just 2.86 seconds—a $\approx13.2\times$ speedup. Notably, the greatest performance gain occurs at $L=800$, where the CUDA \texttt{--maxStep} implementation completes the full simulation in 123.80 seconds, compared to the single-threaded version’s 3511.17 seconds, representing a $\approx28.4\times$ speedup. This dramatic increase highlights the scalability and computational advantage of batching thousands of Monte Carlo steps per kernel invocation on CUDA-enabled hardware. \\

In contrast, the Metal implementation on the M1 Pro MacBook Pro required 60.18 seconds for the same task, slower than the single-threaded benchmark, primarily due to the substantial overhead of initialising Metal pipeline objects and dispatching frequent small kernel invocations. However, the Metal \texttt{--maxStep} mode showed improved efficiency, completing the simulation in 18.02 seconds. This is attributed to the small system size, which allowed the 100,000 MCS to be completed using fewer kernel invocations, thereby reducing overhead and enabling better amortisation of GPU setup costs. \\

\begin{figure}[h]
    \centering
    \begin{subfigure}[t]{0.3\textwidth}
        \centering
        \includegraphics[width=\linewidth]{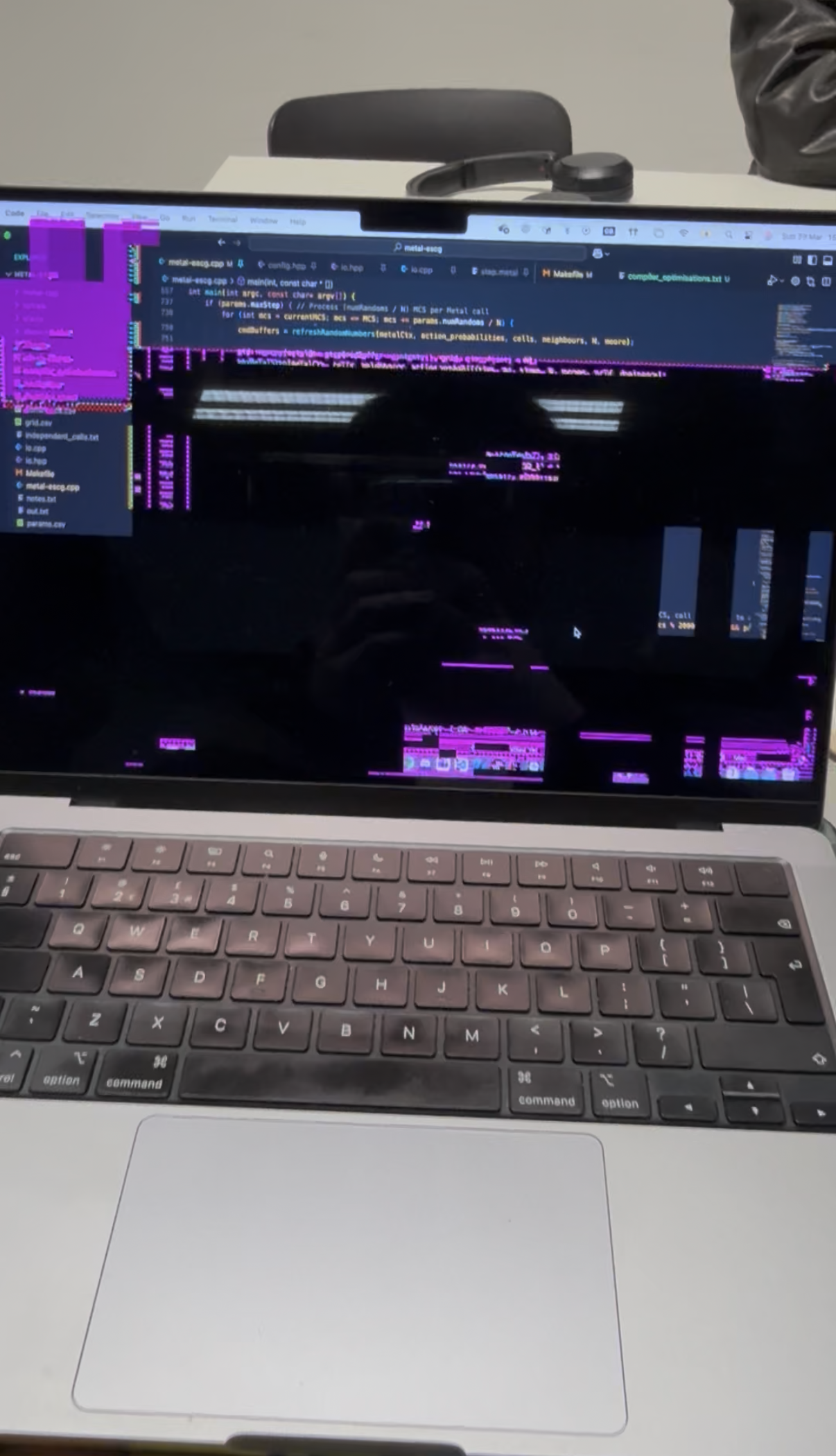}
        \label{fig:metal-bug1}
    \end{subfigure}
    \hspace{0.06\textwidth} 
    \begin{subfigure}[t]{0.3\textwidth}
        \centering
        \includegraphics[width=\linewidth]{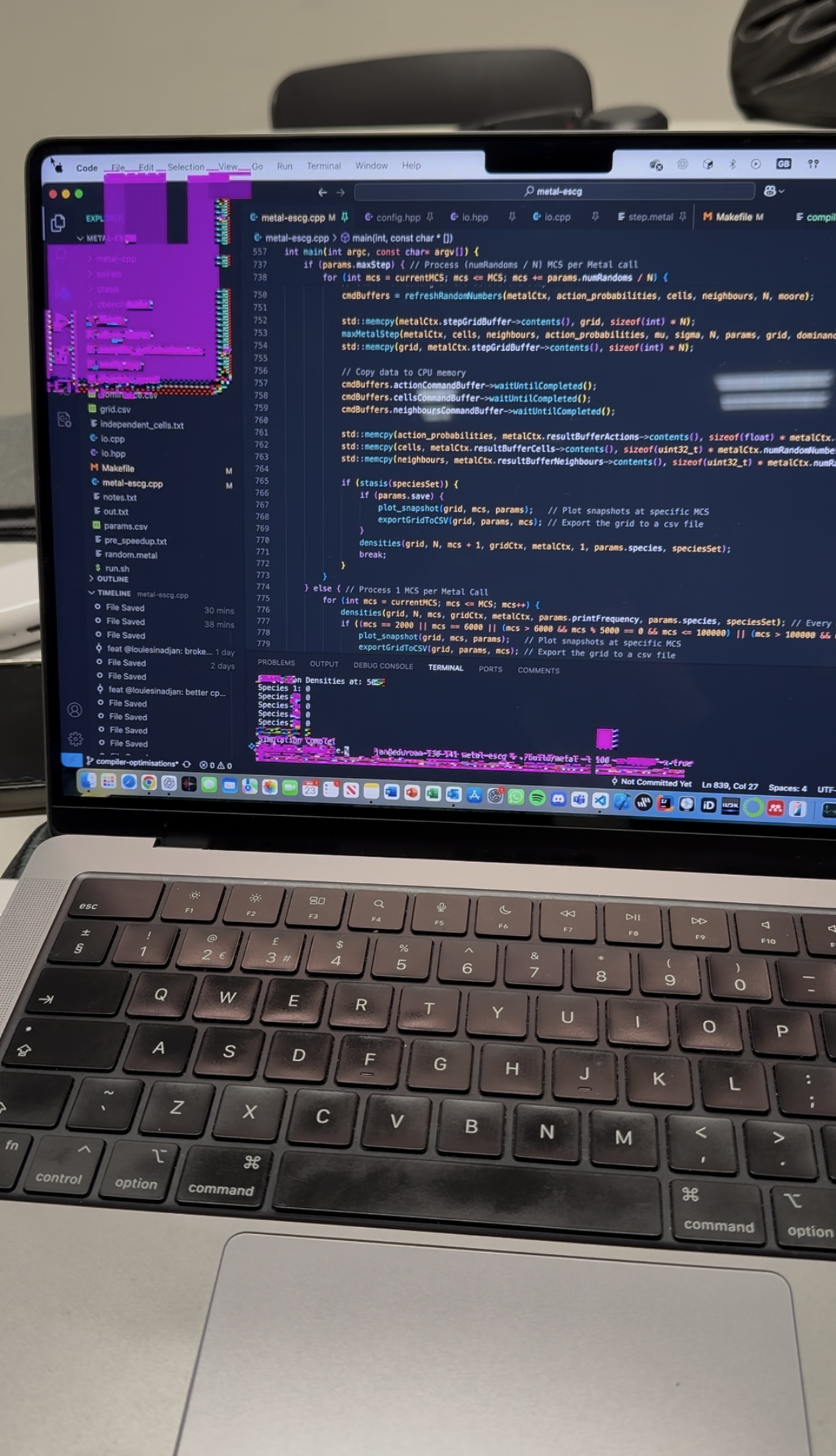}
        \label{fig:metal-bug2}
    \end{subfigure}
    \caption{Visual corruption observed when running large-lattice Metal simulations on the M1 Pro MacBook Pro. As shown, extended dispatching of high-thread-count kernels caused graphics driver instability, indicating the system’s difficulty in sustaining Metal-based compute workloads at scale.}
    \label{fig:metal-crash}
\end{figure}

Although the Metal \texttt{--maxStep} implementation consistently outperforms its single-MCS counterpart at smaller lattice sizes, both variants converge in execution time as $L$ increases, as shown in \textbf{Figure~\ref{fig:comparisons}} and detailed in \textbf{Table~\ref{tab:averages}}. This plateau in performance gain suggests that Metal's inherent architectural overhead, such as command buffer synchronisation and memory transfers, becomes increasingly significant with larger system sizes, limiting its scalability. Moreover, the Metal API's strategy of aggressively utilising the entire GPU for thread dispatching exposes limitations at large workloads. \textbf{Figure~\ref{fig:metal-crash}} demonstrates visual corruption and system instability observed during large-scale Metal simulations on Apple Silicon. These symptoms included rendering failures, unresponsiveness, and eventually forced restarts—signalling that the GPU was overwhelmed by excessive dispatch demands. \\

In contrast, the CUDA-based implementations exhibit more favourable scaling behaviour. The CUDA \texttt{--maxStep} version not only outpaces its single-MCS counterpart at all tested lattice sizes, but its advantage widens with increasing $L$. This is a direct result of CUDA’s superior parallel execution capabilities, efficient kernel batching, and hardware-level concurrency features. Notably, the CUDA \texttt{--maxStep} implementation was the only system capable of completing simulations at lattice sizes up to $L=3200$ within reasonable runtime constraints. At these extreme sizes, simulations using Metal or the single-threaded approach would be computationally infeasible due to polynomial increases in both step execution and memory latency. These findings further underscore the CUDA platform’s suitability for large-scale stochastic modelling tasks and highlight the practical advantages of GPU-accelerated ESCG systems in enabling complex ecological simulations previously restricted by computational limitations. \\

Furthermore, Metal was unable to scale to large $L$ as its design in utilising all of the GPU for thread dispatching quickly showed flaws. \textbf{Figure \ref{fig:metal-crash}} illustrates unintended consequences of overusing the GPU on Apple Silicon-leading to unresponsiveness, rendering crashes and soon after a forced restart by the system. \\

While the Metal-based ESCG implementation exhibited modest speedups over the single-threaded baseline on Apple Silicon devices, these gains remained comparatively limited when benchmarked against the CUDA-based systems. This constrained acceleration is primarily attributable to suboptimal thread dispatching and the overhead of repeated memory transfers between the CPU and GPU. In the current Metal and CUDA single MCS implementations, \texttt{StepContext} arrays containing action probabilities, cell indices, and neighbour directions are computed on the host and copied into the GPU before every kernel invocation. A more efficient alternative would involve keeping the precomputed large random number buffers and storing them in device-local memory, allowing shaders to access values via an index offset. This approach minimises memory transfer overhead and takes advantage of GPU-side caching to improve access times. \\

In the \texttt{--maxStep} mode, where the entire buffer is consumed in a single kernel dispatch, the benefits of device-only memory are even more pronounced. By eliminating unnecessary round-trip transfers to the host, this design reduces latency and improves throughput. Although this optimisation is already realised in the CUDA \texttt{--maxStep} implementation, where random numbers are preallocated in device memory and accessed directly, the Metal version still incurs the cost of copying input buffers per dispatch. This may introduce substantial bottlenecks, especially at larger system sizes or higher MCS counts, and restricts Metal’s ability to fully utilise the GPU’s concurrency and caching capabilities. \\

Importantly, these memory residency strategies apply not only to Metal but also to CUDA. Extending device-resident memory allocation to the single-step CUDA kernels would further reduce bandwidth contention and unlock additional performance gains, particularly in large-scale simulations where memory movement is a dominant cost. This strategy would be particularly effective in accelerating simulations where per-MCS data granularity is essential for analysis. \\

Another notable observation from the benchmarking results arises when examining the Metal simulation trials, as illustrated in \textbf{Figure~\ref{fig:warmups}} and corroborated by the large standard deviation observed at $L = 300$ in \textbf{Table~\ref{tab:averages}}. The early trials of the single-MCS Metal implementation exhibit higher execution times and greater variability, a trend attributable to initial pipeline state object (PSO) compilation and shader caching mechanisms. These one-time setup costs are progressively amortised in subsequent runs, resulting in improved and more consistent performance. By contrast, the \texttt{metal-max} implementation maintains a consistently lower variance across all trials, owing to its reduced number of kernel invocations and more efficient utilisation of GPU initialisation overhead. \\

Meanwhile, the CUDA trials, shown in \textbf{Figure~\ref{fig:warmups}}, demonstrate a markedly different behaviour. When configured with equivalent parameters, the CUDA implementation displays low inter-trial variance and stable performance from the outset. This lack of a noticeable warm-up phase highlights the efficiency of CUDA’s runtime and driver-level kernel management, which appears to incur minimal initial overhead. This contrast underscores the importance of accounting for platform-specific startup behaviours when evaluating the runtime performance of GPU-accelerated systems. \\

Additionally, it is worth noting that the relatively large standard deviation observed at $L = 400$ for the single-threaded implementation may be attributed to competing CPU resource contention during benchmarking. In contrast, for the \texttt{metal-max} implementation at $L = 400$, the comparatively lower variability is likely a result of accumulated caching effects and Metal runtime optimisations, further emphasising the benefits of reduced kernel dispatch frequency in improving consistency and throughput. However, it remains unclear why such optimisation effects are sometimes prominently observed (e.g., at $L = 300$) but not consistently at all lattice sizes. This inconsistency suggests that factors such as GPU load balancing, background operating system processes, and runtime heuristics may also influence the extent of Metal’s caching and pipeline reuse during execution. \\

Overall, it can be concluded that a GPU-extended ESCG implementation that accelerates simulation has been successfully achieved. The CUDA implementation, particularly in \texttt{--maxStep} mode, consistently outperformed both the Metal and single-threaded systems across all tested lattice sizes. The most substantial improvement was observed at $L=800$, where CUDA \texttt{--maxStep} yielded a $\approx28.4\times$ speedup over the single-threaded baseline. These results demonstrate that parallelisation on general-purpose GPUs not only accelerates individual simulations but also enables large-scale experimentation previously limited by compute constraints. This directly fulfils the third objective outlined in the introduction: to evaluate and quantify the performance gains achieved through GPU acceleration compared to conventional single-threaded simulations. The benchmarks presented confirm improved scalability, significantly higher throughput, and practical viability of GPU-based ESCG simulations across varying system sizes and configurations.

\newpage
\section{Extending Current Literature}

The primary motivation behind developing GPU-accelerated ESCG systems was to provide a research platform for this type of ecological modelling, enabling the replication and expansion of key studies at larger scales and in less time. With enhanced performance, my implementations allow researchers to simulate larger systems, explore more intricate interaction rules, and push simulations to greater temporal extents—all of which are often prohibitive on conventional CPU-based platforms. 

\subsection{Reichenbach, Mobilia and Frey}

As a first demonstration, I revisited and extended the influential findings of Reichenbach, Mobilia, and Frey~\cite{Reichenbach2007MobilityGames}, whose seminal exploration of mobility-driven pattern formation in three-species ESCGs has become a cornerstone in the field. The emergence of spiral domains and their dependence on mobility thresholds bears strong resemblance to phenomena observed in real-world ecosystems~\cite{Siegert1995SpiralMounds, Igoshin2004WavesAggregation}. However, the tractability of three-species models makes them a simplification of more complex ecological systems—generalising to larger numbers of interacting species remains a key challenge.

\begin{figure}[h]
    \centering
    \begin{subfigure}[]{0.30\textwidth}
        \includegraphics[width=\textwidth]{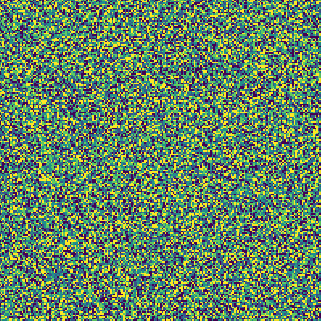}
        \caption{MCS = 0}
        \label{fig:5specie1}
    \end{subfigure}
    \hfill
    \begin{subfigure}[]{0.30\textwidth}
        \includegraphics[width=\textwidth]{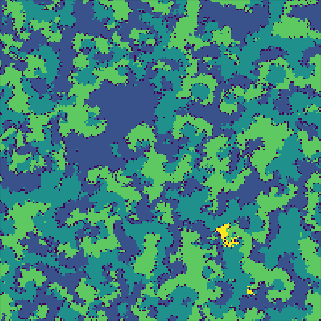}
        \caption{MCS = 55000}
        \label{fig:5specie2}
    \end{subfigure}
    \hfill
    \begin{subfigure}[]{0.30\textwidth}
        \includegraphics[width=\textwidth]{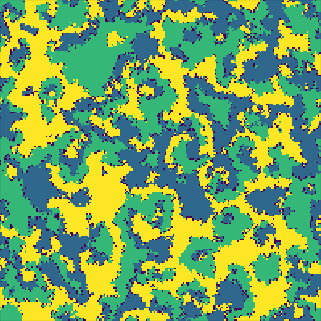}
        \caption{MCS = 100000}
        \label{fig:5specie3}
        \end{subfigure}
    \caption{Snapshots from a 5-species ESCG, circulant dominance matrix where each species dominates two other species, $N = 200\times200$, $M=1\times10^{-6}$.}
    \label{fig:5specie}
\end{figure}

As illustrated in \textbf{Figure~\ref{fig:5specie}}, a representative 5-species ESCG simulation, executed using parameters commonly found in recent literature, highlights the limitations imposed by small system sizes. Despite the symmetric structure of the dominance matrix, one species becomes extinct by MCS = 55,000, resulting in a reduced four-species ecosystem. This early loss can be attributed not to inherent asymmetries in interaction rules, but to the interplay of restricted spatial capacity and mobility-driven fluctuations, which amplify local population imbalances. By MCS = 100,000, these dynamics further simplify the system into a three-species configuration—significantly diminishing the intended ecological complexity. \\

\begin{figure}[h]
    \centering
    \begin{subfigure}[]{0.30\textwidth}
        \includegraphics[width=\textwidth]{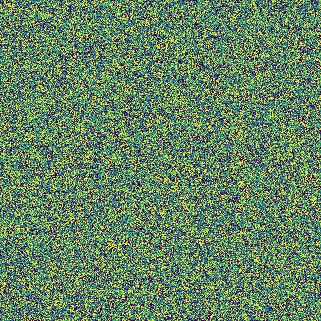}
        \caption{MCS = 0}
    \end{subfigure}
    \hfill
    \begin{subfigure}[]{0.30\textwidth}
        \includegraphics[width=\textwidth]{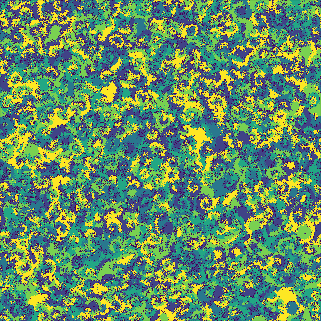}
        \caption{MCS = 2000}
    \end{subfigure}
    \hfill
    \begin{subfigure}[]{0.30\textwidth}
        \includegraphics[width=\textwidth]{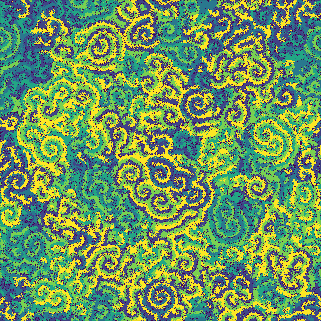}
        \caption{MCS = 100000}
        \end{subfigure}
    \caption{Snapshots from a 5-species ESCG, circulant dominance matrix where each species dominates two other species, $N = 3200\times3200$, $M=1\times10^{-6}$. The full MCS evolution of this grid can be found on GitHub (in the \texttt{cuda/5specie-spiral} directory).}
    \label{fig:3200spiral}
\end{figure}

By leveraging the CUDA-accelerated ESCG implementation and simply increasing the system size to a previously infeasible $N = 3200 \times 3200$, the resulting dynamics exhibit significant qualitative changes, as illustrated in \textbf{Figure~\ref{fig:3200spiral}}. At MCS = 2000, the system remains in a transient state, with localised interactions yet to fully propagate due to the expansive lattice. However, by MCS = 100{,}000, aesthetic and well pronounced spiral structures begin to emerge—visually distinct and spatially coherent. Remarkably, these complex patterns arise from a simple change in lattice dimensions, accentuating the pivotal role of spatial scale in capturing emergent phenomena. These results raise a compelling question: do all circulant-dominance $n$-species ESCGs ultimately give rise to spiral domains, with system size acting as the limiting factor? Such hypotheses, once computationally inaccessible, now become experimentally tractable through the high-throughput capabilities of the GPU-accelerated framework developed here.

\subsection{Park, Chen, Szolnoki}

\begin{figure}[h]
    \centering
    \includegraphics[width=0.5\linewidth]{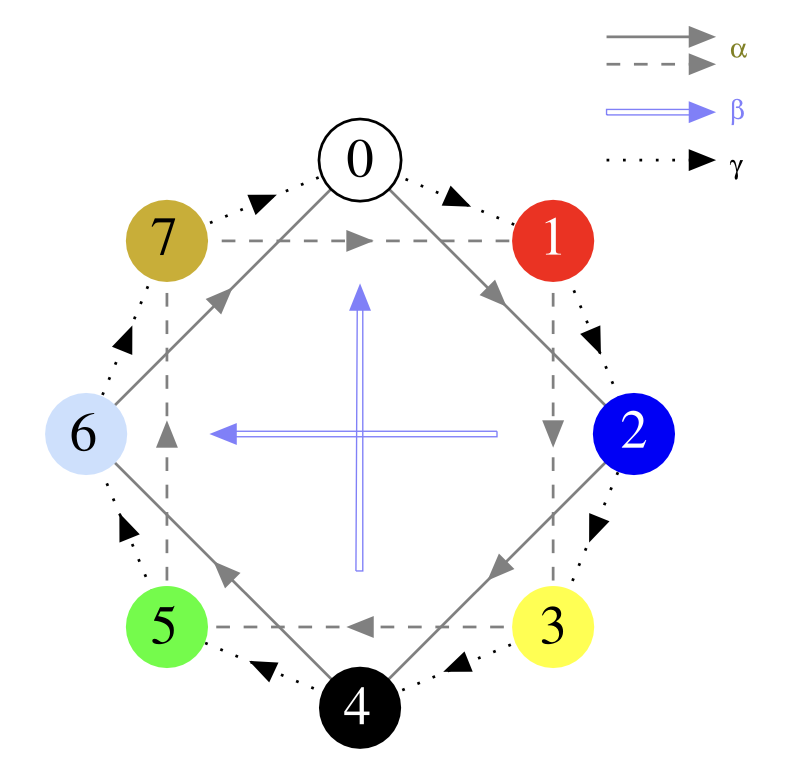}
    \caption{Dominance network structure for the eight-species ESCG proposed by Park, Chen, and Szolnoki, where $\alpha$, $\beta$, and $\gamma$ represent probabilistic rates of competitive interactions.}
    \label{fig:pcs-network}
\end{figure}

Park, Chen, and Szolnoki introduced a functionally distinct variant of the conventional ESCG model~\cite{Park2023CompetitionPopulation}. Unlike traditional ESCGs where species interact via deterministic win–lose relationships and exhibit mobility across the lattice, their model removes species mobility entirely and replaces binary dominance outcomes with probabilistic interaction rates. Specifically, each directed edge in the dominance network is annotated with a probability parameter $\alpha$, $\beta$, or $\gamma$, governing the likelihood of one species dominating another. This probabilistic approach introduces a tunable layer of stochasticity that enables a more nuanced study of species interactions and ecosystem resilience. The 8-species cyclic dominance structure used in their study is shown in \textbf{Figure \ref{fig:pcs-network}}.

While the Metal and CUDA implementations were initially designed for traditional ESCG models, small and straightforward modifications to the codebase enable the replication of more complex models, such as the one described by Park, Chen, and Szolnoki~\cite{Park2023CompetitionPopulation}. In their study, Park et al.\ propose an ESCG framework where mobility is entirely removed, and dominance interactions are governed by probabilistic rates $\alpha$, $\beta$, and $\gamma$, rather than deterministic win/lose rules. Their novel dominance network for an eight-species ecosystem is shown in \textbf{Figure~\ref{fig:pcs-network}}. \\

\begin{figure}[h]
    \centering
    \includegraphics[width=0.8\linewidth]{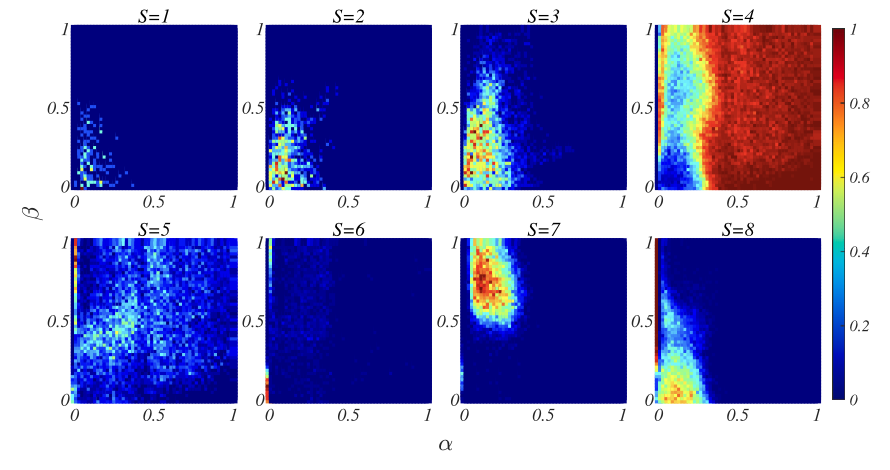}
    \caption{Original results from Park, Chen, and Szolnoki~\cite{Park2023CompetitionPopulation}, showing the probability of different numbers of surviving species for varying $(\alpha, \beta)$ combinations, with $\gamma=1$, for $L = 100$ and terminating after $L^{2}$ MCS.}
    \label{fig:pcs-fig4-og}
\end{figure}

\begin{figure}[h]
    \centering
    \includegraphics[width=0.8\linewidth]{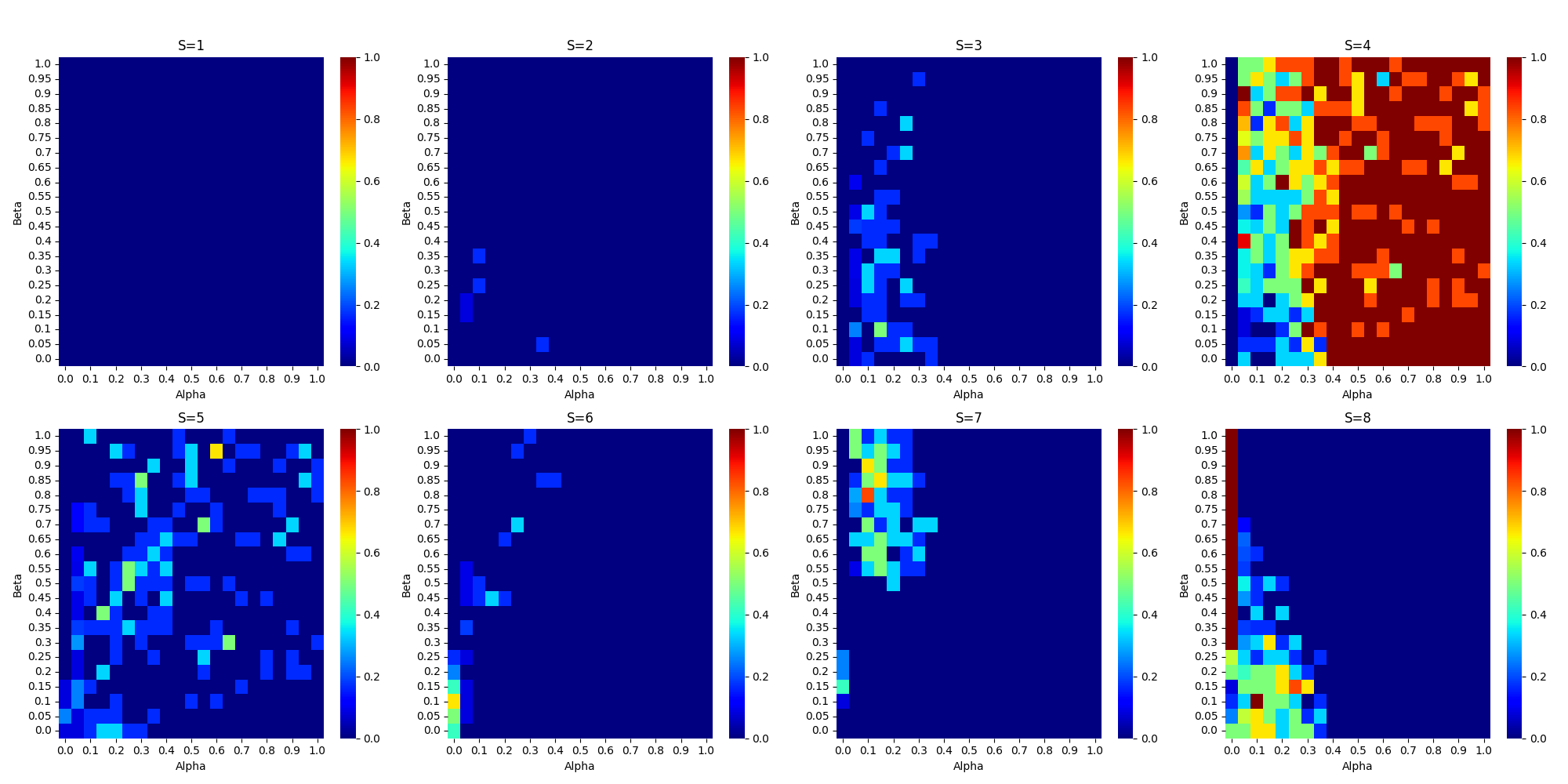}
    \caption{Replication of Park, Chen, and Szolnoki's study using the adapted CUDA-based ESCG framework, validating the probabilistic species survival distributions under varying $(\alpha, \beta)$ configurations with $\gamma=1$, for $L = 100$ and terminating after $L^{2}$ MCS.}
    \label{fig:pcs-replication}
\end{figure}

By extending the GPU-accelerated frameworks with minor adjustments to interaction logic, the simulations were adapted to replicate Park et al.'s results. Specifically, their investigation into the probability distribution of surviving species counts under varying $\alpha$ and $\beta$ parameters (with $\gamma=1$) was reproduced. \textbf{Figures~\ref{fig:pcs-fig4-og}} and \textbf{\ref{fig:pcs-replication}} compare Park et al.'s original findings with those obtained using the modified GPU-accelerated ESCG system.

Although the replication was produced at a lower resolution compared to the original, the key visual patterns are clearly preserved. Notably, the extinction zones and regions of species coexistence emerge in corresponding critical areas of the parameter space, closely matching those identified by Park et al. This strong visual alignment not only validates the adapted ESCG model but also reinforces the reliability of the GPU-accelerated system in accurately capturing complex stochastic dynamics. \\

From this point, the adapted ESCG implementation was further utilised to replicate additional findings from Park, Chen, and Szolnoki. While earlier replications aligned well with the original results, other experiments began to reveal discrepancies. These divergences are likely attributable to incomplete or ambiguous descriptions in the source paper—particularly concerning initialisation protocols, update rules, or the precise runtime parameters used in generating specific results. \\

\begin{figure}[h]
    \centering
    \includegraphics[width=0.5\linewidth]{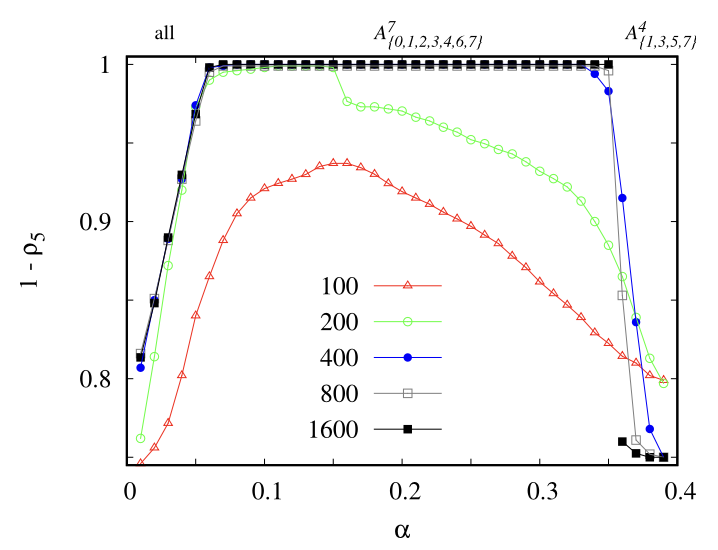}
    \caption{Original Figure 5 from Park, Chen, and Szolnoki~\cite{Park2023CompetitionPopulation}, depicting the survival probability of species 5 across $\alpha$-space with $\beta=0.75$ and $\gamma=1.0$.}
    \label{fig:pcs-fig5-og}
\end{figure}

\begin{figure}[h]
    \centering
    \includegraphics[width=0.7\linewidth]{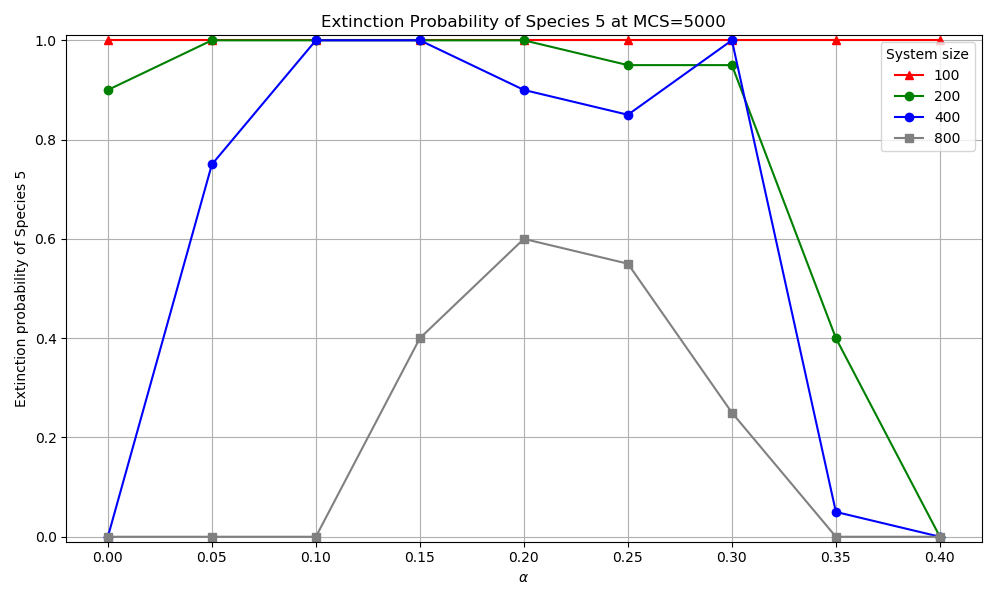}
    \caption{Replicated survival probability of species 5 across $\alpha$-space with $\beta=0.75$ and $\gamma=1.0$ after 5000 MCS, using the CUDA-accelerated ESCG system, currently over 20 IID trials.}
    \label{fig:pcs-fig5-5000}
\end{figure}

\begin{figure}[h]
    \centering
    \includegraphics[width=0.7\linewidth]{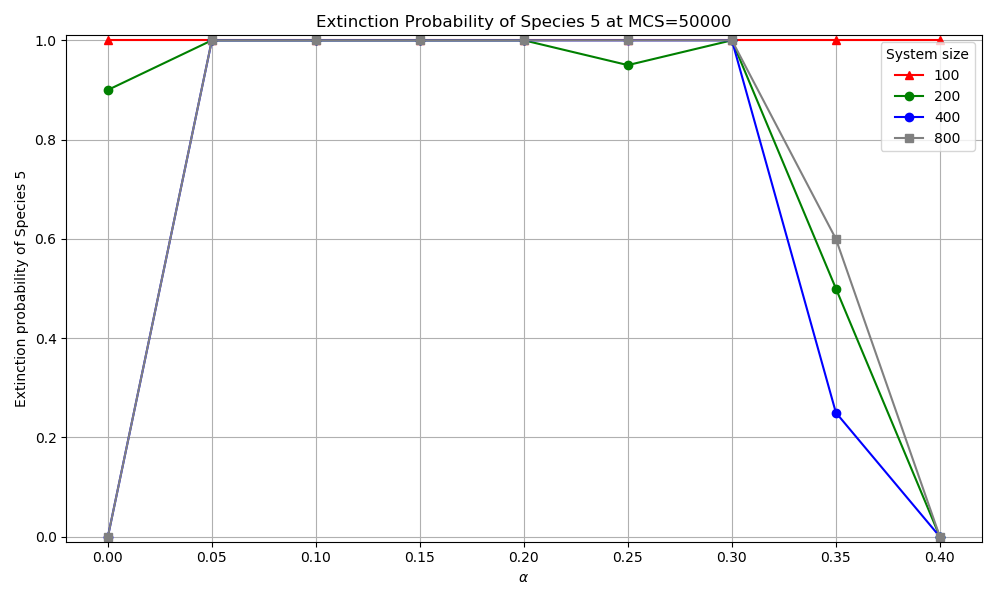}
    \caption{Replicated survival probability of species 5 across $\alpha$-space with $\beta=0.75$ and $\gamma=1.0$ after 50,000 MCS, using the CUDA-accelerated ESCG system, currently over 20 IID trials.}
    \label{fig:pcs-fig5-50000}
\end{figure}

One such discrepancy emerged when attempting to replicate Park et al.'s analysis of the survival probability of species 5, evaluated over a range of $\alpha$ values with $\beta =0.75$ and $\gamma = 1$. The original results, shown in \textbf{Figure~\ref{fig:pcs-fig5-og}}, present a well-defined survival landscape, with extinction and persistence regions clearly structured across the parameter space. However, when the same experimental conditions were applied using the adapted GPU-accelerated ESCG system, the results from my attempted replications shown in \textbf{Figures~\ref{fig:pcs-fig5-5000}} and \textbf{\ref{fig:pcs-fig5-50000}} displayed marked differences, both in survival probability values and spatial distributions across the grid. My replications were from runs ending at $5000$ and $50000$ MCS, respectively. \\

The precise cause of these discrepancies is difficult to determine, largely due to a lack of clarity in Park et al.'s description of their experimental methodology. In particular, they do not explicitly state the number of MCS used to generate the data shown in their Figure~5, and it appears likely that different system sizes ($L$) were simulated for differing numbers of MCS. Earlier in their paper, Park et al.\ state that ``...system size varied between $L=100$ to $L=3200$ and the necessary relaxation steps were between $10^{3}$ to $3\times10^{5}$ MCS,'' suggesting that smaller systems such as $L=100$ may have been run for as few as $1000$ MCS. Further inconsistencies arise in their reporting: within the main body of their text they note that ``the symbols are the average of many independent runs; as an example for $L=100$ we executed 2000 times,'' yet in the caption of their Figure~5 they state that ``the plots are the average of 100--2000 runs depending on the system size.'' These ambiguities make it difficult to precisely replicate their experimental conditions and may explain the observed differences in species survival patterns.

\begin{table}[h]
    \centering
    \caption{Standard deviation of extinction probabilities for Species 5 across system sizes and MCS values, currently over 20 IID trials.}
    \label{tab:std-extinction}
    \begin{tabular}{|c|c|c|c|c|}
        \hline
        \textbf{MCS} & \textbf{100} & \textbf{200} & \textbf{400} & \textbf{800} \\
        \hline
        0     & 0.000 & 0.000 & 0.000 & 0.000 \\
        1000  & 0.194 & 0.176 & 0.000 & 0.000 \\
        2000  & 0.083 & 0.425 & 0.071 & 0.000 \\
        3000  & 0.017 & 0.376 & 0.301 & 0.000 \\
        5000  & 0.000 & 0.356 & 0.458 & 0.256 \\
        7000  & 0.000 & 0.346 & 0.480 & 0.440 \\
        10000 & 0.000 & 0.346 & 0.477 & 0.522 \\
        25000 & 0.000 & 0.346 & 0.464 & 0.457 \\
        50000 & 0.000 & 0.346 & 0.464 & 0.436 \\
        \hline
    \end{tabular}
\end{table}

Furthermore, Park et al.\ acknowledge that their presentation of averaged extinction probabilities in Figure~5 masks the underlying multimodality present in their raw data. Specifically, they note that for $L=100$ and $\alpha=0.15$, the extinction probability of species 5 across independent runs was typically either 1.0 or 0.75, suggesting a bimodal distribution. However, their published figure only shows the mean value of 0.937, without reporting any accompanying measure of variability such as standard deviation, standard error, or confidence intervals. This omission obscures the true spread and uncertainty of the results. \\

In contrast, analysis of my replicated data, summarised in \textbf{Table~\ref{tab:std-extinction}}, reveals a different behaviour. While the number of trials per configuration was relatively small, at MCS $\geq5000$ all trials for $L=100$ resulted in an extinction probability of exactly 1.0 for species 5, yielding a standard deviation of zero. This indicates no multimodal behaviour in the replication at these timescales—contradicting the variability that Park et al.\ describe. While this discrepancy may partly arise from differences in sample size, initialisation, or simulation length, it highlights the necessity of reporting full statistical information when characterising highly stochastic systems such as ESCGs. \\

\textbf{Figure~\ref{fig:pcs-fig5-5000}} further illustrates that system size ($L$) proved to be a significant factor in replication fidelity. Experiments with a relatively small lattice size of $L=100$ often resulted in premature species extinction and unstable spatial domains, particularly in regimes of near-neutral dominance. These behaviours diverged from the original study, where more stable persistence dynamics were reported. However, when the lattice size was increased to $L=800$, these anomalies were considerably mitigated. The survival landscapes generated at $L=800$ bore closer resemblance to those in the original figure, suggesting that system size is a critical determinant of stability and should be clearly specified in future publications to enable accurate reproduction. Furthermore, when simulations were extended to MCS $=50,000$ (see \textbf{Figure~\ref{fig:pcs-fig5-50000}}), species 5 was seen to be extinct across all lattice sizes, a result that again contrasts with the findings reported by Park et al. \\

While broad qualitative similarities exist between the original and replicated results—such as identifiable regions of extinction and survival—the detailed probability gradients and phase boundaries do not align precisely. This reinforces the critical importance of comprehensive methodological transparency when publishing simulation-based ecological models. Without complete disclosure of simulation protocols, including system size, runtime, sampling methodology, and aggregation methods, accurate reproduction and fair evaluation of published findings become increasingly difficult. \\

Nonetheless, the ability to rapidly iterate and explore high-resolution parameter sweeps using GPU acceleration proved instrumental in diagnosing such divergences. The CUDA-based system enabled large-scale parallel experimentation, making it feasible to conduct robustness checks and explore model behaviours under slight perturbations of parameters or configurations. This flexibility reinforces the value of GPU-accelerated ESCG platforms in both validating existing findings and uncovering new insights in ecological game dynamics.


\chapter{Conclusion}
\label{chap:conclusion}

The outcomes of this work, with reference to the aims and objectives in \textbf{Section~\ref{motivation}}, can be summarised as follows:

\begin{enumerate}
    \item The development and successful validation of three distinct ESCG implementations—single-threaded C++, Metal, and CUDA—fulfilled the goal of creating a robust, cross-platform simulation infrastructure. By accurately reproducing well-established ecological dynamics, such as those presented by Reichenbach et al., each system was shown to be consistent with prior literature, thereby confirming the correctness and reliability of the underlying simulation logic across all platforms.
    \item The simulation framework was designed with flexibility and configurability at its core. Command-line parameterisation and modular architecture allow researchers to specify a wide range of ecological parameters, lattice configurations, and simulation behaviours. This adaptability ensures that the system is not only capable of supporting current experiments, but is also well-equipped to handle future extensions, such as higher-order dominance matrices, alternative update rules, and complex network structures.
    \item A comprehensive performance analysis demonstrated significant computational gains achieved through GPU acceleration. Benchmarking trials showed that the CUDA-based implementation achieved up to a $28.4\times$ speedup over the single-threaded baseline at \( L = 800 \), while also supporting system sizes up to \( L = 3200 \), which were otherwise computationally infeasible. These findings satisfy the third objective by quantitatively showcasing the improved scalability, throughput, and execution time enabled by parallel processing, especially when combined with techniques such as the \texttt{--maxStep} optimisation.
    \item The accelerated systems enabled the extension of existing ecological models to unexplored regimes. Reproducing and expanding on studies like those by Park et al. and Reichenbach et al., this work investigated higher-species systems, longer temporal dynamics, and extreme spatial domains. For example, increasing the system size to \( N = 3200 \times 3200 \) revealed emergent spiral formations that were not observable in smaller systems. Such extensions offer the potential to challenge and refine ecological hypotheses, opening new pathways for exploration and insight into spatial cyclic competition dynamics.
\end{enumerate}

\noindent
Overall, this project has successfully achieved its stated objectives by developing a high-performance, reliable, and extensible ESCG simulation platform. By bridging the gap between ecological theory and computational feasibility, it enables researchers to explore previously inaccessible regions of complex ecological dynamics. As outlined in \textbf{Chapter~\ref{chap:context}}, the outcomes of this work are expected to make significant contributions to the field through a forthcoming conference paper in the \emph{Proceedings of the 2025 European Modelling \& Simulation Symposium} and a journal article to be submitted to \emph{Chaos, Solitons \& Fractals}. Together, these publications will share the advances made in this project and further establish its impact within the scientific community. \\

As the field of computational ecological modelling continues to evolve, several improvements have emerged that could significantly enhance the functionality, performance, and scientific value of the systems developed in this dissertation. The following strategies are proposed as potential avenues for future exploration:

\begin{enumerate}
    \item \textbf{Parallel PRNG Optimisation:} Future work could incorporate robust parallel pseudorandom number generation strategies such as those described by Bradley et al.~\cite{Bradley2011ParallelizationGenerators}, to further speedup the parallel PRNG.

    \item \textbf{Integration of PCG:} The Permuted Congruential Generator (PCG) family of PRNGs~\cite{oneill2014pcg} offers compelling advantages over traditional generators, including improved statistical quality, reduced memory footprint, faster performance, and resistance to reverse engineering. Integrating PCG into the ESCG framework may yield both performance and reliability improvements.

    \item \textbf{Enhanced Memory Locality:} Optimising device-side memory allocation for random number buffers could improve cache coherence and reduce access latency. This includes persistent allocation of device-local memory for \texttt{maxStep} simulations and efficient circular buffering for single-step simulations.

    \item \textbf{Adaptive Thread Dispatching:} Implementing dynamic kernel dispatch strategies that tune thread and block configurations based on GPU capabilities could improve load balancing and throughput across a wider range of CUDA-enabled or Metal-compatible devices.

    \item \textbf{Expanded Visualisation Pipeline:} Integrating real-time or deferred GPU-based visualisation using technologies like Vulkan, Metal Performance Shaders, or CUDA-OpenGL interop could enhance interpretability and streamline experiment analysis.
\end{enumerate}


%
%
%

\backmatter

\bibliography{dissertation.bib}

\begin{thebibliography}{10}

\bibitem{Apple2023MetalStorage}
{Apple Developer Documentation}.
\newblock Choosing a resource storage mode for apple gpus, 2023.
\newblock Accessed: 2025-04-28, page 1.
\newblock URL: \url{https://developer.apple.com/documentation/metal/choosing-a-resource-storage-mode-for-apple-gpus?language=objc}.

\bibitem{appleMetalCPP}
{Apple Developer Documentation}.
\newblock Get started with metal-cpp, 2024.
\newblock page 1.
\newblock URL: \url{https://developer.apple.com/metal/cpp/}.

\bibitem{AppleInc.2023Target3}
{Apple Inc.}
\newblock {Target and optimize GPU binaries with Metal 3}, 2023.
\newblock URL: \url{https://developer.apple.com/videos/play/wwdc2022/10102/}.

\bibitem{Banger2013OpenCLExample}
Ravishekhar Banger and Koushik Bhattacharyya.
\newblock {\em {OpenCL Programming by Example}}.
\newblock Packt Publishing, 2013.

\bibitem{Bradley2011ParallelizationGenerators}
Thomas Bradley, Jacques du~Toit, Robert Tong, Mike Giles, and Paul Woodhams.
\newblock {Parallelization techniques for random number generators}.
\newblock In {\em GPU Computing Gems Emerald Edition}, pages 231--246. Elsevier Inc., 2011.
\newblock \href {https://doi.org/10.1016/B978-0-12-384988-5.00016-4} {\path{doi:10.1016/B978-0-12-384988-5.00016-4}}.

\bibitem{CliffGraphicalNetworks}
Dave Cliff.
\newblock {Graphical Abstract On Long-Term Species Coexistence in Five-Species Evolutionary Spatial Cyclic Games with Ablated and Non-Ablated Dominance Networks On Long-Term Species Coexistence in Five-Species Evolutionary Spatial Cyclic Games with Ablated and Non-Ablated Dominance Networks}.
\newblock Technical report.
\newblock Referenced: Section 5.

\bibitem{Cliff2024NeverGames}
Dave Cliff.
\newblock {Never Mind The No-Ops: Faster and Less Volatile Simulation Modelling of Co-Evolutionary Species Interactions via Spatial Cyclic Games}.
\newblock In {\em European Modeling and Simulation Symposium, EMSS}, volume 2024-September. Cal-Tek srl, 2024.
\newblock page 6.
\newblock \href {https://doi.org/10.46354/i3m.2024.emss.011} {\path{doi:10.46354/i3m.2024.emss.011}}.

\bibitem{Cliff2025OnNetworks}
Dave Cliff.
\newblock On long-term species coexistence in five-species evolutionary spatial cyclic games with ablated and non-ablated dominance networks.
\newblock {\em Chaos, Solitons and Fractals}, 190, 1 2025.
\newblock Referenced Section 5.
\newblock \href {https://doi.org/10.1016/j.chaos.2024.115702} {\path{doi:10.1016/j.chaos.2024.115702}}.

\bibitem{Cook2012CUDAGPUs}
Shane Cook.
\newblock {\em {CUDA Programming: A Developer's Guide to Parallel Computing with GPUs}}.
\newblock Morgan Kaufmann, 2012.

\bibitem{Igoshin2004WavesAggregation}
Oleg~A. Igoshin, Roy Welch, Dale Kaiser, and George Oster.
\newblock Waves and aggregation patterns in myxobacteria.
\newblock {\em Proceedings of the National Academy of Sciences (PNAS)}, 101(12):4256--4261, 2004.
\newblock Referenced pages 4259.
\newblock URL: \url{https://www.pnas.org/doi/10.1073/pnas.0400704101}, \href {https://doi.org/10.1073/pnas.0400704101} {\path{doi:10.1073/pnas.0400704101}}.

\bibitem{Laird2006CompetitiveCoexistence}
Robert~A Laird and Brandon~S Schamp.
\newblock {Competitive Intransitivity Promotes Species Coexistence}.
\newblock 168(2):182--193, 2006.
\newblock page 182.

\bibitem{matsumoto1998mersenne}
Makoto Matsumoto and Takuji Nishimura.
\newblock Mersenne twister: A 623-dimensionally equidistributed uniform pseudo-random number generator.
\newblock {\em ACM Transactions on Modeling and Computer Simulation (TOMACS)}, 8(1):3--30, 1998.
\newblock Referenced pages 8--10.

\bibitem{NvidiaCUDAZone}
{NVIDIA Corporation}.
\newblock Cuda zone, 2024.
\newblock Accessed: 2025-04-28. Referenced main page.
\newblock URL: \url{https://developer.nvidia.com/cuda-zone}.

\bibitem{oneill2014pcg}
Melissa~E. O'Neill.
\newblock Pcg: A family of simple fast space-efficient statistically good algorithms for random number generation.
\newblock Technical Report HMC-CS-2014-0905, Harvey Mudd College, Computer Science Department, September 2014.
\newblock Technical Report.
\newblock URL: \url{https://www.cs.hmc.edu}.

\bibitem{Park2023CompetitionPopulation}
Junpyo Park, Xiaojie Chen, and Attila Szolnoki.
\newblock {Competition of alliances in a cyclically dominant eight-species population}.
\newblock {\em Chaos, Solitons and Fractals}, 166, 1 2023.
\newblock Referenced lines 10--12, p.~23, page 3 Fig. 4, page 4 Fig. 5.
\newblock \href {https://doi.org/10.1016/j.chaos.2022.113004} {\path{doi:10.1016/j.chaos.2022.113004}}.

\bibitem{Reichenbach2007MobilityGames}
Tobias Reichenbach, Mauro Mobilia, and Erwin Frey.
\newblock {Mobility promotes and jeopardizes biodiversity in rock-paper-scissors games}.
\newblock {\em Nature}, 448(7157):1046--1049, 8 2007.
\newblock page 1046--1048.
\newblock \href {https://doi.org/10.1038/nature06095} {\path{doi:10.1038/nature06095}}.

\bibitem{Siegert1995SpiralMounds}
Florian Siegert and Cornelis~J. Weijer.
\newblock {Spiral and concentric waves organize multicellular Dictyostelium mounds}.
\newblock {\em Current Biology}, 5(8):937--943, 8 1995.
\newblock page 938.
\newblock \href {https://doi.org/10.1016/S0960-9822(95)00184-9} {\path{doi:10.1016/S0960-9822(95)00184-9}}.

\bibitem{Warren2020AppleYear}
Tom Warren.
\newblock Apple is switching macs to its own processors starting later this year.
\newblock {\em The Verge}, June 2020.
\newblock Accessed: 2025-04-28.
\newblock URL: \url{https://www.theverge.com/2020/6/22/21295475/apple-mac-processors-arm-silicon-chips-wwdc-2020}.

\bibitem{Watt1992AdvancedPractice}
Alan Watt and Mark Watt.
\newblock {\em Advanced Animation and Rendering Techniques: Theory and Practice}.
\newblock Addison-Wesley, Reading, Massachusetts, 1992.
\newblock Referenced Chapters 8 and 9.

\bibitem{Zhong2022SpeciesSpecies}
Linwu Zhong, Liming Zhang, Haihong Li, Qionglin Dai, and Junzhong Yang.
\newblock {Species coexistence in spatial cyclic game of five species}.
\newblock {\em Chaos, Solitons and Fractals}, 156, 3 2022.
\newblock pages 1--3, Fig. 2.
\newblock \href {https://doi.org/10.1016/j.chaos.2022.111806} {\path{doi:10.1016/j.chaos.2022.111806}}.

\end{thebibliography}



\appendix

\chapter{Appendix A: AI Prompts/Tools}
\label{appx:ai_prompt}

\begin{itemize}
    \item I used ChatGPT 4o to learn and understand how to use basic LaTeX syntax and code. Example prompts are:
    \begin{itemize}
        \item How to make an algorithm in LaTeX
        \item How to add/increase line spacing in LaTeX
        \item How to add mathematical notation to this equation
        \item How to increase the gap between subfigures
        \item How to position figures
    \end{itemize}

    \item I used ChatGPT for synonyms to maintain a professional, academic tone. Example usages are: 
    \begin{itemize}
        \item provide a more academic synonym for the word large in this sentence, [sentence].
        \item what would a suitable subsection heading be for these paragraphs? [paragraphs]
    \end{itemize}

    \item I used ChatGPT 4o to summarise web pages into a Mendeley reference. Example usage is - Create a Mendeley reference for this webpage \url{https://developer.Nvidia.com/cuda-zone}

    \item I used ChatGPT 4o to help with visualisation code, example usage is - I am using matplotlib-cpp...
    \begin{itemize}
        \item how do I make a logarithmic graph?
        \item how can I keep my colours consistent across snapshots?
        \item how can I make the font size for axis larger?
        \item I have a .csv file with these headings, how can I create a visualisation of [x].
    \end{itemize}

    \item I used ChatGPT 4o as a debugging tool, when I did not understand error messages or needed guidance, example usages are: 
    \begin{itemize}
        \item CUDA kernel launch failed - what causes this to happen? 
        \item I have the numpy package installed but it isn't linking to my project, why? 
    \end{itemize}

    \item I used ChatGPT 4o to help find functions for specific use cases, to expedite learning the language and speed up reading documentation, example usages are:
    \begin{itemize}
        \item what is the CUDA equivalent to \texttt{commandBuffer -> waitUntilCompleted()} in metal?
        \item how can I parse command line arguments in C++
        \item what library can generate random numbers in C++
    \end{itemize}

    \item I used ChatGPT 4o to explore ideas on how to make a PRNG more random, example usage was - what techniques can I apply to make a random number generator more random ?
\end{itemize}

\chapter{Paper in Preparation for the \emph{European Modelling \& Simulation Symposium}}
\label{emss-paper}

\begin{center}
\textbf{\Large GPU-Based Simulation of Evolutionary Spatial Cyclic Games: Nvidia vs Apple Silicon}

\vspace{0.5cm}

\textit{L. Sinadjan and D. Cliff}
\end{center}

\section*{Abstract}

For almost 20 years, physicists, biologists, applied mathematicians, and complex adaptive systems researchers have studied Evolutionary Spatial Cyclic Games (ESCGs), a class of minimal highly nonlinear agent-based models (ABMs) of multiple biological species interacting within an ecosystem. Simulations of ESCGs are highly compute-intensive and inherently parallelisable, yet the existing research literature contains (as far as we are aware) no publications that systematically exploit modern graphical processing units (GPUs) for performance acceleration. \\

The novel contribution of this paper is the design and comparison of two GPU-accelerated ESCG implementations: one targeted at the GPUs in Apple’s M1 Pro (M-Series Apple Silicon), and the other at Nvidia’s RTX A2000. We present results from a series of performance evaluation tests demonstrating that both GPU implementations offer speedups of $10\times$ or more compared to a traditional single-threaded serial implementation. Notably, the CUDA-based Nvidia implementation outperformed the Metal-based Apple implementation, achieving up to a $28.4\times$ speedup over the serial baseline, compared to more modest gains on Apple Silicon. These findings showcase the effectiveness of GPU acceleration for ESCGs and highlight the importance of hardware choice for scaling complex ecological simulations. All source code developed for this paper is made freely available on GitHub. \\

\section*{Keywords}

Agent-Based Models; Evolutionary Spatial Cyclic Games; GPU Acceleration; CUDA; Metal; Nvidia RTX; Apple M-Series.

\chapter{Paper in Preparation for \emph{Chaos, Solitons \& Fractals}}
\label{csf-paper}

\begin{center}
\textbf{\Large Mobility Matters in a Cyclically Dominant Eight-Species Model of Competing Alliances} 

\vspace{0.5cm}

\textit{D. Cliff and L. Sinadjan}
\end{center}

\section*{Abstract}

We replicate results from simulation studies of an agent-based model involving eight-species evolutionary spatial cyclic games that were presented in a 2023 paper by Park, Chen, and Szolnoki (\textit{Chaos, Solitons and Fractals}, 166: 113004), and then go on to demonstrate that all the results in that paper are consequences of an implicit simplification in their model which, if altered to increase the model’s realism, totally changes every result identified as significant by Park et al., thereby rendering obsolete all of their observations. \\

Park et al. reported results from simulations in which the ``food web'' species-interaction network had two separate four-species cyclic ``alliances'', that could potentially compete via a Rock-Paper-Scissors-style intransitive dominance network in which there were three key parameters: $\gamma$, the probability of dominance in an eight-species Lotka-Volterra ring; $\alpha$, the probability of dominance of another species in the same four-species alliance; and $\beta$, the probability of additional ``symmetry-breaking'' dominance between species in only one of the two alliances. \\

A primary contribution of Park et al.'s paper is the plotting of phase diagrams of the $\beta$--$\alpha$ parameter plane observed at $\gamma=0.5$ and at $\gamma=1.0$, derived from heat maps for species survival probabilities estimated from multiple simulation runs. Here, we empirically replicate those phase diagrams and then demonstrate that they totally change when the realism of the model is increased by allowing the individual agents some nonzero degree of mobility. \\

\section*{Keywords}

Biodiversity, Cyclic Competition, Asymmetric Interaction, Species Coexistence, Evolutionary Spatial Games, Rock-Paper-Scissors, Design of Experiments, Replication.

\chapter{Source Code}

The full source code of each implementation of ESCGs developed for this dissertation can be found at: \url{https://github.com/louiesinadjan/escg}


\end{document}